\documentclass[12pt]{article}

\usepackage{axodraw}
\usepackage{cite}

\makeatletter

\def\paragraph{\@startsection{paragraph}{4}{\z@}{+2.00ex plus
 +1ex minus +.2ex}{1.5ex plus .2ex}{\it\normalsize}}

\def\section{\@startsection {section}{1}{\z@}{+3.0ex plus +1ex minus
  +.2ex}{2.3ex plus .2ex}{\normalsize\bf\boldmath}}
\def\subsection{\@startsection{subsection}{2}{\z@}{+2.5ex plus +1ex
minus +.2ex}{1.5ex plus .2ex}{\normalsize\bf\boldmath}}
\def\subsubsection{\@startsection{subsubsection}{3}{\z@}{+2.5ex plus
 +1ex minus +.2ex}{1.5ex plus .2ex}{\normalsize\it}}

\expandafter\ifx\csname mathrm\endcsname\relax\def\mathrm#1{{\rm #1}}\fi

\renewcommand{\theequation}{\thesection.\arabic{equation}}
\newcounter{saveeqn}

\@addtoreset{equation}{section}

\newcommand{\ks}{k\hspace{-0.52em}/\hspace{0.1em}}
\newcommand{\lslash}{l\hspace{-0.42em}/\hspace{0.1em}}
\newcommand{\ps}{p\hspace{-0.42em}/}
\newcommand{\qs}{q\hspace{-0.48em}/}

\def\beq{\begin{equation}}
\def\eeq{\end{equation}}
\def\beqar{\begin{eqnarray}}
\def\eeqar{\end{eqnarray}}
\def\barr#1{\begin{array}{#1}}
\def\earr{\end{array}}
\def\bfi{\begin{figure}}
\def\efi{\end{figure}}
\def\btab{\begin{table}}
\def\etab{\end{table}}
\def\bce{\begin{center}}
\def\ece{\end{center}}
\def\nn{\nonumber}

\def\text{\textstyle}

\def\al{\alpha}
\def\be{\beta}
\def\ga{\gamma}
\def\de{\delta}
\def\teps{\varepsilon}
\def\veps{\epsilon}

\def\si{\sigma}
\def\om{\omega}

\def\De{\Delta}

\def\refeq#1{\mbox{(\ref{#1})}}

\def\refse#1{\mbox{Sect.~\ref{#1}}}
\def\refses#1{\mbox{Sects.~\ref{#1}}}
\def\refapp#1{\mbox{App.~\ref{#1}}}
\def\refapps#1{\mbox{Apps.~\ref{#1}}}
\def\citere#1{\mbox{Ref.~\cite{#1}}}
\def\citeres#1{\mbox{Refs.~\cite{#1}}}

\def\solid{\raise.9mm\hbox{\protect\rule{1.1cm}{.2mm}}}
\def\dash{\raise.9mm\hbox{\protect\rule{2mm}{.2mm}}\hspace*{1mm}}

\def\nl{\nonumber\\}

\def\nln{\nonumber\\*[-1ex]\phantom{\fbox{\rule{0em}{2ex}}}}

\newcommand{\dsl}[1]{\not \hspace{-0.7mm}#1}
\def\dsl{\mathpalette\make@slash}
\def\make@slash#1#2{\setbox\z@\hbox{$#1#2$}%
  \hbox to 0pt{\hss$#1/$\hss\kern-\wd0}\box0}

\newcommand{\order}{\mathcal{O}}

\def\asymp#1%
{\mathrel{\raisebox{-.4em}{$\widetilde{\scriptstyle #1}$}}}

\def\Nequal#1%
{\mathrel{\raisebox{-.5em}{$\stackrel{=}{\scriptstyle\rm#1}$}}}

\newcommand{\LLA}{\mathrel{\stackrel{\mathrm{LL}}{=}}}
\newcommand{\NLLA}{\mathrel{\stackrel{\mathrm{NLL}}{=}}}

\def\sgn{\mathop{\mathrm{sgn}}\nolimits}

\newcommand{\Tr}{\mathop{\mathrm{Tr}}\nolimits}

\newcommand{\GeV}{\unskip\,\mathrm{GeV}}

\newcommand{\TeV}{\unskip\,\mathrm{TeV}}

\def\mathswitchr#1{\relax\ifmmode{\mathrm{#1}}\else$\mathrm{#1}$\fi}

\newcommand{\PW}{\mathswitchr W}
\newcommand{\PZ}{\mathswitchr Z}

\newcommand{\PH}{\mathswitchr H}

\newcommand{\Pd}{\mathswitchr d}

\newcommand{\Pu}{\mathswitchr u}
\newcommand{\Pp}{\mathswitchr p}

\newcommand{\Pt}{\mathswitchr t}

\newcommand{\PWpm}{\mathswitchr {W^\pm}}

\newcommand{\FA}{A}
\newcommand{\FZ}{Z}
\newcommand{\FW}{W}
\newcommand{\FWpm}{{W^\pm}}

\def\mathswitch#1{\relax\ifmmode#1\else$#1$\fi}

\newcommand{\MW}{\mathswitch {M_\PW}}

\newcommand{\MA}{\mathswitch {M_A}}
\newcommand{\MZ}{\mathswitch {M_\PZ}}
\newcommand{\MH}{\mathswitch {M_\PH}}

\newcommand{\Mt}{\mathswitch {m_\Pt}}

\newcommand{\Mphipm}{M_{\phi^\pm}}%

\newcommand{\scrs}{\scriptscriptstyle}
\newcommand{\sw}{\mathswitch {s_{\scrs\PW}}}
\newcommand{\cw}{\mathswitch {c_{\scrs\PW}}}

\newcommand{\muD}{\mu_{\mathrm{D}}}
\newcommand{\muR}{\mu_\mathrm{R}}

\newcommand{\betacoeff}[1]{b_{#1}}

\newcommand{\vev}{v} 

\def\ie{i.e.\ }

\newcommand{\etal}{{\it et al.}}

\newcommand{\rd}{\mathrm{d}}

\newcommand{\ri}{\mathrm{i}}

\newcommand{\rF}{\mathrm{F}}
\newcommand{\rL}{\mathrm{L}}
\newcommand{\rR}{\mathrm{R}}

\newcommand{\rT}{{\mathrm{T}}}

\newcommand{\SU}{\mathrm{SU}}
\newcommand{\U}{\mathrm{U}}

\newcommand{\fact}{\mathrm{F}}
\newcommand{\nfact}{\mathrm{NF}}
\newcommand{\elm}{\mathrm{em}}
\newcommand{\sew}{\mathrm{sew}}
\newcommand{\QED}{{\mathrm{QED}}}

\newcommand{\M}{{\cal {M}}}

\newcommand{\univfact}[2]{I(#1,#2)}
\newcommand{\FF}[2]{F_{#1}^{#2}}
\newcommand{\GG}[2]{G_{#1}^{#2}}
\newcommand{\Ff}[2]{f_{#1}^{#2}}
\newcommand{\Gg}[2]{g_{#1}^{#2}}
\newcommand{\DD}[1]{D_{#1}}
\newcommand{\deDD}[1]{\Delta D_{#1}}
\let \DDsub \DD 
\let \deDDsub \deDD
\newcommand{\DDUV}[1]{\DD{#1}^{\mathrm{UV}}}
\newcommand{\gb}{g_1}
\newcommand{\gw}{g_2}

\newcommand{\mel}[2]{\M_{#1}^{#2}}
\newcommand{\nmel}[2]{\tilde{\M}_{#1}^{#2}}
\newcommand{\alphaeps}{\alpha_{\veps}}
\newcommand{\alphaseps}{\alpha_{\mathrm{S},\veps}}

\newcommand{\eqdiagl}{&=&\hspace{-3mm}}
\newcommand{\eqdiagr}{\hspace{-1mm}=}

\newcommand{\lr}[1]{l_{#1}} 
\newcommand{\LMZW}{\lr{\mathrm{Z}}}
\newcommand{\LmuR}{\lr{\muR}}
\newcommand{\LrMI}{\lr{1}}
\newcommand{\LrMII}{\lr{2}}
\newcommand{\LrMIII}{\lr{3}}
\newcommand{\Lrij}{\lr{ij}}
\newcommand{\Lrik}{\lr{ik}}

\def\draftdate{\relax}
\def\mda{\relax}
\def\mua{\relax}
\def\mla{\relax}
\def\draft{
\def\thtystars{******************************}
\def\sixtystars{\thtystars\thtystars}
\typeout{}
\typeout{\sixtystars**}
\typeout{* Draft mode!
         For final version remove \protect\draft\space in source file *}
\typeout{\sixtystars**}
\typeout{}
\def\draftdate{\today}
\def\mua{\marginpar[\boldmath\hfil$\uparrow$]%
                   {\boldmath$\uparrow$\hfil}%
                    \typeout{marginpar: $\uparrow$}\ignorespaces}
\def\mda{\marginpar[\boldmath\hfil$\downarrow$]%
                   {\boldmath$\downarrow$\hfil}%
                    \typeout{marginpar: $\downarrow$}\ignorespaces}
\def\mla{\marginpar[\boldmath\hfil$\rightarrow$]%
                   {\boldmath$\leftarrow $\hfil}%
                    \typeout{marginpar: $\leftrightarrow$}\ignorespaces}
\def\Mua{\marginpar[\boldmath\hfil$\Uparrow$]%
                   {\boldmath$\Uparrow$\hfil}%
                    \typeout{marginpar: $\Uparrow$}\ignorespaces}
\def\Mda{\marginpar[\boldmath\hfil$\Downarrow$]%
                   {\boldmath$\Downarrow$\hfil}%
                    \typeout{marginpar: $\Downarrow$}\ignorespaces}
\def\Mla{\marginpar[\boldmath\hfil$\Rightarrow$]%
                   {\boldmath$\Leftarrow $\hfil}%
                    \typeout{marginpar: $\Leftrightarrow$}\ignorespaces}

\overfullrule 5pt
\oddsidemargin -15mm
\marginparwidth 29mm
}

\makeatletter

\def\eqnarray{\stepcounter{equation}\let\@currentlabel=\theequation
\global\@eqnswtrue
\global\@eqcnt\z@\tabskip\@centering\let\\=\@eqncr
$$\halign to \displaywidth\bgroup\hskip\@centering
  $\displaystyle\tabskip\z@{##}$\@eqnsel&\global\@eqcnt\@ne
  \hskip 2\arraycolsep \hfil${##}$\hfil
  &\global\@eqcnt\tw@ \hskip 2\arraycolsep $\displaystyle\tabskip\z@{##}$\hfil
   \tabskip\@centering&\llap{##}\tabskip\z@\cr}

\makeatother

\begin{document}

\newcommand{\leg}[1]{\scriptstyle{#1}}

\newcommand{\wblob}{
\Vertex(-15.9138,3.75675){0.8}
\Vertex(-16.3512,-0.00758122){0.8}
\Vertex(-15.9138,-3.75675){0.8}
\Line(0.,0.)(-19.5,-9.75)
\Line(0.,0.)(-19.5,9.75)
\GCirc(0.,0.){10.9008}{1}
}
\newcommand{\blob}{
\Vertex(-15.9138,3.75675){0.8}
\Vertex(-16.3512,-0.00758122){0.8}
\Vertex(-15.9138,-3.75675){0.8}
\Line(0.,0.)(-19.5,-9.75)
\Line(0.,0.)(-19.5,9.75)
\GCirc(0.,0.){10.9008}{0.5}
}
\newcommand{\factblob}{\wblob \Text(0,0)[]{\scriptsize F}}
\newcommand{\nfactblob}{\wblob \Text(0,0)[]{\scriptsize N}}

\newcommand{\diaggeneric}{
\begin{picture}(120.,104.)(-28.,-52.)
\Gluon(-53.6656,0.)(0.,0.){3.2}{5}
\Line(0.,0.)(24.,12.)
\Line(24.,12.)(48.,24.)
\Line(24.,-12.)(0.,0.)
\Line(48.,-24.)(24.,-12.)
\GCirc(4.47214,0.){17}{0.8}
\Text(84.,42.)[l]{$\leg{i}$}
\Text(84.,-42.)[l]{$\leg{j}$}
\Text(27.668,14.6453)[br]{}
\Text(28.,-14.)[tr]{}
\end{picture}
}

\newcommand{\diagone}[4]{
\begin{picture}(120.,104.)(-28.,-52.)
\Line(0.,0.)(48.,24.)
\Line(48.,24.)(80.,40.)
\Line(48.,-24.)(0.,0.)
\Line(80.,-40.)(48.,-24.)
\Photon(48.,24.)(48.,-24.){2.4}{3.5}
\Vertex(48.,24.){2}
\Vertex(48.,-24.){2}
\Text(84.,42.)[l]{#1}
\Text(84.,-42.)[l]{#2}
\Text(53.6656,0.)[l]{#3}
\Text(27.668,14.6453)[br]{}
\Text(28.,-14.)[tr]{}
{#4}
\end{picture}
}

\newcommand{\diagnewb}[1]{
\begin{picture}(120.,104.)(-28.,-52.)
\Line(0.,0.)(30.,15.)
\Line(30.,-15.)(0.,0.)
{#1}
\end{picture}
}

\newcommand{\diagnewc}[1]{
\begin{picture}(120.,104.)(-28.,-52.)
\Line(0.,0.)(48.,24.)
\Line(48.,24.)(80.,40.)
\Line(48.,-24.)(0.,0.)
\Line(80.,-40.)(48.,-24.)
\Photon(48.,24.)(48.,-24.){2.4}{3.5}
\Vertex(48.,24.){2}
\Vertex(48.,-24.){2}
\Text(84.,42.)[l]{$\leg{i}$}
\Text(84.,-42.)[l]{$\leg{j}$}
\Text(53.6656,0.)[l]{$\scriptscriptstyle{V_1}$}
\Text(27.668,14.6453)[br]{}
\Text(28.,-14.)[tr]{}
{#1}
\end{picture}
}
\newcommand{\diagnewd}[1]{
\begin{picture}(120.,104.)(-28.,-52.)
\Line(0.,0.)(80,0)
\PhotonArc(25.,-8.)(26.2488,17.7447,162.255){2}{5}
\Vertex(50.,0.){2}
\Text(84.,0)[l]{$\leg{i}$}
\Text(23,28.)[l]{{$\scriptstyle{V_\mu}$}}
\Text(27.668,14.6453)[br]{}
\Text(28.,-14.)[tr]{}
{#1}
\end{picture}
}

\newcommand{\diagnewe}[1]{
\begin{picture}(120.,104.)(-28.,-52.)
\Line(0.,0.)(80,0)
\Line(0.,0.)(50,50)
\PhotonArc(0,0)(48,0,45){2}{4}
\Vertex(48.,0.){2}
\Vertex(33.941,33.941){2}
\Text(84.,0)[l]{$\leg{i}$}
\Text(53.,53.)[l]{$\leg{j}$}
\Text(50,22.)[l]{{$\scriptstyle{V_\mu}$}}
\Text(27.668,14.6453)[br]{}
\Text(28.,-14.)[tr]{}
{#1}
\end{picture}
}

\newcommand{\diagoned}[1]{
\begin{picture}(120.,104.)(-28.,-52.)
\Line(0.,0.)(50,0)
\PhotonArc(50.,-50.)(70.71,90.,135.){2}{5}
\Text(54.,0)[l]{$\leg{i}$}
\Text(54,22.)[l]{{$\scriptstyle{V_\mu}$}}
\Text(27.668,14.6453)[br]{}
\Text(28.,-14.)[tr]{}
{#1}
\end{picture}
}

\newcommand{\diagonee}[1]{
\begin{picture}(120.,104.)(-28.,-52.)
\Line(0.,0.)(60,0)
\Line(0.,0.)(50,50)
\Vertex(25,25){2}
\Photon(25,25)(60,25){2}{3}
\Text(64.,0)[l]{$\leg{i}$}
\Text(53.,53.)[l]{$\leg{j}$}
\Text(64,22.)[l]{{$\scriptstyle{V_\mu}$}}
\Text(27.668,14.6453)[br]{}
\Text(28.,-14.)[tr]{}
{#1}
\end{picture}
}

\newcommand{\diagI}[5]{
\begin{picture}(120.,104.)(-28.,-52.)
\Line(0.,0.)(28.,14.)
\Line(28.,14.)(56.,28.)
\Line(56.,28.)(80.,40.)
\Line(28.,-14.)(0.,0.)
\Line(56.,-28.)(28.,-14.)
\Line(80.,-40.)(56.,-28.)
\Photon(56.,28.)(56.,-28.){2.4}{3.5}
\Photon(28.,14.)(28.,-14.){2.4}{2}
\Vertex(56.,28.){2}
\Vertex(56.,-28.){2}
\Vertex(28.,14.){2}
\Vertex(28.,-14.){2}
\Text(84.,42.)[l]{#1}
\Text(84.,-42.)[l]{#2}
\Text(62.6099,0.)[l]{#3}
\Text(31.305,0.)[l]{#4}
\Text(41.5019,21.9679)[br]{}
\Text(42.,-21.)[tr]{}
\Text(19.515,10.9162)[br]{}
\Text(20.,-10.)[tr]{}
{#5}
\end{picture}
}

\newcommand{\diagII}[5]{
\begin{picture}(120.,104.)(-28.,-52.)
\Line(0.,0.)(28.,14.)
\Line(28.,14.)(56.,28.)
\Line(56.,28.)(80.,40.)
\Line(28.,-14.)(0.,0.)
\Line(56.,-28.)(28.,-14.)
\Line(80.,-40.)(56.,-28.)
\Photon(56.,28.)(28.,-14.){-2.4}{3}
\Photon(56.,-28.)(28.,14.){-2.4}{3}
\Vertex(56.,28.){2}
\Vertex(56.,-28.){2}
\Vertex(28.,14.){2}
\Vertex(28.,-14.){2}
\Text(84.,42.)[l]{#1}
\Text(84.,-42.)[l]{#2}
\Text(48.3499,8.07117)[l]{#3}
\Text(48.3499,-9.07117)[l]{#4}
\Text(41.5019,21.9679)[br]{}
\Text(42.,-21.)[tr]{}
\Text(19.515,10.9162)[br]{}
\Text(20.,-10.)[tr]{}
{#5}
\end{picture}
}

\newcommand{\diagIII}[6]{
\begin{picture}(120.,104.)(-28.,-52.)
\Line(0.,0.)(24.,12.)
\Line(24.,12.)(56.,28.)
\Line(56.,28.)(80.,40.)
\Line(49.1935,-24.5967)(0.,0.)
\Line(80.,-40.)(49.1935,-24.5967)
\Photon(56.,28.)(49.1935,0.){2.4}{2}
\Photon(24.,12.)(49.1935,0.){-2.4}{2}
\Photon(49.1935,0.)(49.1935,-24.5967){2.4}{2}
\Vertex(56.,28.){2}
\Vertex(24.,12.){2}
\Vertex(49.1935,0.){2}
\Vertex(49.1935,-24.5967){2}
\Text(84.,42.)[l]{#1}
\Text(84.,-42.)[l]{#2}
\Text(56.5825,13.3573)[l]{#3}
\Text(56.5825,-13.3573)[l]{#4}
\Text(35.307,5.78034)[tr]{#5}
\Text(41.5019,21.9679)[br]{}
\Text(32.,-16.)[tr]{}
\Text(19.515,10.9162)[br]{}
{#6}
\end{picture}
}

\newcommand{\diagV}[5]{
\begin{picture}(120.,104.)(-28.,-52.)
\Line(0.,0.)(16.,8.)
\Line(16.,8.)(44.,22.)
\Line(44.,22.)(64.,32.)
\Line(64.,32.)(80.,40.)
\Line(64.,-32.)(0.,0.)
\Line(80.,-40.)(64.,-32.)
\Photon(64.,32.)(64.,-32.){2.4}{3.5}
\PhotonArc(30.,15.)(15.6525,26.5651,206.565){2}{4}
\Vertex(64.,-32.){2}
\Vertex(16.,8.){2}
\Vertex(44.,22.){2}
\Vertex(64.,32.){2}
\Text(84.,42.)[l]{#1}
\Text(84.,-42.)[l]{#2}
\Text(71.5542,0.)[l]{#3}
\Text(23.8885,35.1332)[br]{#4}
\Text(41.5019,21.9679)[br]{}
\Text(42.,-21.)[tr]{}
\Text(19.515,10.9162)[br]{}
\Text(20.,-10.)[tr]{}
{#5}
\end{picture}
}

\newcommand{\diagVII}[5]{
\begin{picture}(120.,104.)(-28.,-52.)
\Line(0.,0.)(24.,12.)
\Line(24.,12.)(40.,20.)
\Line(40.,20.)(56.,28.)
\Line(56.,28.)(80.,40.)
\Line(40.,-20.)(0.,0.)
\Line(80.,-40.)(40.,-20.)
\Photon(40.,20.)(40.,-20.){2.4}{3.5}
\PhotonArc(40.,20.)(17.8885,26.5651,206.565){2}{4}
\Vertex(40.,-20.){2}
\Vertex(24.,12.){2}
\Vertex(40.,20.){2}
\Vertex(56.,28.){2}
\Text(84.,42.)[l]{#1}
\Text(84.,-42.)[l]{#2}
\Text(49.1935,0.)[l]{#3}
\Text(32.1994,42.9325)[br]{#4}
\Text(41.5019,21.9679)[br]{}
\Text(42.,-21.)[tr]{}
\Text(19.515,10.9162)[br]{}
\Text(20.,-10.)[tr]{}
{#5}
\end{picture}
}

\newcommand{\diagIX}[7]{
\begin{picture}(120.,104.)(-28.,-52.)
\Line(0.,0.)(56.,28.)
\Line(56.,28.)(80.,40.)
\Line(56.,-28.)(0.,0.)
\Line(80.,-40.)(56.,-28.)
\Photon(56.,28.)(56.,10.5064){2.4}{2.5}
\Photon(56.,-10.5064)(56.,-28.){2.4}{2.5}
\ArrowArc(56.9771,0.)(11.7466,-90.,90.)
\ArrowArc(56.9771,0.)(11.7466,90.,270.)
\Vertex(56.,-28.){2}
\Vertex(56.,28.){2}
\Vertex(56.,10.5064){2}
\Vertex(56.,-10.5064){2}
\Text(84.,42.)[l]{#1}
\Text(84.,-42.)[l]{#2}
\Text(64.4276,21.6767)[lu]{#3}
\Text(63.3537,-24.6395)[lb]{#4}
\Text(74.2375,0.)[l]{#5}
\Text(41.1437,0.)[r]{#6}
\Text(41.5019,21.9679)[br]{}
\Text(42.,-21.)[tr]{}
\Text(19.515,10.9162)[br]{}
\Text(20.,-10.)[tr]{}
{#7}
\end{picture}
}

\newcommand{\diagself}[3]{
\begin{picture}(120.,104.)(-28.,-52.)
\Line(0.,0.)(56.,28.)
\Line(56.,28.)(80.,40.)
\Line(56.,-28.)(0.,0.)
\Line(80.,-40.)(56.,-28.)
\Photon(56.,28.)(56.,9){2.4}{2.5}
\Photon(56.,-9)(56.,-28.){2.4}{2.5}
\Vertex(56.,-28.){2}
\Vertex(56.,28.){2}
\Text(84.,42.)[l]{$\leg{i}$}
\Text(84.,-42.)[l]{$\leg{j}$}
\Text(64.4276,21.6767)[lu]{#1}
\Text(63.3537,-24.6395)[lb]{#2}
\Text(41.5019,21.9679)[br]{}
\Text(42.,-21.)[tr]{}
\Text(19.515,10.9162)[br]{}
\Text(20.,-10.)[tr]{}
\GCirc(56,0.){9}{0.9}
{#3}
\end{picture}
}

\newcommand{\diagX}[1]{
\begin{picture}(120.,104.)(-28.,-52.)
\Line(0.,0.)(56.,28.)
\Line(56.,28.)(80.,40.)
\Line(56.,-28.)(0.,0.)
\Line(80.,-40.)(56.,-28.)
\Photon(56.,28.)(56.,10.5064){2.4}{2.5}
\Photon(56.,-10.5064)(56.,-28.){2.4}{2.5}
\PhotonArc(56.9771,0.)(11.7466,-101.459,258.541){-2}{8}
\Vertex(56.,-28.){2}
\Vertex(56.,28.){2}
\Vertex(56.,10.5064){2}
\Vertex(56.,-10.5064){2}
\Text(84.,42.)[l]{$\leg{i}$}
\Text(84.,-42.)[l]{$\leg{j}$}
\Text(64.4276,21.6767)[lu]{$\scriptscriptstyle{V_1}$}
\Text(76.0263,0.)[l]{$\scriptscriptstyle{V_2}$}
\Text(38.4604,0.)[r]{$\scriptscriptstyle{V_3}$}
\Text(63.3537,-24.6395)[lb]{$\scriptscriptstyle{V_4}$}
\Text(41.5019,21.9679)[br]{}
\Text(42.,-21.)[tr]{}
\Text(19.515,10.9162)[br]{}
\Text(20.,-10.)[tr]{}
{#1}
\end{picture}
}

\newcommand{\diagXI}[1]{
\begin{picture}(120.,104.)(-28.,-52.)
\Line(0.,0.)(56.,28.)
\Line(56.,28.)(80.,40.)
\Line(56.,-28.)(0.,0.)
\Line(80.,-40.)(56.,-28.)
\Photon(56.,28.)(56.,10.5064){2.4}{2.5}
\Photon(56.,-10.5064)(56.,-28.){2.4}{2.5}
\DashArrowArcn(56.9771,0.)(11.7466,90.,-90.){1}
\DashArrowArcn(56.9771,0.)(11.7466,270.,90.){1}
\Vertex(56.,-28.){2}
\Vertex(56.,28.){2}
\Vertex(56.,10.5064){2}
\Vertex(56.,-10.5064){2}
\Text(84.,42.)[l]{$\leg{i}$}
\Text(84.,-42.)[l]{$\leg{j}$}
\Text(64.4276,21.6767)[lu]{$\scriptscriptstyle{V_1}$}
\Text(73.343,1.)[l]{$\scriptstyle{u^{V_2}}$}
\Text(41.1437,1.)[r]{$\scriptstyle{u^{V_3}}$}
\Text(63.3537,-24.6395)[lb]{$\scriptscriptstyle{V_4}$}
\Text(41.5019,21.9679)[br]{}
\Text(42.,-21.)[tr]{}
\Text(19.515,10.9162)[br]{}
\Text(20.,-10.)[tr]{}
{#1}
\end{picture}
}

\newcommand{\diagXII}[1]{
\begin{picture}(120.,104.)(-28.,-52.)
\Line(0.,0.)(56.,28.)
\Line(56.,28.)(80.,40.)
\Line(56.,-28.)(0.,0.)
\Line(80.,-40.)(56.,-28.)
\Photon(56.,28.)(56.,10.5064){2.4}{2.5}
\Photon(56.,-10.5064)(56.,-28.){2.4}{2.5}
\DashCArc(56.9771,0.)(11.7466,-90.,90.){3}
\DashCArc(56.9771,0.)(11.7466,90.,270.){3}
\Vertex(56.,-28.){2}
\Vertex(56.,28.){2}
\Vertex(56.,10.5064){2}
\Vertex(56.,-10.5064){2}
\Text(84.,42.)[l]{$\leg{i}$}
\Text(84.,-42.)[l]{$\leg{j}$}
\Text(64.4276,21.6767)[lu]{$\scriptscriptstyle{V_1}$}
\Text(71.5542,0.)[l]{$\scriptscriptstyle{\Phi_{i_2}}$}
\Text(42.9325,0.)[r]{$\scriptscriptstyle{\Phi_{i_3}}$}
\Text(63.3537,-24.6395)[lb]{$\scriptscriptstyle{V_4}$}
\Text(41.5019,21.9679)[br]{}
\Text(42.,-21.)[tr]{}
\Text(19.515,10.9162)[br]{}
\Text(20.,-10.)[tr]{}
{#1}
\end{picture}
}

\newcommand{\diagXV}[1]{
\begin{picture}(120.,104.)(-28.,-52.)
\Line(0.,0.)(56.,28.)
\Line(56.,28.)(80.,40.)
\Line(56.,-28.)(0.,0.)
\Line(80.,-40.)(56.,-28.)
\Photon(56.,28.)(56.,10.5064){-2.4}{2.5}
\Photon(56.,-10.5064)(56.,-28.){-2.4}{2.5}
\PhotonArc(56.9771,0.)(11.7466,90.,270.){-2}{3.5}
\DashCArc(56.9771,0.)(11.7466,-90.,90.){3}
\Vertex(56.,-28.){2}
\Vertex(56.,28.){2}
\Vertex(56.,10.5064){2}
\Vertex(56.,-10.5064){2}
\Text(84.,42.)[l]{$\leg{i}$}
\Text(84.,-42.)[l]{$\leg{j}$}
\Text(64.4276,21.6767)[lu]{$\scriptscriptstyle{V_1}$}
\Text(76.0263,0.)[l]{$\scriptscriptstyle{\Phi_{i_2}}$}
\Text(38.4604,0.)[r]{$\scriptscriptstyle{V_3}$}
\Text(63.3537,-24.6395)[lb]{$\scriptscriptstyle{V_4}$}
\Text(41.5019,21.9679)[br]{}
\Text(42.,-21.)[tr]{}
\Text(19.515,10.9162)[br]{}
\Text(20.,-10.)[tr]{}
{#1}
\end{picture}
}

\newcommand{\diagXVI}[1]{
\begin{picture}(120.,104.)(-28.,-52.)
\Line(0.,0.)(48.,24.)
\Line(48.,24.)(80.,40.)
\Line(48.,-24.)(0.,0.)
\Line(80.,-40.)(48.,-24.)
\Photon(48.,24.)(58.1378,0.){-2.4}{2.5}
\Photon(48.,-24.)(58.1378,0.){2.4}{2.5}
\PhotonArc(67.082,0.)(8.94427,-177.135,182.865){-2}{7}
\Vertex(48.,-24.){2}
\Vertex(48.,24.){2}
\Vertex(58.1378,0.){2}
\Text(84.,42.)[l]{$\leg{i}$}
\Text(84.,-42.)[l]{$\leg{j}$}
\Text(55.9503,18.8245)[lu]{$\scriptscriptstyle{V_1}$}
\Text(80.4984,0.)[l]{$\scriptscriptstyle{V_2}$}
\Text(55.0177,-21.3975)[lb]{$\scriptscriptstyle{V_3}$}
\Text(41.5019,21.9679)[br]{}
\Text(42.,-21.)[tr]{}
\Text(19.515,10.9162)[br]{}
\Text(20.,-10.)[tr]{}
{#1}
\end{picture}
}

\newcommand{\diagXVII}[1]{
\begin{picture}(120.,104.)(-28.,-52.)
\Line(0.,0.)(48.,24.)
\Line(48.,24.)(80.,40.)
\Line(48.,-24.)(0.,0.)
\Line(80.,-40.)(48.,-24.)
\Photon(48.,24.)(58.1378,0.){-2.4}{2.5}
\Photon(48.,-24.)(58.1378,0.){2.4}{2.5}
\DashCArc(67.082,0.)(8.94427,-177.135,182.865){3}
\Vertex(48.,-24.){2}
\Vertex(48.,24.){2}
\Vertex(58.1378,0.){2}
\Text(84.,42.)[l]{$\leg{i}$}
\Text(84.,-42.)[l]{$\leg{j}$}
\Text(55.9503,18.8245)[lu]{$\scriptscriptstyle{V_1}$}
\Text(80.4984,0.)[l]{$\scriptscriptstyle{\Phi_{i_2}}$}
\Text(55.0177,-21.3975)[lb]{$\scriptscriptstyle{V_3}$}
\Text(41.5019,21.9679)[br]{}
\Text(42.,-21.)[tr]{}
\Text(19.515,10.9162)[br]{}
\Text(20.,-10.)[tr]{}
{#1}
\end{picture}
}

\newcommand{\diagAZ}[1]{
\begin{picture}(120.,104.)(-28.,-52.)
\Line(0.,0.)(56.,28.)
\Line(56.,28.)(80.,40.)
\Line(56.,-28.)(0.,0.)
\Line(80.,-40.)(56.,-28.)
\Photon(56.,28.)(56.,10.5064){-2.4}{2.5}
\Photon(56.,-10.5064)(56.,-28.){-2.4}{2.5}
\GCirc(56.,0.){11}{0.8}
\Vertex(56.,-28.){2}
\Vertex(56.,28.){2}
\Text(84.,42.)[l]{$\leg{i}$}
\Text(84.,-42.)[l]{$\leg{j}$}
\Text(64.4276,21.6767)[lu]{$\scriptscriptstyle{A}$}
\Text(76.0263,0.)[l]{}
\Text(38.4604,0.)[r]{}
\Text(63.9079,-23.1643)[lb]{$\scriptscriptstyle{Z}$}
\Text(41.5019,21.9679)[br]{}
\Text(42.,-21.)[tr]{}
\Text(19.515,10.9162)[br]{}
\Text(20.,-10.)[tr]{}
{#1}
\end{picture}
}

\newcommand{\diagXX}[6]{
\begin{picture}(120.,104.)(-28.,-52.)
\Line(0.,0.)(65.,32.5)
\Line(0.,0.)(65.,-32.5)
\Line(0.,0.)(72.6722,0.)
\Photon(55.25,27.625)(55.25,0.){1.95}{2.5}
\Photon(37.05,-18.525)(37.05,0.){-1.95}{2}
\Vertex(55.25,27.625){2}
\Vertex(37.05,-18.525){2}
\Vertex(55.25,0.){2}
\Vertex(37.05,0.){2}
\Text(68.25,34.125)[l]{#1}
\Text(76.3058,0.)[l]{#2}
\Text(68.25,-34.125)[l]{#3}
\Text(60.4318,13.7948)[l]{#4}
\Text(42.4369,-13.018)[l]{#5}
\Text(46.5102,0.)[b]{}
\Text(32.1146,16.999)[br]{}
\Text(22.75,-11.375)[tr]{}
\Text(25.4353,0.)[b]{}
{#6}
\end{picture}
}

\newcommand{\diagXXd}[6]{
\begin{picture}(120.,104.)(-28.,-52.)
\Line(0.,0.)(65.,32.5)
\Line(0.,0.)(65.,-32.5)
\Line(0.,0.)(72.6722,0.)
\PhotonArc(30.,15.)(15.6525,26.5651,206.565){2}{4}
\Vertex(16.,8.){2}
\Vertex(44.,22.){2}
\Photon(50,-25)(50,0.){-1.95}{2}
\Vertex(50,-25){2}
\Vertex(50,0.){2}
\Text(68.25,34.125)[l]{#1}
\Text(76.3058,0.)[l]{#2}
\Text(68.25,-34.125)[l]{#3}
\Text(10.4318,30.7948)[b]{#4}
\Text(55,-15)[l]{#5}
\Text(46.5102,0.)[b]{}
\Text(32.1146,16.999)[br]{}
\Text(22.75,-11.375)[tr]{}
\Text(25.4353,0.)[b]{}
{#6}
\end{picture}
}

\newcommand{\diagXXc}[6]{
\begin{picture}(120.,104.)(-28.,-52.)
\Line(0.,0.)(65.,32.5)
\Line(0.,0.)(65.,-32.5)
\Line(0.,0.)(72.6722,0.)
\Photon(37.05,18.525)(37.05,0.){1.95}{2}
\Photon(55.25,-27.625)(55.25,0.){-1.95}{2.5}
\Vertex(37.05,18.525){2}
\Vertex(55.25,-27.625){2}
\Vertex(37.05,0.){2}
\Vertex(55.25,0.){2}
\Text(68.25,34.125)[l]{#1}
\Text(76.3058,0.)[l]{#2}
\Text(68.25,-34.125)[l]{#3}
\Text(60.1189,-14.1922)[l]{#4}
\Text(42.4369,10.018)[l]{#5}
\Text(46.5102,0.)[t]{}
\Text(32.1146,16.999)[br]{}
\Text(22.75,-11.375)[tr]{}
\Text(25.4353,0.)[t]{}
{#6}
\end{picture}
}

\newcommand{\diagXXI}[7]{
\begin{picture}(120.,104.)(-28.,-52.)
\Line(0.,0.)(65.,32.5)
\Line(0.,0.)(65.,-32.5)
\Line(0.,0.)(72.36,-6.72921)
\Photon(45.5,22.75)(50.3792,7.05301){1.95}{2}
\Photon(39.,-19.5)(50.3792,7.05301){-1.95}{3}
\Photon(62.9532,-5.85441)(50.3792,7.05301){1.95}{2}
\Vertex(45.5,22.75){2}
\Vertex(62.9532,-5.85441){2}
\Vertex(39.,-19.5){2}
\Vertex(50.3792,7.05301){2}
\Text(68.25,34.125)[l]{#1}
\Text(75.978,-7.06567)[l]{#2}
\Text(68.25,-34.125)[l]{#3}
\Text(52.0477,16.1784)[l]{#4}
\Text(61.6758,3.43505)[l]{#5}
\Text(46.4209,-13.9719)[l]{#6}
\Text(26.,-13.)[tr]{}
\Text(28.9031,15.2991)[br]{}
\Text(29.0689,0.)[b]{}
{#7}
\end{picture}
}

\newcommand{\diagXXII}[7]{
\begin{picture}(120.,104.)(-28.,-52.)
\Line(0.,0.)(65.,32.5)
\Line(0.,0.)(65.,-32.5)
\Line(0.,0.)(72.36,6.72921)
\Line(0.,0.)(72.36,-6.72921)
\Photon(52.,26.)(57.888,5.38336){1.95}{2}
\Photon(52.,-26.)(57.888,-5.38336){-1.95}{2}
\Vertex(52.,26.){2}
\Vertex(52.,-26.){2}
\Vertex(57.888,5.38336){2}
\Vertex(57.888,-5.38336){2}
\Text(68.25,34.125)[l]{#1}
\Text(75.978,7.06567)[l]{#2}
\Text(75.978,-7.06567)[l]{#3}
\Text(68.25,-34.125)[l]{#4}
\Text(59.3965,16.9633)[l]{#5}
\Text(59.3965,-16.9633)[l]{#6}
{#7}
\end{picture}
}

\newcommand{\diagonenf}[3]{
\begin{picture}(120.,76.)(-28.,-38.)
\Line(0.,0.)(80,0)
\PhotonArc(25.,-8.)(26.2488,17.7447,162.255){2}{5}
\Vertex(50.,0.){2}
\Text(84.,0)[l]{#1}
\Text(23,29)[l]{#2}
\Text(27.668,14.6453)[br]{}
\Text(28.,-14.)[tr]{}
{#3}
\end{picture}
}

\newcommand{\diagoneselfnf}[3]{
\begin{picture}(120.,76.)(-28.,-38.)
\Line(0.,0.)(80,0)
\PhotonArc(40.,0)(20,0,180){2}{5}
\Vertex(60.,0.){2}
\Vertex(20.,0.){2}
\Text(84.,0)[l]{#1}
\Text(40,29.)[b]{#2}
\Text(27.668,14.6453)[br]{}
\Text(28.,-14.)[tr]{}
{#3}
\end{picture}
}

\newcommand{\diagonefact}[4]{
\begin{picture}(120.,104.)(-28.,-52.)
\Line(0.,0.)(80,0)
\Line(0.,0.)(50,50)
\PhotonArc(0,0)(48,0,45){2}{4}
\Vertex(48.,0.){2}
\Vertex(33.941,33.941){2}
\Text(84.,0)[l]{#1}
\Text(53.,53.)[l]{#2}
\Text(50,22.)[l]{#3}
\Text(27.668,14.6453)[br]{}
\Text(28.,-14.)[tr]{}
{#4}
\end{picture}
}

\newcommand{\diagInf}[6]{
\begin{picture}(120.,104.)(-28.,-52.)
\Line(0.,0.)(28.,14.)
\Line(28.,14.)(56.,28.)
\Line(56.,28.)(80.,40.)
\Line(28.,-14.)(0.,0.)
\Line(56.,-28.)(28.,-14.)
\Line(80.,-40.)(56.,-28.)
\Photon(56.,28.)(56.,0.){2.4}{2}
\Photon(56.,0.)(56.,-28.){2.4}{2}
\Photon(56.,0.)(0,0){2.4}{4}
\Vertex(56.,28.){2}
\Vertex(56.,0){2}
\Vertex(56.,-28.){2}
\Text(84.,42.)[l]{#1}
\Text(84.,-42.)[l]{#2}
\Text(62.6099,14.)[l]{#3}
\Text(62.6099,-14.)[l]{#4}
\Text(35.305,-5.)[t]{#5}
\Text(41.5019,21.9679)[br]{}
\Text(42.,-21.)[tr]{}
\Text(19.515,10.9162)[br]{}
\Text(20.,-10.)[tr]{}
{#6}
\end{picture}
}
\newcommand{\diagIInf}[5]{
\begin{picture}(120.,104.)(-28.,-52.)
\Line(0.,0.)(16.,8.)
\Line(16.,8.)(44.,22.)
\Line(44.,22.)(64.,32.)
\Line(64.,32.)(80.,40.)
\Line(64.,-32.)(0.,0.)
\Line(80.,-40.)(64.,-32.)
\Photon(64.,32.)(64.,-32.){2.4}{3.5}
\PhotonArc(17.9434,8.83368)(20.,26.2114,206.211){2}{5} 
\Vertex(35.8868,17.6674){2}
\Vertex(64.,-32.){2}
\Vertex(64.,32.){2}
\Text(84.,42.)[l]{#1}
\Text(84.,-42.)[l]{#2}
\Text(71.5542,0.)[l]{#3}
\Text(20.8885,35.1332)[br]{#4}
\Text(41.5019,21.9679)[br]{}
\Text(42.,-21.)[tr]{}
\Text(19.515,10.9162)[br]{}
\Text(20.,-10.)[tr]{}
{#5}
\end{picture}
}

\newcommand{\diagIIInf}[5]{
\begin{picture}(120.,104.)(-28.,-52.)
\Line(0.,0.)(24.,12.)
\Line(24.,12.)(40.,20.)
\Line(40.,20.)(56.,28.)
\Line(56.,28.)(80.,40.)
\Line(40.,-20.)(0.,0.)
\Line(80.,-40.)(40.,-20.)
\Photon(40.,20.)(40.,-20.){2.4}{3.5}
\PhotonArc(44.6515,-11.4562)(46.0977,66.8127,165.61){2}{5}
\Vertex(62.8019,30.9179){2}
\Vertex(40.,-20.){2}
\Vertex(40.,20.){2}
\Text(84.,42.)[l]{#1}
\Text(84.,-42.)[l]{#2}
\Text(49.1935,0.)[l]{#3}
\Text(30.1994,40.9325)[br]{#4}
\Text(41.5019,21.9679)[br]{}
\Text(42.,-21.)[tr]{}
\Text(19.515,10.9162)[br]{}
\Text(20.,-10.)[tr]{}
{#5}
\end{picture}
}

\newcommand{\diagIVnf}[6]{
\begin{picture}(120.,104.)(-28.,-52.)
\Line(0.,0.)(65.,32.5)
\Line(0.,0.)(65.,-32.5)
\Line(0.,0.)(72.6722,0.)
\PhotonArc(23.7198,2.76053)(23.8799,45.7845,186.638){2}{5}
\Vertex(40.3727,19.8758){2}
\Photon(50,-25)(50,0.){-1.95}{2}
\Vertex(50,-25){2}
\Vertex(50,0.){2}
\Text(68.25,34.125)[l]{#1}
\Text(76.3058,0.)[l]{#2}
\Text(68.25,-34.125)[l]{#3}
\Text(10.4318,30.7948)[b]{#4}
\Text(55,-15)[l]{#5}
\Text(46.5102,0.)[b]{}
\Text(32.1146,16.999)[br]{}
\Text(22.75,-11.375)[tr]{}
\Text(25.4353,0.)[b]{}
{#6}
\end{picture}
}

\newcommand{\diagIwi}[3]{
\begin{picture}(120.,64.)(-28.,-32.)
\Line(0.,0.)(50,0)
\PhotonArc(50.,-50.)(70.71,90.,135.){2}{5}
\Text(54.,0)[l]{#1}
\Text(54,22.)[l]{#2}
\Text(27.668,14.6453)[br]{}
\Text(28.,-14.)[tr]{}
{#3}
\end{picture}
}
\newcommand{\diagIIwi}[3]{
\begin{picture}(120.,64.)(-28.,-32.)
\Line(0.,0.)(50,0)
\Photon(25.,0)(50,20){-2}{3}
\Vertex(25,0.){2}
\Text(54.,0)[l]{#1}
\Text(54,22.)[l]{#2}
\Text(27.668,14.6453)[br]{}
\Text(28.,-14.)[tr]{}
{#3}
\end{picture}
}

\newcommand{\diagIIIwi}[2]{
\begin{picture}(120.,104.)(-28.,-52.)
\Line(0.,0.)(50,0)
\Text(54.,0)[l]{#1}
\Text(27.668,14.6453)[br]{}
\Text(28.,-14.)[tr]{}
{#2}
\end{picture}
}

\newcommand{\diagIVwi}[4]{
\begin{picture}(120.,104.)(-28.,-52.)
\Line(0.,0.)(70,0)
\Line(0.,0.)(50,50)
\Photon(25,25)(70,25){2}{5}
\Vertex(25,25){2}
\Text(74.,0)[l]{#1}
\Text(53.,53.)[l]{#2}
\Text(75,25)[l]{#3}
\Text(27.668,14.6453)[br]{}
\Text(28.,-14.)[tr]{}
{#4}
\end{picture}
}

\newcommand{\subloopI}[4]{
\begin{picture}(120.,104.)(-28.,-52.)
\Line(0.,0.)(16.,8.)
\Line(16.,8.)(44.,22.)
\Line(44.,22.)(64.,32.)
\Line(64.,32.)(80.,40.)
\Line(64.,-32.)(0.,0.)
\Line(80.,-40.)(64.,-32.)
\Photon(64.,32.)(64.,-32.){2.4}{3.5}
\GCirc(35.,17.5){10}{0.8}
\Vertex(64.,-32.){2}
\Vertex(64.,32.){2}
\Text(84.,42.)[l]{#1}
\Text(84.,-42.)[l]{#2}
\Text(71.5542,0.)[l]{#3}
\Text(41.5019,21.9679)[br]{}
\Text(42.,-21.)[tr]{}
\Text(19.515,10.9162)[br]{}
\Text(20.,-10.)[tr]{}
{#4}
\end{picture}
}

\newcommand{\subloopII}[4]{
\begin{picture}(120.,104.)(-28.,-52.)
\Line(0.,0.)(24.,12.)
\Line(24.,12.)(40.,20.)
\Line(40.,20.)(56.,28.)
\Line(56.,28.)(80.,40.)
\Line(40.,-20.)(0.,0.)
\Line(80.,-40.)(40.,-20.)
\Photon(40.,20.)(40.,-20.){2.4}{3.5}
\GCirc(40.,20){10}{0.8}
\Vertex(40.,-20.){2}
\Text(84.,42.)[l]{#1}
\Text(84.,-42.)[l]{#2}
\Text(49.1935,0.)[l]{#3}
\Text(41.5019,21.9679)[br]{}
\Text(42.,-21.)[tr]{}
\Text(19.515,10.9162)[br]{}
\Text(20.,-10.)[tr]{}
{#4}
\end{picture}
}

\newcommand{\diagIcollin}[4]{
\begin{picture}(100.,88.)(-15.,-44.)
\Line(0.,0.)(90,0)
\Text(94.,0)[l]{#1}
\Photon(0,0)(0,-30){2}{4}
\Text(0,-34)[t]{#2}
\Photon(55,0)(55,-30){2}{4}
\Vertex(55,0){2}
\Text(55,-34)[t]{#3}
{#4}
\end{picture}
}

\newcommand{\diagIIcollin}[6]{
\begin{picture}(175.,88.)(-5.,-44.)
\Photon(0,0)(-15,-30){2}{4}
\Text(0,-34)[t]{#1}
\Photon(0,0)(15,-30){2}{4}
\Text(15,-34)[t]{#2}
\Line(0.,0.)(80,0)
\Text(90,0)[l]{$\dots$}
\Line(120.,0.)(170,0)
\Text(174.,0)[l]{#3}
\Photon(65,0)(65,-30){2}{4}
\Vertex(65,0){2}
\Text(65,-34)[t]{#4}
\Photon(135,0)(135,-30){2}{4}
\Vertex(135,0){2}
\Text(135,-34)[t]{#5}
{#6}
\end{picture}
}

\newcommand{\Frac}{\frac}
\newcommand{\Epsinv}[1]{\veps^{- #1}}
\newcommand{\Eps}[1]{\veps^{#1}}
\newcommand{\Right}{\right}
\newcommand{\EG}{\gamma_{\mathrm{E}}}


\thispagestyle{empty} 


\thispagestyle{empty}
\def\thefootnote{\fnsymbol{footnote}}
\setcounter{footnote}{1}
\null
\draftdate\hfill  PSI-PR-06-10 \\
\strut\hfill  MPP-2006-109 \\
\strut\hfill hep-ph/0608326
\vskip 0cm
\vfill
\begin{center}
{\Large \bf
Two-loop electroweak next-to-leading logarithmic \\ corrections
to massless fermionic processes
\par}
 \vskip 1em
{\large
{\sc A.\ Denner$^1$\footnote{Ansgar.Denner@psi.ch}, 
B.\ Jantzen$^1$\footnote{physics@bernd-jantzen.de},
and S.\ Pozzorini$^2$\footnote{pozzorin@mppmu.mpg.de} }}
\\[.5cm]
$^1$ {\it Paul Scherrer Institut\\
CH-5232 Villigen PSI, Switzerland}
\\[0.3cm]
$^2$ {\it 
Max-Planck-Institut f\"ur Physik,
F\"ohringer Ring 6\\
D-80805 M\"unchen, Germany}
\par
\end{center}\par
\vskip 1.0cm \vfill {\bf Abstract:} \par We consider two-loop leading
and next-to-leading logarithmic virtual corrections to arbitrary
processes with external massless fermions in the electroweak Standard
Model at energies well above the electroweak scale. Using the
sector-decomposition method and alternatively the strategy of regions
we calculate the mass singularities that arise as logarithms of
$Q^2/\MW^2$, where $Q$ is the energy scale of the considered process,
and $1/\teps$ poles in $D=4-2\teps$ dimensions, to one- and two-loop
next-to-leading logarithmic accuracy. The derivations are performed
within the complete electroweak theory with spontaneous symmetry
breaking.  Our results indicate a close analogy between the form of
two-loop electroweak logarithmic corrections and the singular
structure of scattering amplitudes in massless QCD.  We find agreement
with the resummation prescriptions that have been proposed in the
literature based on a symmetric $\SU(2)\times\U(1)$ theory matched
with QED at the electroweak scale and provide new next-to-leading
contributions proportional to $\ln(\MZ^2/\MW^2)$.
\par
\vskip 1cm
\noindent
August 2006
\par
\null
\setcounter{page}{0}
\clearpage
\def\thefootnote{\arabic{footnote}}
\setcounter{footnote}{0}

\section{Introduction}
\label{se:intro}

The electroweak radiative corrections to high-energy processes are
characterized by the presence of logarithms of the type
$\ln(Q^2/M^2)$, which involve the ratio of the typical scattering
energy $Q$ over the gauge-boson mass scale $M=\MW\sim\MZ$
\cite{Kuroda:1991wn}.
These logarithmic corrections affect every reaction that involves
electroweakly interacting particles and has a characteristic scale
$Q\gg M$.  They start to be sizable at energies of the order of a few
hundred GeV and their impact increases with energy.
At the LHC \cite{Haywood:1999qg} and future lepton colliders
\cite{Aguilar-Saavedra:2001rg,Abe:2001wn,Abe:2001gc}, where scattering
energies of the order of $1\TeV$ will be reached, logarithmic
electroweak effects can amount to tens of per cent at one loop and
several per cent at two loops.
Thus, this class of corrections will be very important for the
interpretation of precision measurements at future high-energy
colliders.
%

For sufficiently high $Q$, terms of $\order(M^2/Q^2)$ become
negligible and, at $l$ loops,
the electroweak  corrections assume the form of a tower of logarithms,
\beq\label{logform}
\alpha^l\ln^{j}{\left(\frac{Q^2}{M^2}\right)},
\qquad\mbox{with}\quad
\quad 0\le j \le 2l,\textmd{}
\eeq
where the leading logarithms (LLs), also known as Sudakov logarithms
\cite{Sudakov:1954sw}, have power $j=2l$, and the subleading terms
with $j=2l-1,2l-2,\dots$ are denoted as next-to-leading logarithms
(NLLs), next-to-next-to-leading logarithms (NNLLs), and so on.  The
complete asymptotic limit includes all logarithmic contributions
$(j>0)$ as well as terms that are not logarithmically enhanced at high
energy ($j=0$).%
\footnote{ In the literature, such terms are sometimes denoted as
  ``constants'' since they do not involve logarithms that grow with
  energy.  However, in general they are functions of the ratios of
  kinematical invariants, which depend on the scattering angles.  At
  one loop such terms are formally classified as NNLLs.  }
%

Electroweak logarithmic corrections have a twofold origin.
On the one hand they can appear as terms of the form $\ln(Q^2/\muR^2)$
resulting from the renormalization of ultraviolet (UV) singularities at
the renormalization scale $\muR\sim M$.  These logarithms can
easily be controlled through the running of the coupling constants.
The other source of electroweak logarithms are mass singularities, \ie
logarithms of the form $\ln(Q^2/M^2)$ that are formally singular in
the limit where the gauge-boson masses tend to zero.  As is well
known, in gauge theories, mass singularities result from the
interactions of the initial- and final-state particles with soft
and/or collinear gauge bosons.
This permits, in principle, to treat mass singularities in a
process-independent way and to derive universal properties of
mass-singular logarithmic corrections.
%

At one loop, it was proven that the electroweak LLs and NLLs are
universal, and a general formula was derived that expresses the
logarithmic corrections to arbitrary processes in terms of the
electroweak quantum numbers of the initial- and final-state particles
\cite{Denner:2001jv,Pozzorini:rs}.
In recent years, the impact of one-loop electroweak corrections was
studied in detail for various specific processes at high-energy
colliders
\cite{Beccaria:2000fk,Layssac:2001ur,Accomando:2001fn,Accomando:2005ra,Maina:2003is,Kuhn:2004em,Beccaria:2005sv}.
At the LHC, in general, large negative corrections are observed that
appear at transverse momenta $p_\rT$ around $100\GeV$ and grow with
$p_\rT$.  Depending on the process, the size of these corrections can
reach up to 10--40\% at $p_\rT\sim$ 500--1000 GeV.
In \citeres{Accomando:2005ra,Kuhn:2004em}, the predictions based on
NLL and NNLL high-energy approximations were compared with exact
one-loop calculations.  In both cases it turned out that in the
high-$p_\rT$ region, where the corrections are large, the NNLL
predictions deviate from the exact calculation by much less than 1\%,
\ie the numerical effect of $\order(M^2/Q^2)$ contributions is
negligible.  Also the NLL approximation provides a correct description
of the bulk of the corrections and their energy dependence.  However,
it was found that the actual precision of the NLL approximation,
better than 1\% for $\Pp\Pp \to \mathrm{jet} + \PZ/\gamma$
\cite{Kuhn:2004em} and about 5\% for pp $\to$ W$\gamma$
\cite{Accomando:2005ra}, depends relatively strongly on the process.
As it was shown for $\Pp\Pp\to\mathrm{jet}+\PZ/\gamma$, also the
two-loop electroweak logarithms can have an impact at the several
per-cent level on high-$p_\rT$ measurements at the LHC
\cite{Kuhn:2004em}.

In the recent years, the properties of electroweak logarithmic
corrections beyond one loop have been studied with two complementary
approaches.%
\footnote{For recent developments in the exact numerical calculation
  of complete two-loop integrals see, for instance, \citere{Passarino:2006gv}.} 
On the one hand, evolution equations, which are well known in QED and
QCD, have been applied to the electroweak theory in order to obtain the
higher-order terms through a resummation of the one-loop logarithms%
\footnote{In order to resum the LLs and NLLs it is sufficient to
  determine the kernel of the evolution equations to one-loop
  accuracy.  However, starting from the NNLLs also the two-loop
  contributions to the $\beta$-function and to the anomalous
  dimensions are needed.}
\cite{Fadin:2000bq,Melles:2001gw,Melles:2001ia,Melles:2001mr,Melles:2001dh,Kuhn:2000nn,Kuhn:2001hz}.
On the other hand, explicit diagrammatic calculations based on the
electroweak Feynman rules have been performed
\cite{Melles:2000ed,Hori:2000tm,Beenakker:2000kb,Denner:2003wi,Pozzorini:2004rm,Feucht:2003yx,Jantzen:2005xi,Jantzen:2005az}.

Fadin \etal\ \cite{Fadin:2000bq} have resummed the electroweak LL
corrections to arbitrary matrix elements by means of the infrared
evolution equation (IREE), and this approach has been extended by
Melles to the NLL terms for arbitrary processes
\cite{Melles:2001gw,Melles:2001ia,Melles:2001mr,Melles:2001dh}.  In
\citeres{Kuhn:2000nn,Kuhn:2001hz} K\"uhn \etal\ have resummed the
logarithmic corrections to neutral-current massless \mbox{4-fermion}
processes up to the NNLL level.
These resummations
\cite{Fadin:2000bq,Melles:2001gw,Melles:2001ia,Melles:2001mr,Melles:2001dh,Kuhn:2000nn,Kuhn:2001hz} are based
on the IREE \cite{Fadin:2000bq}, which describes the all-order LL
dependence of matrix elements with respect to the cut-off parameter
$\mu_\perp$. This cut-off fixes the minimal transverse momentum for
the gauge bosons that couple to the initial- and final-state particles
and acts as a regulator of mass singularities.
The IREE, which was originally derived within symmetric gauge theories
(QED and QCD), has been {applied} to the spontaneously broken
electroweak theory under the assumption that this latter can be split
into two regimes with exact gauge symmetry.
In practice, in the regime $Q\ge\mu_\perp\ge M$ the masses of the
electroweak gauge bosons, which result from the breaking of gauge
symmetry, are supposed to be negligible and exact SU(2)$\times$U(1)
symmetry is assumed.
Instead, in the regime $M \ge \mu_\perp$ the weak gauge bosons are
supposed to be frozen owing to their masses such that only photons
contribute to the evolution, and this is characterized by exact
U(1)$_\elm$ symmetry.

As a result, the resummed electroweak corrections factorize into two
parts corresponding to these two regimes: (i) a symmetric-electroweak
part that can be computed within a symmetric SU(2)$\times$U(1) gauge
theory using the same cut-off parameter $M$ to regularize the mass
singularities that result from all gauge bosons, \ie assuming that the
photon is as heavy as the weak gauge bosons,
{and}  (ii) an electromagnetic
part that originates from the mass gap between photons and massive
gauge bosons and can be computed within QED.
The electromagnetic part contains divergences that are due to massless
photons and depend on the scheme adopted for their regularization.
These divergences cancel when the contributions of virtual and real
photons are combined, and then the electromagnetic contribution
depends on the cuts imposed on real photons.
Instead the symmetric-electroweak part involves logarithms of the form
$\ln(Q^2/M^2)$ which are
formally mass singular but numerically finite since $M$ does not
vanish.
These logarithms are present in all physical observables that are
exclusive with respect to real radiation of Z and W bosons.  Moreover,
the electroweak logarithms remain present even in inclusive
observables due to a lack of cancellation between virtual and real
contributions from electroweak gauge bosons.  This is due to the fact
that the conditions of the Bloch--Nordsieck theorem are not fulfilled
since fermions and gauge bosons carry non-abelian weak-isospin charges
\cite{Ciafaloni:2000df}.


The resummations
\cite{Fadin:2000bq,Melles:2001gw,Melles:2001ia,Melles:2001mr,Melles:2001dh,Kuhn:2000nn,Kuhn:2001hz} rely on
the assumption that all relevant implications of electroweak symmetry
breaking are correctly taken into account by simply splitting the
evolution equation into two regimes with exact gauge symmetry.
In particular, the following assumptions are explicitly or implicitly
made.  (i) In the massless limit $M/Q\to 0$, all couplings with mass
dimension, which originate from symmetry breaking, are neglected.
(ii) The weak-boson masses are introduced in the corresponding
propagators as regulators of soft and collinear singularities from W
and Z bosons without spontaneous symmetry breaking. Since these masses
are of the same order, one uses $\MW=\MZ$ as an approximation.  (iii)
The regimes above and below the electroweak scale are treated as an
unmixed SU(2)$\times$U(1) theory and QED, respectively, and mixing
effects in the gauge sector, \ie the interaction of photons with W
bosons, are neglected.

It is important to understand to which extent the above assumptions
are legitimate and whether the resulting resummation prescriptions are
correct. This can be achieved by explicit diagrammatic two-loop
calculations based on the electroweak Lagrangian, where all effects
related to spontaneous symmetry breaking are consistently taken into
account.
At the LL level, these checks have already been completed.  A
calculation of the massless fermionic singlet form factor
\cite{Melles:2000ed,Hori:2000tm} and then a Coulomb-gauge calculation
for arbitrary processes \cite{Beenakker:2000kb}
have supported the exponentiation of the electroweak LLs predicted by
the IREE.
Also the angular-dependent subset of the NLL corrections for arbitrary
processes \cite{Denner:2003wi} has been shown to be consistent with
the exponentiation anticipated in
\citeres{Melles:2001dh,Kuhn:2000nn,Kuhn:2001hz}.
At present the complete set of electroweak NLLs is known only for the
so-called fermionic form factor, which corresponds to the
gluon--fermion--antifermion vertex \cite{Pozzorini:2004rm}.  This
calculation has confirmed that the IREE approach provides a correct
description of the terms that do not depend on the difference between
the masses of the Z and the W bosons and has provided first insights
into the behaviour of the NLL terms of the type $\alpha^2
\ln(\MZ^2/\MW^2)\ln^3(Q^2/M^2)$.
The complete tower of two-loop logarithms for the fermionic form
factor has been computed in \citeres{Feucht:2003yx,Jantzen:2005xi}
adopting an unmixed SU(2)$\times$U(1) theory with $\MZ=\MW$ as
approximation.
Within this framework it was found that the soft--collinear
singularities resulting from massless photons factorize as suggested
by the IREE and that, in absence of mixing, symmetry breaking is
negligible up to the NNLL level and its first non-trivial effects
appear through a Higgs-mass dependence of the NNNLL terms.
Using QCD resummation techniques, the NNNLL two-loop results for the
fermionic form-factor have been extended to the amplitude for
neutral-current four-fermion scattering
\cite{Jantzen:2005xi,Jantzen:2005az}.  Here terms of the type
$\ln(\MZ^2/\MW^2)$ were included through an expansion in
$\sw^2=1-\MW^2/\MZ^2$ up to the first order in $\sw^2$.
%

In this paper we develop a formalism to derive virtual electroweak
two-loop LL and NLL corrections for arbitrary processes and apply
it to the case of massless \mbox{$n$-fermion} reactions.
The calculation is performed diagrammatically using the electroweak
Feynman rules in the 't~Hooft--Feynman gauge.
The mass-singular logarithms are extracted from the relevant Feynman
diagrams by means of a soft--collinear approximation for the
interaction of initial- and final-state fermions with gauge bosons.
Using Ward identities we prove that the LL and NLL mass singularities
for generic \mbox{$n$-fermion} amplitudes factorize from the
corresponding Born amplitudes.
All relevant loop integrals are evaluated in the high-energy region
$Q\gg M$ to NLL accuracy using an automatized algorithm based on the
sector-decomposition technique \cite{Denner:2004iz} and alternatively
the method of expansion by regions combined with Mellin--Barnes
representations \cite{Jantzen:2006jv}.
We do not assume $\MZ=\MW$ and we include all relevant contributions
depending on the difference $\MZ -\MW$.
In addition to the logarithms of the type \refeq{logform}, which arise
from massive virtual particles, we include also mass singularities
from massless virtual photons.  The latter are regularized dimensionally
and arise as $1/\veps$ poles in $D=4-2\veps$ dimensions.
In our result the photonic singularities are factorized in a
gauge-invariant electromagnetic term.  The remaining part of the
corrections, which is also gauge invariant and does not depend on the
scheme adopted to regularize photonic singularities, contains only
finite $\log(Q^2/M^2)$ terms.  The divergences contained in the
electromagnetic term cancel if real-photon emission is included.


The paper is organized as follows.  Section \ref{se:definitions}
contains definitions and conventions used in the calculation.  In
\refse{se:singtreatment} we identify the diagrams that produce
ultraviolet and mass singularities, split them into factorizable and
non-factorizable parts and discuss our method to treat these
contributions.  Using collinear Ward identities, we prove in
\refse{se:cwi} that all non-factorizable contributions cancel.
Explicit expressions for the factorizable contributions and for the
counterterms needed for the renormalization are provided in
\refses{se:factcont} and \ref{se:ren}, respectively.  The complete
one- and two-loop results are presented in \refse{se:results} and
discussed in \refse{se:disc}.  Our conclusions are contained in
\refse{se:conc}, and various appendices are devoted to technical
aspects of the calculation.

\section{Definitions and conventions}
\label{se:definitions}

We consider a generic $n\to 0$ process involving an even number $n$ of
polarized fermionic particles,
\beq\label{genproc}
\varphi_{1}(p_1)\dots
\varphi_{n}(p_{n})\to 0.
\eeq
The symbols $\varphi_{i}$ represent $n/2$ antifermions and $n/2$
fermions: $\varphi_{i}={\bar f}_{\si_i}^{\kappa_i}$ for
$i=1,\dots,n/2$ and $\varphi_{i}={f}_{\si_i}^{\kappa_i}$ for
$i=n/2+1,\dots,n$.  These particles are massless chiral eigenstates
with $p_k^2=m_k^2=0$ and chirality $\kappa_i=\rR$ or $\rL$.  In
practice every fermion or antifermion can be a lepton or a light
quark.  The indices $\si_i$ characterize the lepton/quark nature, the
isospin and the generation of the fermions,
$f_{\si_i}=\nu_e,e,u,d,\dots$.

The matrix element for the process \refeq{genproc} reads
\beq\label{matel}
\M^{\varphi_{1}\dots\varphi_{n}}
=
\left[\prod_{i=1}^{n/2} {\bar v}(p_i,\kappa_i)\right]
G^{\underline{\varphi}_{1}\dots\underline{\varphi}_{n}}
(p_1,\dots,p_{n})
\left[\prod_{j=n/2+1}^{n} u(p_j,\kappa_j)\right]
,
\eeq
where $G^{\underline{\varphi}_{1}\dots\underline{\varphi}_{n}}$ is the
corresponding truncated Green function.  The spinors for chiral
fermions and antifermions fulfil
\beqar\label{spinors}
 \omega_{\rho} u(p,\kappa)&=&  \de_{\kappa\rho} u(p,\kappa),
\qquad
{\bar  v}(p,\kappa) {\bar \omega}_{\rho}=   \de_{\kappa\rho}\bar {v}(p,\kappa)
\eeqar
for $\rho,\kappa=\rR,\rL$, and
\beqar
 \omega_{\rR}&=&{\bar \omega}_\rL=\frac{1}{2}(1+\gamma^5)
,\qquad
 \omega_{\rL}={\bar \omega}_\rR=\frac{1}{2}(1-\gamma^5)
.
\eeqar
The matrix elements \refeq{matel} are often abbreviated
as  $\M\equiv \M^{\varphi_{1}\dots\varphi_{n}}$.

The amplitudes for physical scattering processes, \ie $\,2\to n-2\,$
reactions, are easily obtained from our results for $n\to0$ reactions
using crossing symmetry.

\subsection{Perturbative and asymptotic expansions}

For the perturbative expansion of the matrix elements in
$\alpha=e^2/(4\pi)$, where $e$ is the electromagnetic coupling
constant, we write
\beqar\label{pertserie1a}
\M&=&
\sum_{l=0}^\infty
\mel{l}{}
,
\eeqar
with
\beqar\label{pertserie1b}
\mel{l}{}=
\left(\frac{\alphaeps}{4\pi}\right)^l
\nmel{l}{}
,
\eeqar
and
\beqar\label{pertserie1}
\alphaeps=
\left(\frac{4\pi\muD^2}{ \mathrm{e}^{\gamma_{\mathrm{E}}}Q^2}\right)^{\veps}
\alpha
.
\eeqar
Note that, for convenience, in $D=4-2\veps$ dimensions we include in
the definition of $\alphaeps$ a normalization factor depending on
$\veps$, the scale $\muD$ of dimensional regularization, the
characteristic energy $Q$ of the scattering process and Euler's
constant $\gamma_{\mathrm{E}}$.  For
$\muD^2=Q^2\mathrm{e}^{\gamma_{\mathrm{E}}}/(4\pi)$ we have
$\alphaeps=\alpha$.

The electroweak corrections are evaluated in the region where all
kinematical invariants, $r_{jk}=(p_j+p_k)^2$, are much larger than the
squared masses of the heavy particles that enter the loops,
\beqar\label{asymptoticregion}
|r_{jk}|\sim Q^2 \gg \MW^2\sim\MZ^2\sim\Mt^2\sim\MH^2.
\eeqar
This means that we consider a situation with essentially two different
scales, one fixed by $Q$, which can be identified with (the square
root of the absolute value of) one of the kinematical invariants,
$Q=\sqrt{|r_{jk}|}$, and one fixed by the heavy masses of the Standard
Model.  In this region, the electroweak corrections are dominated by
mass-singular logarithms,
\beqar\label{logsymbol}
L&=&\ln\left(\frac{Q^2}{\MW^2}\right)
,
\eeqar
and logarithms of UV origin.  Mass singularities that originate from
soft and collinear massless photons and UV singularities are
regularized dimensionally and give rise to $1/\veps$ poles in $D=4-2
\veps$ dimensions.  Therefore, the $l$-loop contributions to the
polarized matrix elements can be written as an expansion in $L$ and
$\veps$,
\beqar\label{logexpansion}
\mel{l}{}
=\sum_{m=0}^{2l}\sum_{n=-m}^{\infty}\mel{l,m,n}{}
\,\veps^{n}
L^{m+n}
.
\eeqar
The logarithmic terms in \refeq{logexpansion} are classified according
to their degree of singularity, defined as as the total power $m$ of
logarithms $L$ and $1/\veps$ poles.  The maximal degree of singularity
at $l$-loop level is $m=2l$ and the corresponding terms are denoted as
leading logarithms (LLs).  The terms with $m= 2l-1, 2l-2,\dots$
represent the next-to-leading logarithms (NLLs),
next-to-next-to-leading logarithms (NNLLs), and so on.

In this paper we systematically neglect mass-suppressed corrections of
order $\MW^2/Q^2$ and we calculate the one- and two-loop corrections
to NLL accuracy, \ie including LLs and NLLs.  For this approximation
we use the symbol $\NLLA$.  The two-loop corrections are expanded in
$\veps$ up to the finite terms, \ie contributions of order $\veps^0
L^4$ and $\veps^0 L^3$.  Instead, the one-loop corrections are
expanded up to order $\veps^2$, \ie including terms of order $\veps^2
L^4$ and $\veps^2 L^3$.  These higher-order terms in the
$\veps$-expansion must be taken into account when expressing two-loop
mass singularities in terms of one-loop ones.

Since the loop corrections depend on various masses,
$\MW\sim\MZ\sim\Mt\sim\MH$, and different invariants $r_{jk}$, the
coefficients $\mel{l,m,n}{}$ in \refeq{logexpansion}
depend on mass ratios and ratios of invariants.%
\footnote{
This dependence only appears at the next-to-leading level, \ie for $m=2l-1$.
}
Actually, as we will see in the final result, 
they depend only on logarithms of these ratios,
\beqar\label{anglogsymbol}
\lr{i}=\ln\left(\frac{M_i^2}{\MW^2}\right),
\qquad
\lr{jk}&=&\ln\left(\frac{-r_{jk}}{Q^2}\right)
.
\eeqar
The logarithms of $-r_{jk}$ are extracted in the region $r_{jk}<0$,
where the corrections are real. The imaginary parts that arise in the
physical region can be obtained via analytic continuation replacing
$r_{jk}$ by $r_{jk}+\ri 0$.

\subsection{Symmetry breaking and mixing}

In the electroweak Standard Model, the physical gauge bosons result
from the gauge bosons $W^1,W^2,W^3$ and $B$ associated with the
$\SU(2)$ and $\U(1)$ gauge groups via the unitary transformation
\beq\label{gbmixing}
W^\pm_\mu=\frac{1}{\sqrt{2}}\left(W_\mu^1\mp\ri W_\mu^2\right)
,\quad\
Z_\mu=\cw W^3_\mu+\sw B_\mu 
,\quad\
A_\mu=-\sw W^3_\mu+\cw B_\mu 
,
\eeq
where $\cw=\cos\theta_\mathrm{w}$, $\sw=\sin\theta_\mathrm{w}$, and
$\theta_\mathrm{w}$ is the weak mixing angle.  The generators
associated with the physical gauge bosons are related to the weak
isospin generators $T^i$ and the weak hypercharge $Y$ through
\beqar\label{genmixing}
e {I}^{W^\pm} &=& \frac{\gw }{\sqrt{2}}\left(T^1\pm\ri T^2\right)
,\quad\
e I^Z=
\gw \cw T^3- \gb \sw \frac{Y}{2}
,\quad\
e I^A=
-\gw \sw T^3- \gb \cw \frac{Y}{2} 
,\nln
\eeqar
where $\gb$ and $\gw$ are the coupling constants associated with the
U(1) and SU(2) groups, respectively.
The electromagnetic charge operator is defined as
\beq
Q=-I^A,
\eeq
and the generator associated with the Z~boson can be expressed as
\beq
e I^Z=
\frac{\gw}{\cw}  T^3 - e\frac{\sw}{\cw}  Q
.  
\eeq
The SU(2)$\times$U(1) Casimir operator is given by
\beq\label{casimir}
\sum_{V=A,Z,W^\pm} I^{\bar V} I^V = 
\frac{\gb^2}{e^2} \left(\frac{Y}{2}\right)^2+\frac{\gw^2}{e^2} C,
\eeq
where $\bar V$ denotes the complex conjugate of $V$, and
$C=\sum_{i=1}^3 (T^i)^2$ is the Casimir operator of the SU(2) group
with eigenvalues $3/4$ and $0$ for left- and right-handed fermions,
respectively.  The generators \refeq{genmixing} obey the commutation
relations
\beqar\label{commrel}
e \left[I^{V_1},I^{V_2}\right]
=
\ri \gw \,
\sum_{V_3=A,Z,W^\pm}
\teps^{V_1V_2V_3} I^{\bar{V}_3}
,
\eeqar
with
\beq\label{phystotalantitens}
\teps^{V_1 V_2V_3}=
\ri \times \left\{\begin{array}{c@{\quad}l} 
 (-1)^{p+1} \,\cw & \mbox{if}\quad V_1V_2V_3= \pi(Z W^+W^-), \\  
 (-1)^p \,\sw & \mbox{if}\quad V_1V_2V_3= \pi(A W^+W^-), \\  
0 & \mbox{otherwise,} 
\end{array}\right.
\eeq
and $(-1)^p$ represents the sign of the permutation $\pi$.

The $\SU(2)\times\U(1)$ symmetry is spontaneously broken by a scalar
Higgs doublet which acquires a vacuum expectation value $\vev$. We
parametrize the Higgs doublet in terms of the four degrees of freedom
$\Phi_i=H,\chi,\phi^+,\phi^-$, where $\phi^-=(\phi^+)^+$. In this
representation the gauge-group generators are $4\times 4$ matrices
with components ${I}^{{V}}_{\Phi_i\Phi_j}$.
The gauge-boson masses and the mixing parameters are fixed by the
condition that the gauge-boson mass matrix is diagonal,
\beqar\label{massmatrix}
M^2_{{V}\bar{V}'}=\frac{1}{2}e^2 \vev^2 
\left\{{I}^{{V}},{I}^{\bar{V}'}\right\}_{HH} 
= \de_{VV'} M_{V}^2,
\eeqar
for $V,V'=A,Z,W^\pm$.
The curly brackets in \refeq{massmatrix} 
denote an anticommutator. 
The mass eigenvalues are given by
\beqar\label{massspectrum}
M_{\PWpm}&=&\frac{1}{2} \gw \vev
,\qquad
\MZ=\frac{1}{2\cw} \gw \vev
,\qquad
\MA=0.
\eeqar
The weak mixing angle $\theta_\mathrm{w}$ is related to the
gauge-boson masses via
\beqar\label{massrelation}
\cw =\frac{\MW}{\MZ}.
\eeqar
The vanishing mass of the photon is connected to the fact that the
electric charge of the vacuum expectation value
 is zero.  This provides the relation
\beqar\label{neutralhiggs}
\cw \gb Y_\Phi =\sw \gw,
\eeqar
between the weak mixing angle, the coupling constants $\gb$, $\gw$ and
the hypercharge $Y_\Phi$ of the Higgs doublet.  Identifying $e=\cw
\gb$, the Gell-Mann--Nishijima relation for general $Y_\Phi$ reads
$Q=Y/2+Y_\Phi T^3$.  In the calculation we keep $Y_\Phi$ as a free
parameter, which determines the degree of mixing in the gauge sector.
In this way our analysis applies to the Standard Model case,
$Y_\Phi=1$, as well as to the case $Y_\Phi=0$ corresponding to an
unmixed theory with
\beqar\label{unmixed}
\sw=0
,\qquad
\cw=1
,\qquad
Z_\mu=W^3_\mu
,\qquad
A_\mu=B_\mu
,\qquad
\MW=\MZ.
\eeqar

\subsection{Gauge interactions of massless fermions}

For the Feynman rules and various group-theoretical quantities we
adopt the formalism of \citere{Pozzorini:rs} (see Apps.~A and B
therein).  The Feynman rules for the
vector-boson--fermion--antifermion vertices read
\beqar\label{chiralcouplingsdef}
\vcenter{\hbox{
\begin{picture}(110,100)(-50,-50)
\Text(-45,5)[lb]{$V^{\mu}$}
\Text(35,30)[cb]{$\bar{f}_{\si'}$}
\Text(35,-30)[ct]{$f_{\si}$}
\Vertex(0,0){2}
\ArrowLine(0,0)(35,25)
\ArrowLine(35,-25)(0,0)
\Photon(0,0)(-45,0){2}{3}
\end{picture}}}
&=&\ri e \gamma^\mu \sum_{\kappa=\rR,\rL}
\omega_\kappa 
I^{V}_{f^\kappa_{\si'} f^\kappa_{\si}}
\,,
\eeqar
where $V=A,Z,\FWpm$, and $I^{V}_{f^\kappa_{\si'} f^\kappa_{\si}}$ are
the generators that describe SU(2)$\times$U(1) transformations of
fermions in the fundamental ($\kappa=\rL$) or trivial ($\kappa=\rR$)
representation.  For antifermions we have
\beq
I^V_{\bar f^\kappa_{\si'} \bar f^\kappa_\si}
=
- I^V_{f^\kappa_{\si} f^\kappa_{\si'}}\,.
\eeq
The chiral projectors $\omega_\kappa$ that are associated with the
gauge couplings \refeq{chiralcouplingsdef} can easily be shifted along
the fermionic lines using anticommutation relations%
\footnote{ In our calculation we use $\{\gamma^\mu,\gamma^5\}=0$ in
  $D=4-2\veps$ dimensions.  }  until they can be eliminated using
\refeq{spinors}.  As an example, for the coupling of a vector boson
$V$ to an incoming fermion $\varphi_i=f^{\kappa_i}_{\sigma_i}$ one
obtains
\beq
\frac{\ri(\ps_i+\qs)}{(p_i+q)^2}
\ri e \gamma^\mu
I^V_{\varphi'_i\varphi_i}
u(p_i,\kappa_i)
\qquad\mbox{with}\quad
I^V_{\varphi'_i\varphi_i}=
I^V_{f^{\kappa_i}_{\sigma'_i}
f^{\kappa_i}_{\sigma_i}}
\,,
\eeq
where the gauge coupling $I^V_{\varphi'_i\varphi_i}$ depends on the
chirality $\kappa_i$ of the fermion.  Similarly, for the coupling of a
vector boson to an incoming antifermion $\varphi_i={\bar
  f}^{\kappa_i}_{\sigma_i}$ one obtains
\beq
{\bar v}(p_i,\kappa_i)
\ri e \gamma^\mu
I^V_{\varphi'_i\varphi_i}
\frac{\ri(\ps_i+\qs)}{(p_i+q)^2}
\qquad\mbox{with}\quad
I^V_{\varphi'_i\varphi_i}
=
I^V_{\bar f^{\kappa_i}_{\sigma'_i}
\bar f^{\kappa_i}_{\sigma_i}}
=
- I^V_{f^{\kappa_i}_{\sigma_i}
f^{\kappa_i}_{\sigma'_i}}
\,,
\eeq
where the minus sign of the propagator is absorbed in the coupling of
the antifermion.

In our results, the matrix elements \refeq{matel} are often multiplied
by various matrices resulting from the couplings of gauge bosons to
the external fermions.  For such expressions we introduce the
shorthands
\beqar\label{abbcouplings}
\M I_k^{V_1}&=&\sum_{\varphi'_{k}}\M^{\varphi_{1}\dots\varphi'_{k}\dots\varphi_{n}} 
I^{V_1}_{\varphi'_{k}\varphi_{k}}
,\nl
\M I_k^{V_1}I_k^{V_2}&=&\sum_{\varphi'_{k},\varphi''_{k}}
\M^{\varphi_{1}\dots\varphi''_{{k}}\dots\varphi_{n}} 
I^{V_1}_{\varphi''_{k}\varphi'_{k}}
I^{V_2}_{\varphi'_{k}\varphi_{k}}
,
\eeqar
etc.  The gauge couplings in \refeq{abbcouplings} satisfy the
commutation relations
\beqar\label{commrela}
e \left[I_k^{V_1},
I_{k'}^{V_2}
\right]
=
\ri \gw \,
\delta_{kk'} 
\sum_{V_3=A,Z,W^\pm}
\teps^{V_1V_2V_3} I_k^{\bar{V}_3}
,
\eeqar
and the $\teps$-tensor is defined in \refeq{phystotalantitens}.

Global gauge invariance implies the charge-conservation relation 
\beqar\label{chargeconservation}
\M \sum_{k=1}^{n} I_k^{V}&=&0,
\eeqar
which is fulfilled up to mass-suppressed terms in the high-energy limit.

\section{Treatment of ultraviolet and mass singularities}
\label{se:singtreatment}

Large logarithms and $1/\veps$ poles originate from UV and from mass
singularities.  In this section we present the technique that we use
to extract these singularities from one- and two-loop Feynman
diagrams.  In \refse{se:msing} we identify the diagrams that are
responsible for mass singularities within the 't~Hooft--Feynman gauge
and classify their contributions into factorizable and
non-factorizable ones.  In \refse{se:softcollapp} we introduce an
approximation that describes the exchange of virtual gauge bosons in
the soft and collinear limit.  Our treatment of UV singularities is
discussed in \refse{se:uvsing}.

\subsection{Mass singularities}
\label{se:msing}
In gauge theories, as is well known, mass singularities appear in loop
diagrams involving virtual gauge bosons that couple to the on-shell
external legs.%
\footnote{ In principle it is possible to adopt particular gauge
  fixings in order to isolate mass singularities in a smaller subset
  of diagrams.  In the Coulomb gauge, for instance, collinear
  singularities appear only in external-leg self-energy corrections
  \cite{Beenakker:2000kb}.  However, we prefer to use the
  't~Hooft--Feynman gauge for our analysis of mass singularities. Thus,
  we must consider the most general set of Feynman diagrams which
  gives rise to mass singularities.}  
These singularities originate from the integration over the loop
momenta in the regions where the momenta of the virtual gauge bosons
become soft and/or collinear to the external momenta.

\subsubsection{Mass-singular diagrams at one loop}
At one loop, the mass singularities of the \mbox{$n$-fermion}
amplitude originate from diagrams of the type 
{ \unitlength 0.6pt\SetScale{0.6}
\beqar\label{oneloopdiag1}
&&
\vcenter{\hbox{\diagonenf{$\leg{i}$}{$\scriptscriptstyle{V}$}{\blob}}}
,
\eeqar
}%
where an electroweak gauge boson, $V=A,Z,W^\pm$, couples to one of the
external fermions, $i=1,\dots,n$, and to any other line of the
diagram.  These diagrams can be classified into three types:
\begin{itemize}
\item[(i)] The diagrams where the virtual gauge boson $V$ couples to
  the external leg $i$ and to another external leg $j$ with $j\neq i$,
{ \unitlength 0.6pt\SetScale{0.6}
\beqar\label{oneloopdiag2}
&&
\vcenter{\hbox{\diagone{$\leg{i}$}{$\leg{j}$}{$\scriptscriptstyle{V}$}{\blob}}}
.
\eeqar
}%
These diagrams produce single- and double-logarithmic mass
singularities that originate from the regions where the gauge-boson
momentum becomes soft and/or collinear to the momentum of the external
leg $i$ or $j$.

\item[(ii)] The external-leg self-energy insertions 
{\unitlength0.6pt\SetScale{0.6} 
\beqar\label{oneloopdiag3}
\vcenter{\hbox{\diagoneselfnf{$\scriptstyle{i}$}{$\scriptscriptstyle{V}$}{\blob}}}
.
\eeqar 
}%
These diagrams constitute a subset of the general mass-singular
diagrams depicted in \refeq{oneloopdiag1}. However, they have to be
omitted in our calculation since we express \mbox{$S$-matrix} elements
in terms of truncated Green functions and on-shell renormalized
fields.
 
\item[(iii)] The diagrams where the virtual gauge boson $V$ couples to
  the external line $i$ and to an internal line, \ie an internal
  propagator of the tree subdiagram that is represented as the grey
  blob in \refeq{oneloopdiag1}.  These diagrams produce only
  single-logarithmic mass singularities that originate from the region
  where the momenta of the gauge boson and the external fermion $i$
  become collinear.

\end{itemize}

\newcommand{\factdiag}[3]{
\begin{picture}(60.,64.)(-14.,-32.)
\Line(0.,0.)(30.,15.)
\Line(30.,-15.)(0.,0.)
\Text(34.,17.)[l]{#1}
\Text(34.,-17.)[l]{#2}
{#3}
\end{picture}
}

\subsubsection{Factorizable and non-factorizable contributions at one loop}
\label{se:factoneloop}

The diagrams of type (i) involve an \mbox{$n$-fermion} tree
subdiagram.  It is thus possible to extract from the diagrams
\refeq{oneloopdiag2} mass-singular contributions that factorize from
the \mbox{$n$-fermion} Born matrix element. Such contributions are called
factorizable and are defined as%
\footnote{Here we consider the case where the particles $i$ and $j$
  are a fermion and an antifermion, respectively.  The generalization
  to other combinations of particles or antiparticles is obvious.  }
{ \unitlength 0.6pt\SetScale{0.6}
\beqar\label{oneloopdiag4a}
&&
\vcenter{\hbox{\diagone{$\leg{i}$}{$\leg{j}$}{$\scriptscriptstyle{V}$}{\factblob}}}
=-\ri e^2 \muD^{4-D}
\sum_{\varphi'_i,\varphi'_j}
I^{ V}_{\varphi'_i\varphi_i}
I^{\bar V}_{\varphi'_j\varphi_j}
\int\frac{\rd^D q}{({2\pi})^D}
\frac{
1}{
(q^2-M_V^2)
(p_i+q)^2
(p_j-q)^2
}
\nl&&\quad{}\times
{\bar v(p_j,\kappa_j)\gamma^\mu 
(\ps_j-\qs)
G^{\underline{\varphi}_1\dots
\underline{\varphi}'_i\dots
\underline{\varphi}'_j\dots
\underline{\varphi}_{n}}
(p_1,\dots,p_i,\dots,p_j,\dots, p_{n})
(\ps_i+\qs)
\gamma_\mu u(p_i,\kappa_i)
}
.
\nln
\eeqar
}%
Here $G^{\underline{\varphi}_1\dots \underline{\varphi}'_i\dots
  \underline{\varphi}'_j\dots \underline{\varphi}_{n}} $ denotes the
truncated Green function corresponding to the \mbox{$n$-fermion} tree
subdiagram.  By definition, in the factorizable contributions (F) we
include only those parts of the above diagrams that are obtained by
performing the loop integration with the momentum $q$ of the gauge
boson $V$ set to zero in the tree subdiagram. This prescription is
indicated by the label F in the tree subdiagram.  In
\refeq{oneloopdiag4a} the spinors of the external fermions
$k=1,\dots,n$ with $k\neq i,j$ are implicitly understood.

By construction, the factorizable terms
\refeq{oneloopdiag4a} contain all one-loop soft singularities, since
the soft singularities originate only from the diagrams of type
\refeq{oneloopdiag2} in the region where the gauge-boson momentum
tends to zero.  Actually, as we will show, the factorizable
contributions \refeq{oneloopdiag4a} contain all one-loop mass
singularities, \ie not only all soft singularities but also all
collinear singularities.

The combination of all factorizable one-loop contributions is obtained
by summing over all gauge bosons $V=A,Z,W^\pm$ and external legs $i,j$
in \refeq{oneloopdiag4a}, { \unitlength 0.6pt\SetScale{0.6}
\beqar\label{oneloopdiag4}
\mel{1}{\fact}
&=&
\frac{1}{2}
\sum_{i=1}^{n}
\sum_{j=1 \atop j\neq i}^{n}
\sum_{V=A,Z,W^\pm}
\left[
\vcenter{\hbox{\diagone{$\leg{i}$}{$\leg{j}$}{$\scriptscriptstyle{V}$}{\factblob}}}
\right]_{q^\mu\to x p_i^\mu, x p_j^\mu}
.
\eeqar
}%
Note that the symmetry factor $1/2$ avoids double counting of the
pairs of external legs $i,j$.  The limit $q^\mu\to x p_i^\mu, x
p_j^\mu$ in \refeq{oneloopdiag4} indicates that the above diagrams are
evaluated in the approximation where the four-momentum $q^\mu$ of the
gauge boson $V$ is soft and/or collinear to one of the momenta of the
external legs $i$ and $j$.  This approximation, which is defined in
\refse{se:softcollapp}, permits to extract all mass singularities,
which result from soft and/or collinear regions.%
\footnote{Here we adopt a different approach as compared to
  \citere{Denner:2001jv}.  There the diagrams of type
  \refeq{oneloopdiag4} were evaluated using the eikonal approximation,
  which extracts only soft singularities, and the collinear
  singularities where treated separately.  }  As we will see, these
terms can be expressed as products of \mbox{$n$-fermion} Born matrix
elements and one-loop integrals.  The factorizable one-loop terms
\refeq{oneloopdiag4} are computed in \refse{se:oneloop}.

The remaining one-loop mass singularities are obtained by subtracting
the factorizable contributions \refeq{oneloopdiag4a}
and the external
self-energy contributions \refeq{oneloopdiag3} from the diagrams
\refeq{oneloopdiag1}, 
{ \unitlength 0.6pt\SetScale{0.6}
\beqar\label{oneloopdiag5}
\mel{1}{\nfact}
=
\sum_{i=1}^{n}
\sum_{V=A,Z,W^\pm}
&&
\left[
\vcenter{\hbox{
\diagonenf{$\leg{i}$}{$\scriptscriptstyle{V}$}{\blob}
}}
\right.
-
\vcenter{\hbox{
\diagoneselfnf{$\scriptstyle{i}$}{$\scriptscriptstyle{V}$}{\blob}
}}
\nl&&{}
-\sum_{j=1\atop j\neq i}^{n}
\left.
\vcenter{\hbox{
\diagonefact{$\scriptstyle{i}$}{$\scriptstyle{j}$}{$\scriptscriptstyle{V}$}{\factblob}
}}
\right]_{q^\mu\to x p_i^\mu}
.
\eeqar
}%
These contributions are called non-factorizable (NF).  We note that
these NF terms are free of soft singularities since all soft
singularities are contained in the factorizable parts and are
subtracted in \refeq{oneloopdiag5}.  We also observe that, in contrast
to \refeq{oneloopdiag4}, the sum over pairs of external legs $i,j$
appearing in front of the factorizable contributions that are
subtracted in \refeq{oneloopdiag5} is not multiplied by a symmetry
factor $1/2$.  This is due to the fact that in the subtraction terms
in \refeq{oneloopdiag5} we include only the contribution of the
collinear region $q^\mu\to x p_i^\mu$, whereas in \refeq{oneloopdiag4}
every diagram contains the mass singularities resulting from the two
regions $q^\mu\to x p_i^\mu$ and $q^\mu\to x p_j^\mu$.

In \refse{se:cwi} we will prove that the non-factorizable one-loop
terms \refeq{oneloopdiag5} vanish.

\subsubsection{Mass-singular diagrams at two loops}
The diagrams that give rise to NLL mass singularities at two loops,
\ie terms with triple and quartic logarithmic singularities, can be
obtained from the one-loop diagrams of type \refeq{oneloopdiag2},
which produce double logarithms in the soft--collinear region, by
inserting
\begin{itemize}
\item a second soft and/or collinear gauge boson that couples to an
  external line or to the virtual gauge boson in \refeq{oneloopdiag2}
  providing an additional single or double logarithm,
\item or a self-energy subdiagram in the propagator of the virtual
  gauge boson in \refeq{oneloopdiag2}, which provides an additional
  single logarithm.
\end{itemize}
There are five types of such diagrams:
{
\unitlength 0.6pt\SetScale{0.6}
\beq\label{twoloopdiag2}
\!\!
\vcenter{\hbox{\diagIInf{$\leg{i}$}{$\leg{j}$}{$\scriptscriptstyle{V_1}$}{$\scriptscriptstyle{V_2}$}{\blob}}}
,
\vcenter{\hbox{\diagIIInf{$\leg{i}$}{$\leg{j}$}{$\scriptscriptstyle{V_1}$}{$\scriptscriptstyle{V_2}$}{\blob}}}
,
\vcenter{\hbox{\diagIVnf{$\leg{j}$}{$\leg{i}$}{$\leg{k}$}{$\scriptscriptstyle{V_2}$}{$\scriptscriptstyle{V_1}$}{\blob}}}
,
\vcenter{\hbox{\diagInf{$\leg{i}$}{$\leg{j}$}{$\scriptscriptstyle{V_1}$}{$\scriptscriptstyle{V_3}$}{$\scriptscriptstyle{V_2}$}{\blob}}}
,
\vcenter{\hbox{\diagself{$\scriptscriptstyle{V_1}$}{$\scriptscriptstyle{V_2}$}{\blob}}}
.
\eeq
}%
Here the NLL mass singularities originate from the regions where the
gauge boson $V_1$, which couples to two external legs or to an
external leg and another virtual gauge boson, is soft and collinear
and the gauge bosons $V_2$ and $V_3$, in the first four diagrams in
\refeq{twoloopdiag2}, are soft and/or collinear.

\subsubsection{Factorizable and non-factorizable contributions at two loops}
\label{se:facttwoloop}
The two-loop mass singularities resulting from the diagrams
\refeq{twoloopdiag2} are split into factorizable and non-factorizable
contributions.
The factorizable contributions result from the diagrams that contain
\mbox{$n$-fermion} tree subdiagrams, 
{\unitlength 0.6pt \SetScale{0.6}
\beqar\label{twoloopdiag3}
\lefteqn{
\mel{2}{\fact}
=
\sum_{i=1}^{n}
\sum_{j=1\atop j\neq i}^{n}
\sum_{V_{m}=A,Z,W^\pm}
\left\{
\frac{1}{2}
\left[
\vcenter{\hbox{
\diagI{$\leg{i}$}{$\leg{j}$}{$\scriptscriptstyle{V_1}$}{$\scriptscriptstyle{V_2}$}{\factblob}
}}
+\vcenter{\hbox{
\diagII{$\leg{i}$}{$\leg{j}$}{$\scriptscriptstyle{V_1}$}{$\scriptscriptstyle{V_2}$}{\factblob}
}}
\right]
\right.
}\quad&&\nl&&{}
+
\vcenter{\hbox{
\diagIII{$\leg{i}$}{$\leg{j}$}{$\scriptscriptstyle{V_1}$}{$\scriptscriptstyle{V_3}$}{$\scriptscriptstyle{V_2}$}{\factblob}}}
+\vcenter{\hbox{\diagV{$\leg{i}$}{$\leg{j}$}{$\scriptscriptstyle{V_1}$}{$\scriptscriptstyle{V_2}$}
{\factblob}}}
+\vcenter{\hbox{\diagVII{$\leg{i}$}{$\leg{j}$}{$\scriptscriptstyle{V_1}$}{$\scriptscriptstyle{V_2}$}{\factblob}}}
+\frac{1}{2}
\vcenter{\hbox{\diagself{$\scriptscriptstyle{V_1}$}{$\scriptscriptstyle{V_2}$}{\factblob}}}
\nl&&{}
+
\sum_{k=1\atop k\neq i,j}^{n}
\left[
\vcenter{\hbox{\diagXX{$\leg{j}$}{$\leg{i}$}{$\leg{k}$}{$\scriptscriptstyle{V_1}$}{$\scriptscriptstyle{V_2}$}{\factblob}}}
\right.
+\frac{1}{6}
\left.
\vcenter{\hbox{\diagXXI{$\leg{j}$}{$\leg{i}$}{$\leg{k}$}{$\scriptscriptstyle{V_2}$}{$\scriptscriptstyle{V_1}$}{$\scriptscriptstyle{V_3}$}{\factblob}}}
\right]
+\frac{1}{8}
\sum_{k=1\atop k\neq i,j}^{n}
\sum_{l=1\atop l\neq i,j,k}^{n}
\left.
\vcenter{\hbox{\diagXXII{$\leg{i}$}{$\leg{j}$}{$\leg{k}$}{$\leg{l}$}{$\scriptscriptstyle{V_1}$}{$\scriptscriptstyle{V_2}$}{\factblob}}}
\right\}_{q^\mu \to x p^\mu}
\!\!.
\nln
\eeqar
}%
As in the one-loop case, these factorizable contributions are defined
as those parts of the diagrams \refeq{twoloopdiag3} that are obtained
by performing the loop integrations with the momenta of the gauge
bosons $V_1,V_2,V_3$ set to zero in the tree subdiagrams. This
prescription is indicated by the label F in the tree subdiagrams in
\refeq{twoloopdiag3}.

By construction, the factorizable terms \refeq{twoloopdiag3} contain
all two-loop soft--soft singularities, \ie the singularities that
originate from the region where all gauge bosons $V_1,V_2,V_3$ are
soft.  Actually, as we will show, the factorizable contributions
\refeq{twoloopdiag3} contain all two-loop NLL mass singularities, \ie
not only all soft singularities but also all collinear singularities.

The symmetry factors $1/2$, $1/6$ and $1/8$ in \refeq{twoloopdiag3}
avoid double counting in the sums over combinations of external legs
$i,j,k,l$.  The limit $q^\mu\to x p^\mu$ in \refeq{twoloopdiag3}
indicates that the above diagrams are evaluated in the approximation
where each of the four-momenta $q^\mu$ of the various gauge bosons is
collinear to one of the momenta $p^\mu$ of the external legs and/or
soft. Where relevant, also the contributions of hard regions are taken
into account (see \refse{se:softcollapp}).  The factorizable two-loop
terms \refeq{twoloopdiag3} are computed in \refse{se:twoloop}.

The remaining two-loop NLL mass singularities, which we call
non-factorizable (NF), are obtained by subtracting from the diagrams
of type \refeq{twoloopdiag2} the factorizable terms
\refeq{twoloopdiag3} and external self-energy diagrams.  There are
four sets of non-factorizable contributions that are associated with
the first four diagrams in \refeq{twoloopdiag2}, whereas the last
diagram in \refeq{twoloopdiag2} does not produce non-factorizable NLL
terms.  This is due to the fact that all NLL mass singularities
resulting from the last diagram in \refeq{twoloopdiag2} originate from
the region where the momenta of the gauge bosons $V_1,V_2$ are soft
and are thus included in the factorizable part of this diagram.

The non-factorizable terms associated with the first diagram in
\refeq{twoloopdiag2} read {\unitlength 0.6pt\SetScale{0.6}
\beqar\label{twoloopnf1}
\mel{2}{\nfact,A}
&=&
\sum_{i=1}^{n}
\sum_{j=1\atop j\neq i}^{n}
\sum_{V_{m}=A,Z,W^\pm}
\left[
\vcenter{\hbox{\diagIInf{$\leg{i}$}{$\leg{j}$}{$\scriptscriptstyle{V_1}$}{$\scriptscriptstyle{V_2}$}{\blob}}}
-
\vcenter{\hbox{\diagV{$\leg{i}$}{$\leg{j}$}{$\scriptscriptstyle{V_1}$}{$\scriptscriptstyle{V_2}$}
{\factblob}}}
-
\vcenter{\hbox{
\diagI{$\leg{i}$}{$\leg{j}$}{$\scriptscriptstyle{V_1}$}{$\scriptscriptstyle{V_2}$}{\factblob}
}}
\right.\nl&&\left.{}
-\sum_{k=1\atop k\neq i,j}^{n}
\vcenter{\hbox{\diagXX{$\leg{j}$}{$\leg{i}$}{$\leg{k}$}{$\scriptscriptstyle{V_1}$}{$\scriptscriptstyle{V_2}$}{\factblob}}}
\right]_{q_2^\mu\to x p_i^\mu \atop q^\mu_1 \to 0}
.
\eeqar
}%
The non-factorizable contributions associated with the second diagram
in \refeq{twoloopdiag2} are { \unitlength 0.6pt\SetScale{0.6}
\beqar\label{twoloopnf2}
\mel{2}{\nfact,B}
&=&
\sum_{i=1}^{n}
\sum_{j=1\atop j\neq i}^{n}
\sum_{V_{m}=A,Z,W^\pm}
\left[
\vcenter{\hbox{\diagIIInf{$\leg{i}$}{$\leg{j}$}{$\scriptscriptstyle{V_1}$}{$\scriptscriptstyle{V_2}$}{\blob}}}
-
\vcenter{\hbox{\diagVII{$\leg{i}$}{$\leg{j}$}{$\scriptscriptstyle{V_1}$}{$\scriptscriptstyle{V_2}$}{\factblob}}}
-
\vcenter{\hbox{
\diagII{$\leg{i}$}{$\leg{j}$}{$\scriptscriptstyle{V_2}$}{$\scriptscriptstyle{V_1}$}{\factblob}
}}
\right.\nl&&{}\left.
-\sum_{k=1\atop k\neq i,j}^{n}
\vcenter{\hbox{\diagXX{$\leg{k}$}{$\leg{i}$}{$\leg{j}$}{$\scriptscriptstyle{V_2}$}{$\scriptscriptstyle{V_1}$}{\factblob}}}
\right]_{q_2^\mu\to x p_i^\mu \atop q^\mu_1 \to 0}
.
\eeqar
}%
The non-factorizable contributions associated with the third diagram
in \refeq{twoloopdiag2} are { \unitlength 0.6pt\SetScale{0.6}
\beqar\label{twoloopnf3}
\mel{2}{\nfact,C}
&=&
\frac{1}{2}
\sum_{i=1}^{n}
\sum_{j=1\atop j\neq i}^{n}
\sum_{k=1\atop k\neq i,j}^{n}
\sum_{V_{m}=A,Z,W^\pm}
\left[
\vcenter{\hbox{\diagIVnf{$\leg{i}$}{$\leg{j}$}{$\leg{k}$}{$\scriptscriptstyle{V_2}$}{$\scriptscriptstyle{V_1}$}{\blob}}}
-
\vcenter{\hbox{\diagXXd{$\leg{i}$}{$\leg{j}$}{$\leg{k}$}{$\scriptscriptstyle{V_2}$}{$\scriptscriptstyle{V_1}$}{\blob}}}
-
\vcenter{\hbox{\diagXX{$\leg{k}$}{$\leg{j}$}{$\leg{i}$}{$\scriptscriptstyle{V_1}$}{$\scriptscriptstyle{V_2}$}{\factblob}}}
\right.\nl&&{}\left.
-
\vcenter{\hbox{\diagXX{$\leg{j}$}{$\leg{k}$}{$\leg{i}$}{$\scriptscriptstyle{V_1}$}{$\scriptscriptstyle{V_2}$}{\factblob}}}
-\sum_{l=1\atop l\neq i,j,k}^{n}
\vcenter{\hbox{\diagXXII{$\leg{l}$}{$\leg{i}$}{$\leg{j}$}{$\leg{k}$}{$\scriptscriptstyle{V_2}$}{$\scriptscriptstyle{V_1}$}{\factblob}}}
\right]_{q_2^\mu\to x p_i^\mu \atop q^\mu_1 \to 0}
.
\eeqar
}%
Here the first subtraction term represents an external-leg self-energy
insertion and the symmetry factor $1/2$ is introduced in order to
avoid double counting of the terms resulting from the permutation of
the external legs $j$ and $k$.  The non-factorizable contributions
associated with the fourth diagram in \refeq{twoloopdiag2} read {
  \unitlength 0.6pt\SetScale{0.6}
\beqar\label{twoloopnf4}
\mel{2}{\nfact,D}
&=&
\sum_{i=1}^{n}
\sum_{j=1\atop j\neq i}^{n}
\sum_{V_{m}=A,Z,W^\pm}
\left[
\vcenter{\hbox{\diagInf{$\leg{i}$}{$\leg{j}$}{$\scriptscriptstyle{V_2}$}{$\scriptscriptstyle{V_1}$}{$\scriptscriptstyle{V_3}$}{\blob}}}
-
\vcenter{\hbox{\diagIII{$\leg{i}$}{$\leg{j}$}{$\scriptscriptstyle{V_2}$}{$\scriptscriptstyle{V_1}$}{$\scriptscriptstyle{V_3}$}{\factblob}}}
-
\vcenter{\hbox{\diagIII{$\leg{j}$}{$\leg{i}$}{$\scriptscriptstyle{V_1}$}{$\scriptscriptstyle{V_2}$}{$\scriptscriptstyle{V_3}$}{\factblob}}}
\right.\nl&&{}\left.
-\sum_{k=1\atop k\neq i,j}^{n}
\vcenter{\hbox{\diagXXI{$\leg{k}$}{$\leg{i}$}{$\leg{j}$}{$\scriptscriptstyle{V_3}$}{$\scriptscriptstyle{V_2}$}{$\scriptscriptstyle{V_1}$}{\factblob}}}
\right]_{q_2^\mu\to x p_i^\mu \atop q^\mu_1 \to 0}
.
\eeqar
}%
As indicated by the limits $q_2^\mu\to x p_i^\mu$, $q^\mu_1 \to 0$ in
\refeq{twoloopnf1}--\refeq{twoloopnf4}, we extract the NLL mass
singularities that appear in the above combinations of diagrams in the
regions where the momentum $q_1^\mu$ of the gauge boson $V_1$ is soft
and the momentum $q_2^\mu$ of the gauge boson $V_2$ is collinear to
the external momentum $p_i^\mu$.  The terms
\refeq{twoloopnf1}--\refeq{twoloopnf4} are free of singularities in
the soft limit $x\to0$ since all soft--soft singularities are included
in the factorizable parts that are subtracted.

The subtraction of the factorizable terms must be performed without
double-counting of diagrams and regions, \ie the terms that are
subtracted in \refeq{twoloopnf1}--\refeq{twoloopnf4} must correspond
exactly to the ones that are included in \refeq{twoloopdiag3}.  This
correspondence is not obvious at first sight, since certain diagrams
appear a different number of times or with different symmetry factors
in \refeq{twoloopdiag3} and \refeq{twoloopnf1}--\refeq{twoloopnf4}.
This is due to the fact that in \refeq{twoloopnf1}--\refeq{twoloopnf4}
the contributions of certain diagrams are split into various terms
that result from different singular regions, whereas in
\refeq{twoloopdiag3} every diagram includes the contributions of all
singular regions.  For instance, let us consider the last diagram in
\refeq{twoloopnf1}.  This diagram appears only once in
\refeq{twoloopdiag3} but is subtracted four times in
\refeq{twoloopnf1}--\refeq{twoloopnf4}.  These four subtraction terms
can be rewritten as { \unitlength 0.6pt\SetScale{0.6}
\beqar\label{twoloopnf5}
\sum_{i=1}^{n}
\sum_{j=1\atop j\neq i}^{n}
\sum_{k=1\atop k\neq i,j}^{n}
\sum_{V_{m}=A,Z,W^\pm}
&&
\left\{
\left[
\vcenter{\hbox{\diagXX{$\leg{j}$}{$\leg{i}$}{$\leg{k}$}{$\scriptscriptstyle{V_a}$}{$\scriptscriptstyle{V_b}$}{\factblob}}}
\right]_{q_a^\mu\to 0 \atop q_b^\mu \to x p_i^\mu }
\right.
+
\left[
\vcenter{\hbox{\diagXX{$\leg{j}$}{$\leg{i}$}{$\leg{k}$}{$\scriptscriptstyle{V_a}$}{$\scriptscriptstyle{V_b}$}{\factblob}}}
\right]_{q_a^\mu\to  x p_i^\mu \atop q_b^\mu \to 0 }
\nl&&{}+
\left.
\left[
\vcenter{\hbox{\diagXX{$\leg{j}$}{$\leg{i}$}{$\leg{k}$}{$\scriptscriptstyle{V_a}$}{$\scriptscriptstyle{V_b}$}{\factblob}}}
\right]_{q_a^\mu\to 0 \atop q_b^\mu \to x p_k^\mu }
\right\}
,
\eeqar
}%
where the last term in \refeq{twoloopnf5} corresponds to the two terms
that are multiplied with a symmetry factor $1/2$ in
\refeq{twoloopnf3}.  As can easily be seen in \refeq{twoloopnf5},
these three contributions are associated with the three
non-overlapping singular regions of the diagram that give rise to NLL
terms,%
\footnote{The region ${q_a^\mu\to x p_j^\mu}$, ${ q_b^\mu \to 0 }$
  does not give rise to NLL mass singularities.  } and their sum
corresponds to the complete contribution included in
\refeq{twoloopdiag3}.  Similarly, one can verify that all other
subtraction terms in \refeq{twoloopnf1}--\refeq{twoloopnf4} correspond
exactly to the factorizable contributions included in
\refeq{twoloopdiag3}.

In \refse{se:cwi} we will prove that the non-factorizable
contributions \refeq{twoloopnf1}--\refeq{twoloopnf4} cancel.

\subsection{Soft--collinear approximation for virtual gauge-boson exchange}
\label{se:softcollapp}
As discussed in the previous section, mass singularities appear when
the external fermions emit virtual gauge bosons in the soft and
collinear regions.  Here we introduce a soft--collinear approximation
that describes the gauge-boson--fermion--antifermion vertices in these
regions.  This approximation permits to derive the soft and collinear
singularities of generic \mbox{$n$-fermion} amplitudes in a
process-independent way.

Let us first consider an \mbox{$n$-fermion} diagram involving the exchange of
a virtual gauge boson $V=A,Z,W^\pm$ between the external leg $i$ and
some other (external or internal) line,%
\footnote{For the moment, we consider the case where the external
  particle $i$ is a fermion.  The generalization to antifermions is
  discussed below.   }
\beq\label{scapp1}
\vcenter{\hbox{
\unitlength 0.6pt \SetScale{0.6}
\diagIcollin{$\scriptstyle{i}$}{$\scriptstyle{\bar V^\mu}$}{$\scriptstyle{V^\mu}$}{\blob}
}}
=
\sum_{\varphi'_i}
G_\mu^{\bar{\underline{V}}\underline{\varphi}'_i}(-q,p_i+q)
\frac{\ri(\ps_i+\qs)}{(p_i+q)^2}
\ri e \gamma^\mu 
I^V_{\varphi'_i\varphi_i}
u(p_i,\kappa_i)
.
\eeq
Here $G_\mu^{\bar{\underline{V}}\underline{\varphi}'_i}(-q,p_i+q)$
represents the truncated Green function corresponding to the internal
part of the diagram, which is depicted as a grey blob.  The
contraction with the spinors of the external fermions $j=1,\dots,n$
with $j\neq i$ is implicitly understood.  The loop momentum of the
gauge boson is denoted as $q$ and the propagator that connects the
gauge boson lines $V^\mu$ and $\bar{V}^\mu$ has been omitted.
Commuting the Dirac matrices associated with the fermion propagator
and the gauge-boson--fermion vertex and using the massless Dirac
equation we have
\beq\label{collappid}
\ri(\ps_i+\qs) \ri e \gamma^\mu  u(p_i,\kappa_i)
= -e\left[2(p_i+q)^\mu - \gamma^\mu \qs \right]u(p_i,\kappa_i).
\eeq
In the collinear limit $q^\mu\to x p_i^\mu$ the $\qs u(p_i,\kappa_i)$
term vanishes owing to the Dirac equation. Thus, we obtain
\beq\label{collapp0}
\lim_{q^\mu\to x p_i^\mu }
\vcenter{\hbox{
\unitlength 0.6pt \SetScale{0.6}
\diagIcollin{$\scriptstyle{i}$}{$\scriptstyle{\bar V^\mu}$}{$\scriptstyle{V^\mu}$}{\blob}
}}
=
G_\mu^{\bar{\underline{V}}\,\underline{i}}(-q,p_i+q)
u(p_i,\kappa_i)
\frac{-2 e I^V_{i}
(p_i+q)^\mu
}{(p_i+q)^2},
\eeq
where we have introduced the shorthand
$G_\mu^{\bar{\underline{V}}\,\underline{i}}(-q,p_i+q) I^V_{i} =
\sum_{\varphi'_i}
G_\mu^{\bar{\underline{V}}\underline{\varphi}'_i}(-q,p_i+q)
I^V_{\varphi'_i\varphi_i} $.
The approximation \refeq{collapp0} is applicable also to the case
where the gauge boson $V$ becomes collinear to another leg $j\neq i$.
In this case, the term $\gamma^\mu\qs$ on the right-hand side of
\refeq{collappid} is contracted with another soft--collinear factor
$(p_j-q)_\mu$ resulting from the coupling of the gauge boson $V$ to
the leg $j$ and
\beq
\lim_{q^\mu\to x p_j^\mu}
(p_j-q)_\mu \gamma^\mu \qs=(1-x)x \ps_j^2=0.
\eeq
Similarly, \refeq{collapp0} applies also to the case where the gauge
boson $V$ splits into two gauge bosons, $V'$ and $V''$, with $V'$
being collinear to an external leg $j\neq i$ and $V''$ soft. Here, the
combination of the soft--collinear factor associated with the external
leg $j$, the triple gauge-boson vertex and the term $\gamma^\mu\qs$ on
the right-hand side of \refeq{collappid} yields
\beq
\lim_{q^\mu\to x p_j^\mu}
(p_j-q)^{\mu'} 
\left[
-2g_{\mu\mu'} q_{\mu''}
+g_{\mu'\mu''} q_{\mu}
+g_{\mu\mu''} q_{\mu'}
\right]
\gamma^\mu \qs
=0.
\eeq
For the multiple emission of collinear gauge bosons $V_1 \dots V_n$ by
an external fermion $i$ we obtain
\beqar\label{collapp1}
\lim_{q_k^\mu\to x_k p_i^\mu }
\vcenter{\hbox{
\unitlength 0.6pt \SetScale{0.6}
\diagIIcollin{$\scriptstyle{\bar V_n^{\mu_n}}$\footnotesize{\dots}
$\scriptstyle{\bar V_1^{\mu_1}}$}{}{$\scriptstyle{i}$}%
{$\scriptstyle{V_n^{\mu_n}}$}{$\scriptstyle{V_1^{\mu_1}}$}{\blob}
}}
&=&
G_{\mu_1 \dots \mu_n}^{\bar{\underline{V}}_1 \dots
  \bar{\underline{V}}_n\,\underline{i}}(-q_1,\dots,-q_n, p_i+\tilde q_n)\,
u(p_i,\kappa_i)
\nl&&{}\times
\frac{-2 e I^{V_n}_i
(p_i+\tilde q_n)^{\mu_n}
}{(p_i+\tilde q_n)^2}
\cdots
\frac{-2 e I^{V_1}_i
(p_i+q_1)^{\mu_1}
}{(p_i+q_1)^2}
,
\nln
\eeqar
where $\tilde q_j=q_1+\dots+q_j$.  This approximation is also
applicable to all cases where one of the gauge bosons $V_1,\dots, V_n$
becomes collinear to one of the other external legs $j\neq i$ and all
remaining gauge bosons are soft.  Therefore \refeq{collapp1} provides
a correct description of the gauge-boson--fermion couplings in all
regions that are relevant for our NLL analysis.

Also for the case where the external line $i$ represents an
antifermion, the emission of collinear gauge bosons produces factors
$-2e I^{V_k}_i(p_i+\tilde q_k)^{\mu_k}$ and, apart from the obvious
replacement of the spinor $u(p_i,\kappa_i)$ by $\bar v(p_i,\kappa_i)$,
we obtain exactly the same formula as in \refeq{collapp1}.

The soft--collinear approximation \refeq{collapp1} permits to replace
the Dirac matrices associated with each gauge-boson emission by simple
four-vector factors $-2(p_i+\tilde q_k)^{\mu_k}$.  In the soft limit,
$q_j^\mu\to 0$, these factors correspond to the well-known factors
$-2p_i^{\mu_k}$ that are used to derive soft singularities in the
eikonal approximation.  The soft--collinear approximation
\refeq{collapp1} can be regarded as an extension of the eikonal
approximation that permits to describe the emission of gauge bosons in
the soft and the collinear regions.  When applied to the one- and
two-loop factorizable contributions \refeq{oneloopdiag4} and
\refeq{twoloopdiag3}, this approximation permits to factorize the mass
singularities from the \mbox{$n$-fermion} Born amplitude explicitly.

We note that the soft--collinear approximation is not
applicable in the case where NLL two-loop contributions arise as a
combination of logarithmic singularities originating from
soft--collinear and UV regions.  In particular, for the two-loop
factorizable terms of the type {
\beqar\label{oneloopsubd1}
\unitlength 0.6pt\SetScale{0.6}
\vcenter{\hbox{\diagself{$\scriptscriptstyle{V}$}{$\scriptscriptstyle{V'}$}{\factblob}}}
,
\vcenter{\hbox{
\unitlength 0.6pt\SetScale{0.6}
\subloopII{$\leg{i}$}{$\leg{j}$}{$\scriptscriptstyle{V}$}{\factblob}}}
,
\vcenter{\hbox{
\unitlength 0.6pt\SetScale{0.6}
\subloopI{$\leg{i}$}{$\leg{j}$}{$\scriptscriptstyle{V}$}{\factblob}}}
,
\eeqar
}%
where a soft--collinear singularity resulting from the exchange of the
gauge bosons $V$ (and $V'$) appears in combination with an UV
singularity resulting from hard particles in one-loop subdiagrams,
the approximation \refeq{collapp1} can be applied only
to the vertices that occur outside the UV-divergent one-loop
subdiagrams whereas for the vertices and propagators inside the
one-loop subdiagrams we have to apply the usual Feynman rules.

In the case of the last two diagrams in \refeq{oneloopsubd1} this
approximation is not sufficient to eliminate the chain of Dirac
matrices along the external fermionic leg $i$.  However, this chain of
Dirac matrices can be contracted with the spinor of the external
fermion by means of a simple projector.  Let us illustrate how this is
achieved for the second diagram in \refeq{oneloopsubd1}, 
\beqar\label{oneloopsubd2}
\vcenter{\hbox{
\unitlength 0.6pt\SetScale{0.6}
\subloopII{$\leg{i}$}{$\leg{j}$}{$\scriptscriptstyle{V}$}{\factblob}}}
&&
=
-2 e
I^{\bar V}_{\varphi'_{j}\varphi_{j}}
G^{\underline{\varphi}'_i}(p_i)
X^{V\varphi_i\bar \varphi'_i}(p_i,p_j)
u(p_i,\kappa_i)
\eeqar
with
\beqar\label{diracstruct}
X^{V\varphi_i\bar \varphi'_i}(p_i,p_j)
&=&
\muD^{4-D}
\int\frac{\rd^D q}{({2\pi})^D}
\frac{
(p_j-q)^\mu
(\ps_i+\qs)
\ri \Gamma_\mu^{V\varphi_i\bar \varphi'_i}(q,p_i)
}{(q^2-M_{V}^2)(p_i+q)^2(p_j-q)^2}
.
\eeqar
Here the factor $(p_j-q)^\mu$ comes from the soft--collinear vertex
along the leg $j$, and $\ri \Gamma_\mu^{V\varphi_i\bar \varphi'_i}$
represents the one-loop vertex.  The truncated Green function
$G^{\underline{\varphi}'_i}(p_i)$ in \refeq{oneloopsubd2} represents
the internal part of the diagram, where the momentum $q$ of the gauge
boson $V$ is set to zero according to the definition of the
factorizable part.  The contraction with the spinors of the external
fermions $j=1,\dots,n$ with $j\neq i$ is implicitly understood.  If
one contracts this Green function with the spinor of the fermion $i$
one simply obtains the on-shell Born matrix element,
\beq\label{bornterm1}
G^{\underline{\varphi}'_i}(p_i)
u(p_i,\kappa_i)
=
\mel{0}{\varphi_{1}\dots\varphi'_{i}\dots\varphi'_{j}\dots \varphi_{n}} 
.
\eeq
The chain of Dirac matrices occurring in \refeq{diracstruct} can be projected 
on the Dirac spinor using%
\footnote{This identity can easily be verified by means of the general
  decomposition
$$
X^{V\varphi_i\bar \varphi'_i}(p_i,p_j)
\omega_{\kappa}
=\sum_{k=0}^{N}
\sum_{l_1=i,j}
\dots
\sum_{l_{2k}=i,j}
A_{l_1\dots l_{2k}}(p_i,p_j)
\ps_{l_1}\dots\ps_{l_{2k}}
\omega_{\kappa}
,
$$
the Dirac equation, $\ps_i u(p_i,\kappa_i)=0$, the anticommutation
relation $\{\ps_i,\ps_j\}=2 p_i p_j$, the identities
$\ps_i^2=\ps_j^2=0$, and the fact that, for massless fermions, the
subamplitude $X^{V\varphi_i\bar \varphi'_i}(p_i,p_j)$ contains only
Dirac chains with even numbers $2k$ of Dirac matrices.  }
\beqar\label{projection1}
X^{V\varphi_i\bar \varphi'_i}(p_i,p_j)
u(p_i,\kappa_i)
=
\frac{1}{2 p_i p_j}
\Tr\left[
X^{V\varphi_i\bar \varphi'_i}(p_i,p_j)
\omega_{\kappa_i}
\ps_i\ps_j
\right]
u(p_i,\kappa_i),
\eeqar
so that one obtains
\beq\label{projection2}
G^{\underline{\varphi}'_i}(p_i)
X^{V\varphi_i\bar \varphi'_i}(p_i,p_j)
u(p_i,\kappa_i)
=
\frac{1}{2 p_i p_j}
\Tr\left[
X^{V\varphi_i\bar \varphi'_i}(p_i,p_j)
\omega_{\kappa_i}
\ps_i\ps_j
\right]
\mel{0}{\varphi_{1}\dots\varphi'_{i}\dots \varphi'_{j}\dots \varphi_{n}} 
.\quad
\eeq
The same trace projector can be applied to the last diagram in
\refeq{oneloopsubd1}, { \unitlength 0.6pt\SetScale{0.6}
\beqar
&&
\hspace{-10mm}
\vcenter{\hbox{\subloopI{$\leg{i}$}{$\leg{j}$}{$\scriptscriptstyle{V}$}{\factblob}}}
=
2 e^2 
I^{V}_{\varphi'_{i}\varphi_{i}}
I^{\bar V}_{\varphi'_{j}\varphi_{j}}
G^{\underline{\varphi}'_i}(p_i)
X^{\varphi'_i\bar \varphi'_i}(p_i,p_j)
u(p_i,\kappa_i)
,
\eeqar
}%
where 
\beq\label{diracstructb}
X^{\varphi'_i\bar \varphi'_i}(p_i,p_j)
=
\muD^{4-D}
\int\frac{\rd^D q}{({2\pi})^D}
\frac{ 
(p_j-q)^\mu
(\ps_i+\qs)
\ri \Sigma^{\varphi'_i\bar \varphi'_i}(p_i+q)
(\ps_i+\qs)
\gamma_\mu
}{(q^2-M_{V}^2)\left[(p_i+q)^2\right]^2(p_j-q)^2}
,\quad
\eeq
and $\ri \Sigma^{\varphi'_i\bar \varphi'_i}$ represents the fermionic
self-energy.  Again we can project \refeq{diracstructb} on the Dirac
spinor using
\beqar\label{projectionb}
G^{\underline{\varphi}'_i}(p_i)
X^{\varphi_i'\bar \varphi'_i}(p_i,p_j)
u(p_i,\kappa_i)
=
\frac{1}{2 p_i p_j}
\Tr\left[
X^{\varphi'_i\bar \varphi'_i}(p_i,p_j)
\omega_{\kappa_i}
\ps_i\ps_j
\right]
\mel{0}{\varphi_{1}\dots\varphi'_{i}\dots \varphi'_{j}\dots \varphi_{n}}. 
\quad
\eeqar

\subsection{Ultraviolet singularities}
\label{se:uvsing}
Let us first discuss our treatment of UV singularities at one loop.
The UV singularities appearing in the bare loop amplitudes are
cancelled by corresponding singularities provided by the counterterms.
These cancellations give rise to logarithmic contributions of the form
\beq\label{uvsing1}
\left(\frac{\muD^2}{Q^2}\right)^{\veps}
\left[
\frac{1}{\veps}
\left(\frac{Q^2}{\mu_{\mathrm{loop}}^2}\right)^{\veps}
-
\frac{1}{\veps}
\left(\frac{Q^2}{\mu_{\mathrm{R}}^2}\right)^{\veps}
\right]
=
\ln\left(\frac{\mu_{\mathrm{R}}^2}{\mu_{\mathrm{loop}}^2}\right)
+\order(\veps),
\eeq
where the first and the second term between the brackets result from
bare loop diagrams and counterterms, respectively.  Here $\muD$ is the
scale of dimensional regularization and we have factorized the term
$\left({\muD^2}/{Q^2}\right)^{\veps}$ that we always absorb in
$\alphaeps$ [see \refeq{pertserie1}].  The renormalization scale is
denoted as $\mu_{\mathrm{R}}$, and $\mu_{\mathrm{loop}}$ represents
the characteristic scale of the UV-singular loop diagram.  This latter
is related to the momenta of the lines that enter the loop.

At one loop, since in the high-energy limit all combinations of
external momenta are hard, the characteristic scale of UV-divergent
diagrams that contribute to truncated \mbox{$n$-fermion} Green functions is
always of the order $\mu_{\mathrm{loop}}^2\sim Q^2$.  This permits us
to isolate all logarithms that result from UV singularities, \ie terms
of the type \refeq{uvsing1}, in the counterterms. To this end, in our
calculation we perform a minimal subtraction of all UV poles that
appear in the bare loop diagrams and in the counterterms. The
combination of these subtracted terms corresponds to
\beq\label{uvsing2}
\left(\frac{\muD^2}{Q^2}\right)^{\veps}
\left\{
\frac{1}{\veps}
\left[\left(\frac{Q^2}{\mu_{\mathrm{loop}}^2}\right)^{\veps}-1\right]
-
\frac{1}{\veps}
\left[\left(\frac{Q^2}{\mu_{\mathrm{R}}^2}\right)^{\veps}-1\right]
\right\}
\eeq
and is obviously equivalent to \refeq{uvsing1}.  As a result of the
minimal subtraction the logarithmic contributions
${\veps}^{-1}[(Q^2/{\mu_{\mathrm{loop}}^2})^{\veps}-1]$ originating
from bare loop diagrams that are characterized by a hard scale
${\mu_{\mathrm{loop}}^2}\sim Q^2$ vanish.

Thus, at one loop we can restrict ourselves to the calculation of the
mass-singular bare diagrams and the counterterms.  The minimal
subtraction of the UV singularities appearing in these two types of
contributions permits us to ignore any other bare diagram that
produces UV singularities.

At two loops, pure UV singularities produce only NNLL contributions
since every UV-singular loop produces only a single-logarithmic term.
The only NLL two-loop terms resulting from UV singularities are
combinations of one-loop UV logarithms with one-loop double logarithms
resulting from soft--collinear gauge bosons.  These terms originate
from one-loop insertions in the one-loop diagrams
\refeq{oneloopdiag2}.  Here, as in the one-loop case, we perform a
minimal subtraction of the UV singularity, such that the logarithms
associated with one-loop subdivergences from hard subdiagrams with
${\mu_{\mathrm{loop}}^2}\sim Q^2$ are completely isolated in the
counterterms.  As a consequence, the UV contributions associated with
one-loop insertions in the internal part of the one-loop diagram
\refeq{oneloopdiag2} become irrelevant.  In particular the UV
logarithmic contributions resulting from the two-loop ladder diagrams
of type {\unitlength 0.6pt \SetScale{0.6}
\beqar\label{twoloopladder}
\vcenter{\hbox{\diagI{$\leg{i}$}{$\leg{j}$}{$\scriptscriptstyle{V_1}$}{$\scriptscriptstyle{V_2}$}{\factblob}}}
,
\vcenter{\hbox{\diagXX{$\leg{j}$}{$\leg{i}$}{$\leg{k}$}{$\scriptscriptstyle{V_1}$}{$\scriptscriptstyle{V_2}$}{\factblob}}}
,
\vcenter{\hbox{\diagXXII{$\leg{i}$}{$\leg{j}$}{$\leg{k}$}{$\leg{l}$}{$\scriptscriptstyle{V_1}$}{$\scriptscriptstyle{V_2}$}{\factblob}}}
,
\eeqar
}%
in the region where $V_1$ is soft and $V_2$ is hard vanish.  Minimal
subtraction of the UV singularities permits us to use the
soft--collinear approximation for all gauge bosons $V_1,V_2$ in the
above diagrams despite of the fact that this approximation is not
appropriate to describe the hard region and can in principle produce
fake logarithms of UV origin.  Instead, the diagrams of the type
\refeq{oneloopsubd1}, which result from the insertion of one-loop
subdiagrams in the lines that are not hard
(${\mu_{\mathrm{loop}}^2}\ll Q^2$) in \refeq{oneloopdiag2}, give rise
to non-negligible NLL contributions of UV type.  These UV
contributions are correctly taken into account in our calculation as
explained in \refse{se:softcollapp}.

\section{Collinear Ward identities and cancellation of non-factorizable terms}
\label{se:cwi}
In this section we prove that the non-factorizable subset of mass
singularities, \ie the one-loop terms \refeq{oneloopdiag5} and the
two-loop terms \refeq{twoloopnf1}--\refeq{twoloopnf4}, vanish.  The
proof is based on the collinear Ward identities that have been derived
in \citere{Denner:2001jv}.

Let us start with the non-factorizable one-loop terms \refeq{oneloopdiag5}.
This contribution can be written as
\beqar\label{oneloopnonfactpart}
\mel{1}{\nfact}
&=&
\sum_{i=1}^{n}
\sum_{V=A,Z,W^\pm}
\sum_{\varphi'_i}
\muD^{4-D}
\int\frac{\rd^D q}{({2\pi})^D}
\frac{1}{(q^2-M_{V}^2)(p_i-q)^2}
\lim_{q^\mu\to x p_i^\mu}
2(p_i-q)^\mu
\nl&&{}\times
\Biggl\{
G_{\mu}^{[\underline{V} \underline{\varphi}'_i]}(q,p_i-q)u(p_i,\kappa_i)
+\sum_{j=1 \atop j\neq i}^n
\sum_{\varphi'_j}
\frac{2  (p_j+q)_\mu}{(p_j+q)^2}
\mel{0}{\varphi_{1}\dots\varphi'_{i}\dots\varphi'_{j}\dots \varphi_{n}} 
e I^{V}_{\varphi'_{j}\varphi_{j}}
\Biggr\} \ri e I^{\bar V}_{\varphi'_{i}\varphi_{i}}
,
\nl*[-1.5ex]
\eeqar
where we have factorized the two propagators that appear in all three diagrams in \refeq{oneloopdiag5}.
Here $u(p_i,\kappa_i)$ is the spinor of the external fermion $i$
and we have  introduced the abbreviation
{
\beqar\label{treesubd}
 G_{\mu}^{[\underline{V} \,\underline{\varphi}_i]}(q,p_i-q)
&=&
\vcenter{\hbox{
\unitlength 0.6pt \SetScale{0.6}
\diagIwi{$\scriptstyle{i}$}{$\scriptstyle{V_\mu}$}{\blob}
}}
\hspace{-3mm}
-
\vcenter{\hbox{
\unitlength 0.6pt \SetScale{0.6}
\diagIIwi{$\scriptstyle{i}$}{$\scriptstyle{V_\mu}$}{\blob}
}}
\eeqar
}%
for the tree subdiagrams that are associated with the first two terms
on the right-hand side of \refeq{oneloopdiag5}.  In \refeq{treesubd}
the contraction with the spinors of the external fermions
$j=1,\dots,n$ with $j\neq i$ is implicitly understood, and the
incoming momenta associated with the lines $V_\mu$ and $i$ are $q$ and
$p_i-q$, respectively.

The combination of tree subdiagrams \refeq{treesubd} fulfils the
collinear Ward identities \cite{Denner:2001jv}
\beqar\label{collwi1}
\lim_{q^\mu\to x p_i^\mu}
q^\mu G_{\mu}^{[\underline{V} \underline{\varphi}_i]}(q,p_i-q)u(p_i,\kappa_i)
=
\sum_{\varphi'_i}
\mel{0}{\varphi_{1}\dots\varphi'_{i}\dots \varphi_{n}} 
e I^{V}_{{\varphi}'_{i}\varphi_{i}}
.
\eeqar
Using the charge-conservation relation \refeq{chargeconservation}
and 
\beq
\lim_{q^\mu\to x p_i^\mu}
\frac{2 q^\mu (p_j+q)_\mu}{(p_j+q)^2}=1
\qquad \mbox{for } j\neq i, 
\eeq
we can rewrite the identities \refeq{collwi1}
as
\beq\label{collwi2}
\lim_{q^\mu\to x p_i^\mu}
q^\mu 
\Biggl\{
G_{\mu}^{[\underline{V} \underline{\varphi}_i]}(q,p_i-q)u(p_i,\kappa_i)
+\sum_{j=1\atop j\neq i}^{n}
\sum_{\varphi'_j}
\frac{2  (p_j+q)_\mu}{(p_j+q)^2}
\mel{0}{\varphi_{1}\dots\varphi'_{j}\dots \varphi_{n}} 
e I^{V}_{\varphi'_{j}\varphi_{j}}
\Biggr\} 
=0
,
\eeq
where the expression between the curly brackets is identical 
to the one that appears on the right-hand side of
\refeq{oneloopnonfactpart}.
These collinear Ward identities can be represented diagrammatically as
{
\beqar\label{collwi2b}
&&
\lim_{q^\mu\to x p_i^\mu}
{q^\mu}\times
\left[
\vcenter{\hbox{
\unitlength 0.6pt \SetScale{0.6}
\diagIwi{$\scriptstyle{i}$}{$\scriptstyle{V_\mu}$}{\blob}
}}
\hspace{-3mm}
-
\vcenter{\hbox{
\unitlength 0.6pt \SetScale{0.6}
\diagIIwi{$\scriptstyle{i}$}{$\scriptstyle{V_\mu}$}{\blob}
}}
\hspace{-3mm}
-
\sum_{j=1\atop j\neq i}^{n}
\vcenter{\hbox{
\unitlength 0.6pt \SetScale{0.6}
\diagIVwi{$\scriptstyle{i}$}{$\scriptstyle{j}$}{$\scriptstyle{V_\mu}$}{\factblob}
}}
\right]
=0
,
\nln
\eeqar
}%
where the contraction with all fermionic spinors, including
$u(p_i,\kappa_i)$, is implicitly understood.

The cancellation of the non-factorizable terms
\refeq{oneloopnonfactpart} is simply due to the fact that in the
collinear limit $q^\mu \to x p_i^\mu$ the four-vector $(p_i-q)^\mu$ on
the right-hand side of \refeq{oneloopnonfactpart} becomes proportional
to the gauge-boson momentum $q^\mu$ and its contraction with the
expression between the curly brackets vanishes as a result of the
collinear Ward identities \refeq{collwi2}.

Let us now consider the two-loop non-factorizable terms
\refeq{twoloopnf1}--\refeq{twoloopnf4}.  The contribution
\refeq{twoloopnf1} yields
\beqar\label{twoloopnonfactpartA}
\mel{2}{\nfact,A}
&=&
\sum_{i=1}^{n}
\sum_{j=1\atop j\neq i}^{n}
\sum_{V_{m}=A,Z,W^\pm}
\sum_{\varphi'_i,\varphi''_i,\varphi'_j}
\muD^{2(4-D)}
\int\frac{\rd^D q_1}{({2\pi})^D}
\int\frac{\rd^D q_2}{({2\pi})^D}
\frac{ 8 e^3 (p_i-q_1)(p_j+q_1)
}{
(q_1^2-M_{V_1}^2)
(q_2^2-M_{V_2}^2)
}
\nl&&{}\times
\frac{1}
{
(p_i-q_1)^2
(p_j+q_1)^2
(p_i-q_1-q_2)^2
}
\lim_{q_1^\mu \to 0}
\lim_{q_2^\mu\to x p_i^\mu}
(p_i-q_1-q_2)^\mu
\nl&&{}\times
\Biggl\{
G_{\mu}^{[\underline{\bar V_2}\, \underline{\varphi}''_i]}
(q_2,p_i-q_1-q_2)u(p_i,\kappa_i)
\nl&&{}
+
\frac{2 (p_j+q_1+q_2)_\mu}{(p_j+q_1+q_2)^2}
\sum_{\varphi''_j}
\mel{0}{\varphi_{1}\dots\varphi''_{i}\dots\varphi''_{j}\dots \varphi_{n}} 
e I^{\bar V_2}_{\varphi''_{j}\varphi'_{j}}
\nl&&{}
+\sum_{k=1\atop k\neq i,j}^{n}
\frac{2 (p_k+q_2)_\mu}{(p_k+q_2)^2}
\sum_{\varphi'_k}
\mel{0}{\varphi_{1}\dots\varphi''_{i}\dots\varphi'_{j}\dots\varphi'_{k}\dots \varphi_{n}} 
e I^{\bar V_2}_{\varphi'_{k}\varphi_{k}}
\Biggr\} 
I^{V_2}_{\varphi''_{i}\varphi'_{i}}
I^{\bar V_1}_{\varphi'_{j}\varphi_{j}}
I^{V_1}_{\varphi'_{i}\varphi_{i}}
=0
.
\eeqar
This cancellation can easily be verified 
by means of the collinear Ward identities \refeq{collwi2}
by observing that 
in the soft--collinear limit ${q_1^\mu \to 0}$, ${q_2^\mu\to x p_i^\mu}$
the four-vector
$(p_i-q_1-q_2)^\mu$ tends to $(1/x-1)q_2^\mu$ 
and 
the expression within the curly brackets in \refeq{twoloopnonfactpartA}
becomes equivalent to the one in \refeq{collwi2}.
Similarly, for the contributions \refeq{twoloopnf2} and
\refeq{twoloopnf3} we obtain
\beqar\label{twoloopnonfactpartB}
\mel{2}{\nfact,B}
&=&
\sum_{i=1}^{n}
\sum_{j=1\atop j\neq i}^{n}
\sum_{V_{m}=A,Z,W^\pm}
\sum_{\varphi'_i,\varphi''_i,\varphi'_j}
\muD^{2(4-D)}
\int\frac{\rd^D q_1}{({2\pi})^D}
\int\frac{\rd^D q_2}{({2\pi})^D}
\frac{ 8 e^3 (p_i-q_1-q_2)(p_j+q_1)
}{
(q_1^2-M_{V_1}^2)
(q_2^2-M_{V_2}^2)
}
\nl&&{}\times
\frac{1}{
(p_i-q_2)^2
(p_j+q_1)^2
(p_i-q_1-q_2)^2
}
\lim_{q_1^\mu \to 0}
\lim_{q_2^\mu\to x p_i^\mu}
(p_i-q_2)^\mu
\nl&&{}\times
\Biggl\{
G_{\mu}^{[\underline{\bar V_2}\, \underline{\varphi}''_i]}
(q_2,p_i-q_1-q_2)u(p_i,\kappa_i)
\nl&&{}
+
\frac{2 (p_j+q_1+q_2)_\mu}{(p_j+q_1+q_2)^2}
\sum_{\varphi''_j}
\mel{0}{\varphi_{1}\dots\varphi''_{i}\dots\varphi''_{j}\dots \varphi_{n}} 
e I^{\bar V_2}_{\varphi''_{j}\varphi'_{j}}
\nl&&{}
+\sum_{k=1\atop k\neq i,j}^{n}
\frac{2 (p_k+q_2)_\mu}{(p_k+q_2)^2}
\sum_{\varphi'_k}
\mel{0}{\varphi_{1}\dots\varphi''_{i}\dots\varphi'_{j}\dots\varphi'_{k}\dots \varphi_{n}} 
e I^{\bar V_2}_{\varphi'_{k}\varphi_{k}}
\Biggr\} 
I^{\bar V_1}_{\varphi'_{j}\varphi_{j}}
I^{V_1}_{\varphi''_{i}\varphi'_{i}}
I^{V_2}_{\varphi'_{i}\varphi_{i}}
=0
,
\eeqar
and
\beqar\label{twoloopnonfactpartC}
\mel{2}{\nfact,C}
&=&
\frac{1}{2}
\sum_{i=1}^{n}
\sum_{j=1\atop j\neq i}^{n}
\sum_{k=1\atop k\neq i,j}^{n}
\sum_{V_{m}=A,Z,W^\pm}
\sum_{\varphi'_i,\varphi'_j,\varphi'_k}
\muD^{2(4-D)}
\int\frac{\rd^D q_1}{({2\pi})^D}
\int\frac{\rd^D q_2}{({2\pi})^D}
\frac{ 1
}{
(q_1^2-M_{V_1}^2)
}
\nl&&{}\times
\frac{8 e^3 (p_k-q_1)(p_j+q_1)}{
(q_2^2-M_{V_2}^2)
(p_i-q_2)^2
(p_j+q_1)^2
(p_k-q_1)^2
}
\lim_{q_1^\mu \to 0}
\lim_{q_2^\mu\to x p_i^\mu}
(p_i-q_2)^\mu
\nl&&{}\times
\Biggl\{
G_{\mu}^{[\underline{\bar V_2}\, \underline{\varphi}'_i]}
(q_2,p_i-q_2)u(p_i,\kappa_i)
\nl&&{}
+
\frac{2 (p_j+q_1+q_2)_\mu}{(p_j+q_1+q_2)^2}
\sum_{\varphi''_j}
\mel{0}{\varphi_{1}\dots\varphi'_{i}\dots\varphi''_{j}\dots\varphi'_{k}\dots \varphi_{n}} 
e I^{\bar V_2}_{\varphi''_{j}\varphi'_{j}}
\nl&&{}
+
\frac{2 (p_k-q_1+q_2)_\mu}{(p_k-q_1+q_2)^2}
\sum_{\varphi''_k}
\mel{0}{\varphi_{1}\dots\varphi'_{i}\dots\varphi'_{j}\dots\varphi''_{k}\dots \varphi_{n}} 
e I^{\bar V_2}_{\varphi''_{k}\varphi'_{k}}
\nl&&{}
+\sum_{l=1\atop l\neq i,j,k}^{n}
\frac{2 (p_l+q_2)_\mu}{(p_l+q_2)^2}
\sum_{\varphi'_l}
\mel{0}{\varphi_{1}\dots\varphi'_{i}\dots\varphi'_{j}\dots\varphi'_{k}\dots\varphi'_{l}\dots \varphi_{n}} 
e I^{\bar V_2}_{\varphi'_{l}\varphi_{l}}
\Biggr\} 
I^{\bar V_1}_{\varphi'_{j}\varphi_{j}}
I^{V_1}_{\varphi'_{k}\varphi_{k}}
I^{V_2}_{\varphi'_{i}\varphi_{i}}
=0
.
\nln
\eeqar
Finally, for the contribution \refeq{twoloopnf4} we have
\beqar\label{twoloopnonfactpartD}
\mel{2}{\nfact,D}
&=&
\sum_{i=1}^{n}
\sum_{j=1\atop j\neq i}^{n}
\sum_{V_{m}=A,Z,W^\pm}
\sum_{\varphi'_i,\varphi'_j}
\muD^{2(4-D)}
\int\frac{\rd^D q_1}{({2\pi})^D}
\int\frac{\rd^D q_2}{({2\pi})^D}
\frac{4 \ri e^2\gw {\teps^{V_1 V_2 V_3}}
}{
(q_1^2-M_{V_1}^2)
(q_2^2-M_{V_2}^2)
}
\nl&&{}\times
\frac{1}{
(q_3^2-M_{V_3}^2)
(p_i-q_2)^2
(p_j-q_1)^2
}
\lim_{q_1^\mu \to 0}
\lim_{q_2^\mu\to x p_i^\mu}
(p_i-q_2)^{\mu_2}(p_j-q_1)^{\mu_1}
\nl&&{}\times
\biggl[g_{\mu_1\mu_2}(q_1-q_2)^{\mu_3}+g_{\mu_2}^{\mu_3}(q_2+q_3)_{\mu_1}
-g^{\mu_3}_{\mu_1}(q_3+q_1)_{\mu_2}\biggr]
\nl&&{}\times
\Biggl\{
G_{\mu_3}^{[\underline{\bar V_3}\, \underline{\varphi}'_i]}
(q_3,p_i-q_2)u(p_i,\kappa_i)
+
\frac{2 (p_j+q_2)_{\mu_3}}{(p_j+q_2)^2}
\sum_{\varphi''_j}
\mel{0}{\varphi_{1}\dots\varphi'_{i}\dots\varphi''_{j}\dots \varphi_{n}} 
e I^{\bar V_3}_{\varphi''_{j}\varphi'_{j}}
\nl&&{}
+\sum_{k=1\atop k\neq i,j}^{n}
\frac{2 (p_k+q_3)_{\mu_3}}{(p_k+q_3)^2}
\sum_{\varphi'_k}
\mel{0}{\varphi_{1}\dots\varphi'_{i}\dots\varphi'_{j}\dots\varphi'_{k}\dots \varphi_{n}} 
e I^{\bar V_3}_{\varphi'_{k}\varphi_{k}}
\Biggr\} 
I^{\bar V_1}_{\varphi'_{j}\varphi_{j}}
I^{\bar V_2}_{\varphi'_{i}\varphi_{i}}
,
\eeqar
where $q_3=q_1+q_2$.  In the soft--collinear limit we observe that
\beqar
&&\lim_{q_1^\mu \to 0}
\lim_{q_2^\mu\to x p_i^\mu}
(p_i-q_2)^{\mu_2}(p_j-q_1)^{\mu_1}
\biggl[g_{\mu_1\mu_2}(q_1-q_2)^{\mu_3}+g_{\mu_2}^{\mu_3}(q_2+q_3)_{\mu_1}
-g^{\mu_3}_{\mu_1}(q_3+q_1)_{\mu_2}\biggr]
=
\nl&&\quad{}
=
(1-x)(p_i p_j) q_3^{\mu_3},
\eeqar
and again the contraction of this four-vector with the expression 
between the curly brackets in \refeq{twoloopnonfactpartD}
cancels as a result of the collinear Ward identities \refeq{collwi2}.

\section{Factorizable contributions}
\label{se:factcont}

In this section, we present explicit results for the one- and two-loop
factorizable contributions defined in \refses{se:factoneloop} and
\ref{se:facttwoloop}.  These are evaluated within the
't~Hooft--Feynman gauge, where the masses of the Faddeev--Popov ghosts
$u^A,u^\FZ,u^\FWpm$ and would-be Goldstone bosons $\chi,\phi^\pm$ read
$M_{u^A}=\MA=0$, $M_{\chi}=M_{u^\FZ}=\MZ$, and
$M_{\phi^\pm}=M_{u^\FW}=\MW$.
Using the soft--collinear approximation introduced in
\refse{se:softcollapp}, we express the factorizable contributions
resulting from individual diagrams as products of the
\mbox{$n$-fermion} Born amplitude with matrix-valued gauge couplings
and loop integrals.  The definitions of these loop integrals are
provided in \refapp{app:loops}.  The integrals are computed in NLL
accuracy, and the result is expanded in $\veps$ up to
$\order(\veps^2)$ at one loop and $\order(\veps^0)$ at two loops.  The
UV poles are eliminated by means of a minimal subtraction as explained
in \refse{se:uvsing} such that the presented results are UV finite.
All loop integrals have been solved and cross-checked using two
independent methods: an automatized algorithm based on the
sector-decomposition technique \cite{Denner:2004iz} and the method of
expansion by regions combined with Mellin--Barnes representations
\cite{Jantzen:2006jv}.

\subsection{One-loop diagrams}
\label{se:oneloop}
The one-loop factorizable contributions originate only from one type of
diagram,%
\footnote{The $l$-loop diagrams depicted in this section are
  understood without factors $(\alphaeps/4\pi)^l$.}
\beqar
\nmel{1}{ij}
\eqdiagl
\vcenter{\hbox{
\unitlength 0.6pt \SetScale{0.6}
\diagone{$\leg{i}$}{$\leg{j}$}{$\scriptscriptstyle{V_1}$}{\factblob}}}
\eqdiagr
-
\mel{0}{}
\sum_{V_1=A,Z, W^\pm
} 
I_i^{\bar{V}_1}
I_j^{{V}_1}
\DD{0}(M_{V_1};r_{ij}).
\eeqar
The corresponding loop integral $\DD{0}$ 
is defined in \refeq{defint0} and to NLL accuracy yields
\beqar\label{idiag0subt}
\DDsub{0}(\MW;r_{ij})&\NLLA& 
-L^2
-\frac{2}{3}L^3\Eps{}
-\frac{1}{4}L^4\Eps{2}
+ 2\left(2-\Lrij\right) \left(
  L
  +\frac{1}{2}L^2\Eps{}
  +\frac{1}{6}L^3\Eps{2}
  \right)
,\nl
\DDsub{0}(\MZ;r_{ij})&\NLLA& 
\DDsub{0}(\MW;r_{ij})+
\LMZW
\left(
2L
+2L^2\Eps{}
+L^3\Eps{2}
\right)
,\nl
\DDsub{0}(0;r_{ij})&\NLLA& 
-2\Epsinv{2}
-2 \left(2-\Lrij\right)\Epsinv{1}
,
\eeqar
where the UV singularities
\beqar\label{idiag0uv}
\DDUV{0}(\MW;r_{ij}) &\NLLA&
\DDUV{0}(\MZ;r_{ij}) \NLLA
\DDUV{0}(0;r_{ij}) \NLLA
4\Epsinv{1}
\eeqar
have been subtracted.

\subsection{Two-loop diagrams}
\label{se:twoloop}
The two-loop NLL factorizable terms \refeq{twoloopdiag3} involve
fourteen different types of diagrams. The diagrams 1--3, 12, and 14{}
in this section give rise to LLs and NLLs, whereas all other diagrams
yield only NLLs.  The loop integrals associated with the various
diagrams are denoted with symbols of the type
$\DD{h}(m_1,\dots,m_j;r_{kl},\ldots)$ and depend on various
kinematical invariants $r_{kl}$ and masses $m_i$.  The symbols $m_i$
are always used to denote generic mass parameters, which can assume
the values $m_i=\MW,\MZ,\Mt,\MH$ or $m_i=0$.  Instead we use the
symbols $M_i$
to denote non-zero masses, \ie
$M_i=\MW,\MZ,\Mt,\MH$.  The integrals are often singular when certain
mass parameters tend to zero, and the cases where such parameters are
zero or non-zero need to be treated separately.  For every integral we
first evaluate $\DD{h}(\MW,\dots,\MW;r_{kl},\ldots)$, \ie the case
where all mass parameters are equal to $\MW$.  The dependence of the
integral on the various masses is then described by subtracted
functions of the type
\beq\label{eq:subtractedintegral}
\deDD{h}(m_1,\dots,m_j;r_{kl},\ldots)=
\DD{h}(m_1,\dots,m_j;r_{kl},\ldots)-
\DD{h}(\MW,\dots,\MW;r_{kl},\ldots).
\eeq

\subsubsection*{Diagram 1}
\vspace*{-3ex}
\beqar\label{diagram1}
\nmel{2}{1,ij}
\eqdiagl
\vcenter{\hbox{
\unitlength 0.6pt \SetScale{0.6}
\diagI{$\leg{i}$}{$\leg{j}$}{$\scriptscriptstyle{V_1}$}{$\scriptscriptstyle{V_2}$}{\factblob}
}}
\eqdiagr
\mel{0}{}
\sum_{V_1,V_2=A,Z,W^\pm} 
I_i^{\bar{V}_2}
I_i^{\bar{V}_1}
I_j^{{V}_2}
I_j^{{V}_1}
\DD{1}(M_{V_1},M_{V_2};r_{ij})
,\quad
\eeqar
where the loop integral $\DD{1}$ is defined in \refeq{defint1} 
and yields
\beqar\label{idiag1sub}
\DDsub{1}(\MW,\MW;r_{ij})&\NLLA& 
\frac{1}{6}L^4
-\frac{2}{3}\left(2-\Lrij\right)L^3
,\nl
\deDDsub{1}(M_1,M_2;r_{ij})&\NLLA& 
-\frac{2}{3}\LrMI L^3
,
\nl
\deDDsub{1}(0,M_2;r_{ij})&\NLLA& 
2L^2\Epsinv{2}
+\frac{8}{3}L^3\Epsinv{1}
+\frac{11}{6}L^4
- \left(2-\Lrij\right) \left(
  4L\Epsinv{2}
  +4L^2\Epsinv{1}
  +2L^3
  \right)
\nl&&{}
-\LrMII
\left(
  4L\Epsinv{2}
  +8L^2\Epsinv{1}
  +8L^3
  \right)
,\nl
\deDDsub{1}(M_1,0;r_{ij})&\NLLA& 
-\frac{2}{3}\LrMI L^3
,\nl
\deDDsub{1}(0,0;r_{ij})&\NLLA& 
\Epsinv{4}
-\frac{1}{6}L^4
+ \left(2-\Lrij\right) \left(
  2\Epsinv{3}
  +\frac{2}{3}L^3
  \right)
.
\eeqar
Here the UV singularities 
\beqar\label{idiag1sing}
\DDUV{1}(M_1,m_2;r_{ij}) &\NLLA&
-4 L^{2}\Epsinv{1}
-\frac{8}{3} L^{3}
,\nl
\DDUV{1}(0,m_2;r_{ij}) &\NLLA&
-8\Epsinv{3}
\eeqar
have been subtracted.

\subsubsection*{Diagram 2}
\vspace*{-3ex}
\beqar\label{diagram2}
\nmel{2}{2,ij}
\eqdiagl
\vcenter{\hbox{
\unitlength 0.6pt \SetScale{0.6}
\diagII{$\leg{i}$}{$\leg{j}$}{$\scriptscriptstyle{V_1}$}{$\scriptscriptstyle{V_2}$}{\factblob}
}}
\eqdiagr
\mel{0}{}
\sum_{V_1,V_2
=A,Z,W^\pm} 
I_i^{\bar{V}_2}
I_i^{\bar{V}_1}
I_j^{{V}_1}
I_j^{{V}_2}
\DD{2}(M_{V_1},M_{V_2};r_{ij})
,\quad
\eeqar
where the loop integral $\DD{2}$ is defined in \refeq{defint2}.
This integral is free of UV singularities and yields
\beqar\label{idiag2}
\DD{2}(\MW,\MW;r_{ij})&\NLLA& 
\frac{1}{3}L^4
-\frac{4}{3}\left(2-\Lrij\right)L^3
,\nl
\deDD{2}(M_1,M_2;r_{ij})&\NLLA& 
-\frac{2}{3}
\left(\LrMI+\LrMII\right)L^3
,\nl
 \deDD{2}(0,M_2;r_{ij})&\NLLA& 
-\frac{2}{3}L^3\Epsinv{1}
-\frac{7}{6}L^4
+\left(2-\Lrij+\LrMII\right) \left(
  2L^2\Epsinv{1}
  + \frac{10}{3}L^3
  \right)
,\nl
 \deDD{2}(M_1,0;r_{ij})&\NLLA& 
-\frac{2}{3}L^3\Epsinv{1}
-\frac{7}{6}L^4
+\left(2-\Lrij+\LrMI\right) \left(
  2L^2\Epsinv{1}
  + \frac{10}{3}L^3
  \right)
,\nl
 \deDD{2}(0,0;r_{ij})&\NLLA& 
\Epsinv{4}
-\frac{1}{3}L^4
+ \left(2-\Lrij\right) \left(
  2\Epsinv{3}
  +\frac{4}{3}L^3
  \right)
.
\eeqar

\subsubsection*{Diagram 3}
\vspace*{-3ex}
\beqar\label{diagram3}
\nmel{2}{3,ij}
\eqdiagl
\vcenter{\hbox{
\unitlength 0.6pt \SetScale{0.6}
\diagIII{$\leg{i}$}{$\leg{j}$}{$\scriptscriptstyle{V_1}$}{$\scriptscriptstyle{V_3}$}{$\scriptscriptstyle{V_2}$}{\factblob}
}}
\eqdiagr
\nl&=&
-\ri 
\frac{\gw}{e}
\mel{0}{}
\sum_{V_1,V_2,V_3
=A,Z,W^\pm} 
\teps^{V_1 V_2 V_3}
I_i^{\bar{V}_2}
I_i^{\bar{V}_1}
I_j^{\bar{V}_3}
\DD{3}(M_{V_1},M_{V_2},M_{V_3};r_{ij})
,
\eeqar
where the $\teps$-tensor is defined in \refeq{phystotalantitens}.
The loop integral $\DD{3}$ is defined in \refeq{defint3} and yields
\beqar\label{idiag3sub}
\DDsub{3}(\MW,\MW,\MW;r_{ij})&\NLLA&
\frac{1}{6}L^4
-\left( 3-\frac{2\,\Lrij}{3}\right)L^3
,\nl
\deDDsub{3}(M_1,M_2,M_3;r_{ij})&\NLLA&
-\frac{1}{3}
\left(\LrMI+\LrMIII\right)L^3
,\nl
\deDDsub{3}(0,M_2,M_3;r_{ij})&\NLLA&
-\frac{1}{3}L^3\Epsinv{1}
-\frac{7}{12}L^4
+\left(2 - \Lrij + \LrMIII\Right) \left(
  L^2\Epsinv{1}
  + \frac{5}{3}L^3
  \right)
,\nl
\deDDsub{3}(M_1,0,M_3;r_{ij})&\NLLA&
-\frac{1}{3}
\left(\LrMI+\LrMIII\right)L^3
,\nl
\deDDsub{3}(M_1,M_2,0;r_{ij})&\NLLA& 
-\frac{1}{3}L^3\Epsinv{1}
-\frac{7}{12}L^4
-6L\Epsinv{2}
-\left( 2 + \Lrij - \LrMI\Right)L^2\Epsinv{1}
\nl&&{}
+\left(\frac{11}{3} - \frac{5\,\Lrij}{3} + \frac{5\,\LrMI}{3}\Right)L^3
,
\eeqar
where the UV singularities 
\beqar\label{idiag3sing}
\DDUV{3}(m_1,m_2,M_3;r_{ij}) &\NLLA&
-3 L^{2}\Epsinv{1}
-2 L^{3}
,\nl
\DDUV{3}(m_1,m_2,0;r_{ij})
&\NLLA& 
-6\Epsinv{3}
\eeqar
have been subtracted.

\subsubsection*{Diagram 4}
\vspace*{-3ex}
\beqar\label{diagram5}
\nmel{2}{4,ij}
\eqdiagl
\vcenter{\hbox{
\unitlength 0.6pt \SetScale{0.6}
\diagV{$\leg{i}$}{$\leg{j}$}{$\scriptscriptstyle{V_1}$}{$\scriptscriptstyle{V_2}$}
{\factblob}
}}
\eqdiagr
- \mel{0}{}
\sum_{V_1,V_2
=A,Z,W^\pm} 
I_i^{{V}_2}
I_i^{\bar{V}_2}
I_i^{{V}_1}
I_j^{\bar{V}_1}
\DD{4}(M_{V_1},M_{V_2};r_{ij})
,
\nn\\*[-4ex]
\eeqar
where the loop integral $\DD{4}$ is defined in \refeq{defint5} and yields
\beqar\label{idiag5sub}
\DDsub{4}(\MW,\MW;r_{ij})&\NLLA&
\frac{1}{3}L^3
,\nl
\deDDsub{4}(M_1,M_2;r_{ij})&\NLLA&
0
,\nl
 \deDDsub{4}(0,M_2;r_{ij})&\NLLA&
2 L\Epsinv{2}
+2 L^2\Epsinv{1}
+L^3
,\nl
 \deDDsub{4}(M_1,0;r_{ij})&\NLLA&
0
,\nl
 \deDDsub{4}(0,0;r_{ij})&\NLLA&
-\Epsinv{3}
-\frac{1}{3}L^3
.
\eeqar
Here the UV singularities 
\beqar\label{idiag5sing}
\DDUV{4}(M_1,m_2;r_{ij}) &\NLLA&
L^{2}\Epsinv{1}
+\frac{2}{3}L^{3}
,\nl
\DDUV{4}(0,m_2;r_{ij}) &\NLLA&
2\Epsinv{3}
\eeqar
have been subtracted.

\subsubsection*{Diagram 5}
\vspace*{-3ex}
\beqar\label{diagram7}
\nmel{2}{5,ij}
\eqdiagl
\vcenter{\hbox{
\unitlength 0.6pt \SetScale{0.6}
\diagVII{$\leg{i}$}{$\leg{j}$}{$\scriptscriptstyle{V_1}$}{$\scriptscriptstyle{V_2}$}{\factblob}
}}
\eqdiagr
- \mel{0}{}
 \sum_{V_1,V_2
= A,Z,W^\pm} 
I_i^{{V}_2}
I_i^{{V}_1}
I_i^{\bar{V}_2}
I_j^{\bar{V}_1}
\DD{5}(M_{V_1},M_{V_2};r_{ij})
,
\nn\\*[-4ex]
\eeqar
where the loop integral $\DD{5}$ is defined in \refeq{defint7} and yields
\beqar\label{idiag7sub}
\DDsub{5}(\MW,\MW;r_{ij})&\NLLA&
-\frac{1}{3}L^3
,\nl
\deDDsub{5}(M_1,M_2;r_{ij})&\NLLA&
0
,\nl
 \deDDsub{5}(0,M_2;r_{ij})&\NLLA&
-2 L\Epsinv{2}
-2 L^2\Epsinv{1}
-L^3
,\nl
 \deDDsub{5}(M_1,0;r_{ij})&\NLLA&
0
,\nl
 \deDDsub{5}(0,0;r_{ij})&\NLLA&
\Epsinv{3}
+\frac{1}{3}L^3
.
\eeqar
Here the UV singularities 
\beqar\label{idiag7sing}
\DDUV{5}(M_1,m_2;r_{ij}) &\NLLA& 
-L^{2}\Epsinv{1}
-\frac{2}{3}L^{3}
,\nl
\DDUV{5}(0,m_2;r_{ij}) &\NLLA&
-2\Epsinv{3}
\eeqar
have been subtracted.

\subsubsection*{Diagrams 6}
\vspace*{-3ex}
\label{sub:diagten}
\beqar\label{diagram10}
\nmel{2}{6,ij}
\eqdiagl
\vcenter{\hbox{
\unitlength 0.6pt \SetScale{0.6}
\diagX{\factblob}
}}
\quad+\quad
\vcenter{\hbox{
\unitlength 0.6pt \SetScale{0.6}
\diagXI{\factblob}
}}
\eqdiagr
\nl
&=&
\frac{1}{2} 
\frac{\gw^2}{e^2}
\mel{0}{}
\sum_{V_1,V_2,V_3,V_4
= A,Z,W^\pm} 
I_i^{\bar{V}_1}
I_j^{\bar{V}_4}\,
\teps^{V_1 \bar{V}_2 \bar{V}_3}
\teps^{V_4 V_2 V_3}\,
\DD{6}(M_{V_1},M_{V_2},M_{V_3},M_{V_4};r_{ij})
,\nl*[-1.7ex]
\eeqar
where the loop integral $\DD{6}$ is defined in \refeq{defint10} and yields
\beqar\label{idiag10sub}
\DDsub{6}(\MW,\MW,\MW,\MW;r_{ij})
&\NLLA&
\frac{20}{9} L^3
,\nl
\deDDsub{6}(M_1,M_2,M_3,M_4;r_{ij})
&\NLLA&
0
,\nl
\deDDsub{6}(0,M_2,M_3,M_4;r_{ij})
&\NLLA& 
\frac{M_2^2+M_3^2}{{2M_4^2}}
\left[
-16L\Epsinv{2}
-8L^2\Epsinv{1}
+\frac{16}{3}L^3
\right]
,\nl
\deDDsub{6}(M_1,0,M_3,M_4;r_{ij})
&\NLLA& 
0
,\nl
\deDDsub{6}(M_1,M_2,0,M_4;r_{ij})
&\NLLA& 
0
,\nl
\deDDsub{6}(M_1,M_2,M_3,0;r_{ij})
&\NLLA& 
\frac{M_2^2+M_3^2}{{2M_1^2}}
\left[
-16L\Epsinv{2}
-8L^2\Epsinv{1}
+\frac{16}{3}L^3
\right]
,\nl
\deDDsub{6}(0,M_2,M_3,0;r_{ij})
&\NLLA& 
\frac{20}{3}L\Epsinv{2}
+\frac{10}{3}L^2\Epsinv{1}
-\frac{20}{9}L^3
.
\eeqar
Here the UV singularities 
\beqar\label{idiag10sing}
\DDUV{6}(M_1,m_2,m_3,M_4;r_{ij}) &\NLLA&
\frac{10}{3}L^{2}\Epsinv{1}
+\frac{20}{9}L^{3}
,\nl
\DDUV{6}(0,m_2,m_3,M_4;r_{ij}) &\NLLA& 
\frac{10}{3}L^{2}\Epsinv{1}
+\frac{20}{9}L^{3}
\nl&&{}
+\frac{m_2^2+m_3^2}{{2M_4^2}}
  \left[
  -16\Epsinv{3}
  +8 L^2 \Epsinv{1}
  +\frac{16}{3}L^3
  \right]
,\nl
\DDUV{6}(M_1,m_2,m_3,0;r_{ij}) &\NLLA& 
\frac{10}{3}L^{2}\Epsinv{1}
+\frac{20}{9}L^{3}
\nl&&{}
+\frac{m_2^2+m_3^2}{{2M_1^2}}
  \left[
  -16\Epsinv{3}
  +8 L^2 \Epsinv{1}
  +\frac{16}{3}L^3
  \right]
,\nl
\DDUV{6}(0,m_2,m_3,0;r_{ij}) &\NLLA&
\frac{20}{3}\Epsinv{3}
\eeqar
have been subtracted.
We observe that the loop integrals associated with 
$\FA$--$\FZ$ mixing-energy subdiagrams
give rise to the contributions
\beqar\label{lineardep}
\deDDsub{6}(0,\MW,\MW,\MZ;r_{ij})
&=&
\frac{\MW^2}{{\MZ^2}}
\left[
-16L\Epsinv{2}
-8L^2\Epsinv{1}
+\frac{16}{3}L^3
\right],
\eeqar
which depend linearly on the ratio
$\MW^2/\MZ^2$.

\subsubsection*{Diagram 7}
\vspace*{-3ex}
\beqar\label{diagram16}
\nmel{2}{7,ij}
\eqdiagl
\vcenter{\hbox{
\unitlength 0.6pt \SetScale{0.6}
\diagXVI{\factblob}
}}
\eqdiagr
- 
\frac{\gw^2}{e^2}
\mel{0}{}
\sum_{V_1,V_2,V_3
= A,Z,W^\pm} 
I_i^{\bar{V}_1}
I_j^{\bar{V}_3}
\sum_{V= A,Z,W^\pm} \teps^{V_1\bar{V_2}\bar{V}}\teps^{V_3{V_2}{V}}
\nl&&{}\times
(D-1) 
\DD{7}(M_{V_1},M_{V_2},M_{V_3};r_{ij})
,
\eeqar
where 
$D=4-2\veps$.
The loop integral $\DD{7}$ is defined in \refeq{defint16} and yields
\beqar\label{idiag16sub}
\DDsub{7}(\MW,\MW,\MW;r_{ij})
&\NLLA&
0
,\nl
\deDDsub{7}(M_1,M_2,M_3;r_{ij})
&\NLLA&
0
,\nl
\deDDsub{7}(0,M_2,M_3;r_{ij})
&\NLLA&
\frac{M_2^2}{{M_3^2}}
\left[
-2L\Epsinv{2}
-L^2\Epsinv{1}
+\frac{2}{3}L^3
\right]
,\nl
\deDDsub{7}(M_1,0,M_3;r_{ij})
&\NLLA&
0
,\nl
\deDDsub{7}(M_1,M_2,0;r_{ij})
&\NLLA&
\frac{M_2^2}{{M_1^2}}
\left[
-2L\Epsinv{2}
-L^2\Epsinv{1}
+\frac{2}{3}L^3
\right]
,\nl
\deDDsub{7}(0,M_2,0;r_{ij})
&\NLLA&
0
,
\eeqar
where the UV singularities 
\beqar\label{idiag16sing}
\DDUV{7}(0,M_2,M_3;r_{ij})
&\NLLA&
\frac{M_2^2}{{M_3^2}}
\left[
-2\Epsinv{3}
+L^2\Epsinv{1}
+\frac{2}{3}L^3
\right]
,\nl
\DDUV{7}(M_1,M_2,0;r_{ij})
&\NLLA&
\frac{M_2^2}{{M_1^2}}
\left[
-2\Epsinv{3}
+L^2\Epsinv{1}
+\frac{2}{3}L^3
\right]
\eeqar
have been subtracted.

\subsubsection*{Diagram 8}
\vspace*{-3ex}
\beqar\label{diagram15}
\nmel{2}{8,ij}
\eqdiagl
\vcenter{\hbox{
\unitlength 0.6pt \SetScale{0.6}
\diagXV{\factblob}
}}
\eqdiagr
- e^2 \vev^2 \mel{0}{}
\sum_{V_1,V_3,V_4
= A,Z,W^\pm} 
I_i^{\bar{V}_1}
I_j^{\bar{V}_4}
\sum_{\Phi_{i_2}=
H,\chi,\phi^\pm
}
\left\{
I^{{V}_1},
I^{\bar{V}_3}
\right\}_{H\Phi_{i_2}}
\nl&&{}\times
\left\{
I^{{V}_3},
I^{{V}_4}
\right\}_{\Phi_{i_2}H}
\DD{8}(M_{V_1},M_{\Phi_{i_2}},M_{V_3},M_{V_4};r_{ij})
,
\eeqar
where the curly brackets denote anticommutators and $\vev$ is the 
vacuum expectation value.
The loop integral $\DD{8}$ is defined in \refeq{defint15} and yields
\beqar\label{idiag15sub}
\MW^2\DDsub{8}(\MW,\MW,\MW,\MW;r_{ij})
&\NLLA&
0
,\nl
\MW^2\deDDsub{8}(M_1,M_2,M_3,M_4;r_{ij})
&\NLLA&
0
,\nl
\deDDsub{8}(0,M_2,M_3,M_4;r_{ij})
&\NLLA&
\frac{1}{{M_4^2}}
\left[
-2L\Epsinv{2}
-L^2\Epsinv{1}
+\frac{2}{3}L^3
\right]
,\nl
\MW^2 \deDDsub{8}(M_1,M_2,0,M_4;r_{ij})
&\NLLA& 
0
,\nl
 \deDDsub{8}(M_1,M_2,M_3,0;r_{ij})
&\NLLA&
\frac{1}{{M_1^2}}
\left[
-2L\Epsinv{2}
-L^2\Epsinv{1}
+\frac{2}{3}L^3
\right]
,\nl
\MW^2
\deDDsub{8}(0,M_2,M_3,0;r_{ij})
&\NLLA& 
0
.
\eeqar
Here the UV singularities 
\beqar\label{idiag15sing}
\DDUV{8}(0,M_2,M_3,M_4;r_{ij}) &\NLLA&
\frac{1}{M_4^2}
\left[
-2\Epsinv{3}
+L^2\Epsinv{1}
+\frac{2}{3}L^3
\right]
,\nl
\DDUV{8}(M_1,M_2,M_3,0;r_{ij}) &\NLLA&
\frac{1}{M_1^2}
\left[
-2\Epsinv{3}
+L^2\Epsinv{1}
+\frac{2}{3}L^3
\right]
\eeqar
have been subtracted.
The above diagram represents the only contribution involving couplings
proportional to $\vev$, which originate from spontaneous symmetry
breaking.

\subsubsection*{Diagram 9}
\vspace*{-3ex}
\beqar\label{diagram12}
\nmel{2}{9,ij}
\eqdiagl
\vcenter{\hbox{
\unitlength 0.6pt \SetScale{0.6}
\diagXII{\factblob}
}}
\eqdiagr
-\frac{1}{2} \mel{0}{}
\sum_{V_1,V_4
= A,Z,W^\pm} 
I_i^{\bar{V}_1}
I_j^{\bar{V}_4}
\sum_{\Phi_{i_2},\Phi_{i_3}
=H,\chi,\phi^\pm
}
I^{{V}_1}_{\Phi_{i_3}\Phi_{i_2}}
I^{{V}_4}_{\Phi_{i_2}\Phi_{i_3}}
\nl&&{}\times
\DD{9}(M_{V_1},M_{\Phi_{i_2}},M_{\Phi_{i_3}},M_{V_4};r_{ij})
,
\eeqar
where the loop integral $\DD{9}$ is defined in \refeq{defint12} and yields
\beqar\label{massidiag12sub}
\DDsub{9}(\MW,\MW,\MW,\MW;r_{ij})
&\NLLA& 
\frac{2}{9}L^3
,\nl
\deDDsub{9}(M_1,M_2,M_3,M_4;r_{ij})
&\NLLA&
0
,\nl
\deDDsub{9}(0,M_2,M_3,M_4;r_{ij})
&\NLLA& 
\frac{M_2^2+M_3^2}{{2M_4^2}}
\left[
4L\Epsinv{2}
+2L^2\Epsinv{1}
-\frac{4}{3}L^3
\right]
,\nl
\deDDsub{9}(M_1,M_2,M_3,0;r_{ij})
&\NLLA& 
\frac{M_2^2+M_3^2}{{2M_1^2}}
\left[
4L\Epsinv{2}
+2L^2\Epsinv{1}
-\frac{4}{3}L^3
\right]
,\nl
\deDDsub{9}(0,M_2,M_3,0;r_{ij})
&\NLLA&
\frac{2}{3}L\Epsinv{2}
+\frac{1}{3}L^2\Epsinv{1}
-\frac{2}{9}L^3
.
\eeqar
Here the UV singularities 
\beqar\label{idiag12sing}
\DDUV{9}(M_1,M_2,M_3,M_4;r_{ij}) &\NLLA&
\frac{1}{3}L^{2}\Epsinv{1}
+\frac{2}{9}L^{3}
,\nl
\DDUV{9}(0,M_2,M_3,M_4;r_{ij}) &\NLLA& 
\frac{1}{3}L^{2}\Epsinv{1}
+\frac{2}{9}L^{3}
+\frac{M_2^2+M_3^2}{{2M_4^2}}
  \left[
  4\Epsinv{3}
  -2 L^2 \Epsinv{1}
  -\frac{4}{3}L^3
  \right]
,\nl
\DDUV{9}(M_1,M_2,M_3,0;r_{ij}) &\NLLA& 
\frac{1}{3}L^{2}\Epsinv{1}
+\frac{2}{9}L^{3}
+\frac{M_2^2+M_3^2}{{2M_1^2}}
\left[
  4\Epsinv{3}
  -2 L^2 \Epsinv{1}
  -\frac{4}{3}L^3
  \right]
,\nl
\DDUV{9}(0,M_2,M_3,0;r_{ij}) &\NLLA&
\frac{2}{3}\Epsinv{3}
\eeqar
have been subtracted.

\subsubsection*{Diagram 10}
\vspace*{-3ex}
\beqar\label{diagram17}
\nmel{2}{10,ij}
\eqdiagl
\vcenter{\hbox{
\unitlength 0.6pt \SetScale{0.6}
\diagXVII{\factblob}
}}
\eqdiagr
- \frac{1}{2}\mel{0}{}  
\sum_{V_1,V_3
= A,Z,W^\pm} 
I_i^{\bar{V}_1}
I_j^{\bar{V}_3}
\sum_{\Phi_{i_2}
=H,\chi,\phi^\pm
}
\left\{
I^{{V}_1}
,I^{{V}_3}
\right\}_{\Phi_{i_2}\Phi_{i_2}}
\nl&&{}\times
\DD{10}(M_{V_1},M_{\Phi_{i_2}},M_{V_3};r_{ij})
,
\eeqar
where
\beqar
\DD{10}\equiv \DD{7}.
\eeqar
Also this diagram, which yields NLL contributions only through
$\FA$--$\FZ$ mixing-energy subdiagrams, gives rise to a correction
proportional to $\MW^2/\MZ^2$ originating from \linebreak
$\deDD{7}(0,\Mphipm,\MZ;r_{ij})$.  This correction cancels the
contribution proportional to $\MW^2/\MZ^2$ that originates from
diagram 9.

\subsubsection*{Diagram 11}
\label{se:fermions}
For the diagrams involving fermionic self-energy subdiagrams we
consider the contributions of a generic fermionic doublet $\Psi$ with
components $\Psi_i=u,d$.  The sum over the three generations of
leptons and quarks is denoted by $\sum_\Psi$, and colour factors are
implicitly understood.  Assuming that all down-type fermions are
massless, $m_d=0$, and that the masses of up-type fermions are $m_u=
0$ or $\Mt$, we have
\beqar\label{diagram9}
\nmel{2}{11,ij}
\eqdiagl
\vcenter{\hbox{
\unitlength 0.6pt \SetScale{0.6}
\diagIX{$\leg{i}$}{$\leg{j}$}{$\scriptscriptstyle{V_1}$}{$\scriptscriptstyle{V_4}$}{$\scriptscriptstyle{\Psi_{i_2}}$}{$\scriptscriptstyle{\Psi_{i_3}}$}{\factblob}
}}
\eqdiagr
-\frac{1}{2} \mel{0}{}
\sum_{V_1,V_4
= A,Z,W^\pm} 
I_i^{\bar{V}_1}
I_j^{\bar{V}_4}
\nl&&{}\times \sum_\Psi
\Biggl\{
\sum_{\Psi_{i_2},\Psi_{i_3}=u,d}
\;
\sum_{\kappa=\rR,\rL}
I^{{V}_1}_{\Psi_{i_3}^\kappa \Psi_{i_2}^\kappa}
I^{{V}_4}_{\Psi_{i_2}^\kappa \Psi_{i_3}^\kappa}
\,
\DD{11,0}(M_{V_1},m_{i_2},m_{i_3},M_{V_4};r_{ij})
\quad
\nl&&{}-
\left(
  I^{{V}_1}_{u^\rR u^\rR} I^{{V}_4}_{u^\rL u^\rL}
+ I^{{V}_1}_{u^\rL u^\rL} I^{{V}_4}_{u^\rR u^\rR}
\right)
m_u^2 \DD{11,m}(M_{V_1},m_u,m_u,M_{V_4};r_{ij})
\Biggr\}
,
\eeqar
where $\DD{11,m}\equiv -4 \DD{8}$ represents the contribution
associated to the $m_u$-terms in the numerator of the up-type fermion
propagators, whereas the integral $\DD{11,0}$, which is defined in
\refeq{defint9}, accounts for the remaining contributions.  This
latter integral yields
%
\beqar\label{idiag90sub}
\DDsub{11,0}(\MW,\MW,\MW,\MW;r_{ij})&\NLLA&
\frac{8}{9}L^{3}
,\nl
\deDDsub{11,0}(M_1,m_2,m_3,M_4;r_{ij})&\NLLA&
0
,\nl
\deDDsub{11,0}(0,m_2,m_3,M_4;r_{ij})&\NLLA& 
\frac{m_2^2+m_3^2}{{2M_4^2}}
\left[
8L\Epsinv{2}
+4L^2\Epsinv{1}
-\frac{8}{3}L^3
\right]
,\nl
\deDDsub{11,0}(M_1,m_2,m_3,0;r_{ij})&\NLLA& 
\frac{m_2^2+m_3^2}{{2M_1^2}}
\left[
8L\Epsinv{2}
+4L^2\Epsinv{1}
-\frac{8}{3}L^3
\right]
,\nl
\deDDsub{11,0}(0,M_2,M_3,0;r_{ij})&\NLLA&
\frac{8}{3}L\Epsinv{2}
+\frac{4}{3}L^2\Epsinv{1}
-\frac{8}{9}L^3
,\nl
\deDDsub{11,0}(0,0,0,0;r_{ij})&\NLLA&
-2\Epsinv{3}
-\frac{8}{9}L^3
,
\eeqar
where the UV singularities 
\beqar\label{idiag90sing}
\DDUV{11,0}(M_1,m_2,m_3,M_4;r_{ij}) &\NLLA&
\frac{4}{3}L^{2}\Epsinv{1}
+\frac{8}{9}L^{3}
,\nl
\DDUV{11,0}(0,m_2,m_3,M_4;r_{ij}) &\NLLA& 
\frac{4}{3}L^{2}\Epsinv{1}
+\frac{8}{9}L^{3}
+\frac{m_2^2+m_3^2}{{2M_4^2}}
  \left[
  8\Epsinv{3}
  -4 L^2 \Epsinv{1}
  -\frac{8}{3}L^3
  \right]
,\nl
\DDUV{11,0}(M_1,m_2,m_3,0;r_{ij}) &\NLLA& 
\frac{4}{3}L^{2}\Epsinv{1}
+\frac{8}{9}L^{3}
+\frac{m_2^2+m_3^2}{{2M_1^2}}
  \left[
  8\Epsinv{3}
  -4 L^2 \Epsinv{1}
  -\frac{8}{3}L^3
  \right]
,\nl
\DDUV{11,0}(0,m_2,m_3,0;r_{ij}) &\NLLA& 
\frac{8}{3}\Epsinv{3}
\eeqar
have been subtracted.
As a consequence of
\beqar
\deDDsub{11,0}(0,m_2,m_3,M_4;r_{ij})
&=&
\frac{m_2^2+m_3^2}{2} \,
\deDDsub{11,m}(0,m_2,m_3,M_4;r_{ij})
,\nl
\deDDsub{11,0}(M_1,m_2,m_3,0;r_{ij})
&=&
\frac{m_2^2+m_3^2}{2} \,
\deDDsub{11,m}(M_1,m_2,m_3,0;r_{ij})
,
\eeqar
all terms proportional to the fermion masses in
$\nmel{2}{11,ij}$ cancel.

\subsubsection*{Diagram 12}
\vspace*{-3ex}
\beqar\label{diagram20}
\nmel{2}{12,ijk}
\eqdiagl
\vcenter{\hbox{
\unitlength 0.6pt \SetScale{0.6}
\diagXX{$\leg{j}$}{$\leg{i}$}{$\leg{k}$}{$\scriptscriptstyle{V_1}$}{$\scriptscriptstyle{V_2}$}{\factblob}
}}
\eqdiagr
\mel{0}{}
\sum_{V_1,V_2=A,Z,W^\pm} 
I_i^{\bar{V}_2}
I_i^{\bar{V}_1}
I_j^{{V}_1}
I_k^{{V}_2}
\DD{12}(M_{V_1},M_{V_2};r_{ik})
,
\nn\\*[-4ex]
\eeqar
where the loop integral $\DD{12}$ is defined in \refeq{defint1} and yields
\beqar\label{idiag20sub}
\DDsub{12}(\MW,\MW;r_{ik})&\NLLA& 
\frac{1}{2}L^{4}
-2 \left(2-\Lrik\right) L^{3}
,\nl
\deDDsub{12}(M_1,M_2;r_{ik})&\NLLA& 
-\frac{2}{3}\left(2\,\LrMI+\LrMII\right)L^{3}
,
\nl
\deDDsub{12}(0,M_2;r_{ik})&\NLLA& 
2L^{2}\Epsinv{2}
+2 L^{3}\Epsinv{1}
+\frac{2}{3}L^{4}
- \left(2-\Lrik\right) \left(
   4 L^{}\Epsinv{2}
  +2 L^{2}\Epsinv{1}
  -\frac{4}{3}L^{3}
  \right)
\nl&&{}
-\LrMII\left(
4L^{}\Epsinv{2}
+6L^{2}\Epsinv{1}
+\frac{14}{3}L^{3}
\right)
,\nl
\deDDsub{12}(M_1,0;r_{ik})&\NLLA& 
-\frac{2}{3}L^{3}\Epsinv{1}
-\frac{7}{6}L^{4}
+ \left(2-\Lrik\right) \left(
  2L^{2}\Epsinv{1}
  +\frac{10}{3}L^{3}
  \right)
\nl&&{}
+\LrMI\left(
  2L^{2}\Epsinv{1}
  +\frac{8}{3}L^{3}
  \right)
,\nl
\deDDsub{12}(0,0;r_{ik})&\NLLA& 
2\Epsinv{4}
-\frac{1}{2}L^{4}
+ \left(2-\Lrik\right) \left(
  4\Epsinv{3}
  +2L^{3}
  \right)
.
\eeqar
Here the UV singularities 
\beqar\label{idiag20sing}
\DDUV{12}(M_1,m_2;r_{ik}) &\NLLA&
-4 L^{2}\Epsinv{1}
-\frac{8}{3} L^{3}
,\nl
\DDUV{12}(0,m_2;r_{ik}) &\NLLA&
-8\Epsinv{3}
\eeqar
have been subtracted.
Note that, to NLL accuracy, the above diagram does not depend on 
$r_{ij}$ and $r_{jk}$.

\subsubsection*{Diagram 13}
\vspace*{-3ex}
\beqar\label{diagram21}
\nmel{2}{13,ijk}
\eqdiagl
\vcenter{\hbox{
\unitlength 0.6pt \SetScale{0.6}
\diagXXI{$\leg{j}$}{$\leg{i}$}{$\leg{k}$}{$\scriptscriptstyle{V_2}$}{$\scriptscriptstyle{V_1}$}{$\scriptscriptstyle{V_3}$}{\factblob}
}}
\eqdiagr
\nl&=&
-\ri 
\frac{\gw}{e}
\mel{0}{}
\sum_{V_1,V_2,V_3
=A,Z,W^\pm} 
\teps^{V_1 V_2 V_3}
I_i^{\bar{V}_1}
I_j^{\bar{V}_2}
I_k^{\bar{V}_3}
\DD{13}(M_{V_1},M_{V_2},M_{V_3};r_{ij},r_{ik},r_{jk})
,
\nl*[-1.7ex]
\eeqar
where the loop integral $\DD{13}$ is defined in \refeq{defint2}.
This integral is free of UV singularities and yields
\beqar\label{idiag21}
\DD{13}(\MW,\MW,\MW;r_{ij},r_{ik},r_{jk})&\NLLA& 
0
,\nl
\deDD{13}(M_1,M_2,M_3;r_{ij},r_{ik},r_{jk})&\NLLA& 
0
,\nl
 \deDD{13}(0,M_2,M_3;r_{ij},r_{ik},r_{jk})&\NLLA& 
\left(l_{ij}-l_{ik}\right)
\left(
L^2\Epsinv{1}
+\frac{5}{3}L^3
\right)
,\nl
 \deDD{13}(M_1,0,M_3;r_{ij},r_{ik},r_{jk})&\NLLA& 
\left(l_{jk}-l_{ij}\right)
\left(
L^2\Epsinv{1}
+\frac{5}{3}L^3
\right)
,\nl
 \deDD{13}(M_1,M_2,0;r_{ij},r_{ik},r_{jk})&\NLLA& 
\left(l_{ik}-l_{jk}\right)
\left(
L^2\Epsinv{1}
+\frac{5}{3}L^3
\right)
.
\eeqar

\subsubsection*{Diagram 14}%
\vspace*{-3ex}
\beqar\label{diagram22}
\nmel{2}{14,ijkl}
\eqdiagl
\vcenter{\hbox{
\unitlength 0.6pt \SetScale{0.6}
\diagXXII{$\leg{i}$}{$\leg{j}$}{$\leg{k}$}{$\leg{l}$}{$\scriptscriptstyle{V_1}$}{$\scriptscriptstyle{V_2}$}{\factblob}
}}
\eqdiagr
\mel{0}{}
\sum_{V_1,V_2
=A,Z,W^\pm} 
I_i^{\bar{V}_1}
I_j^{{V}_1}
I_k^{\bar{V}_2}
I_l^{{V}_2}
\DD{14}(M_{V_1},M_{V_2};r_{ij},r_{kl})
,
\nl*[-1.7ex]
\eeqar
where the loop integral $\DD{14}$ is simply given by the product of
one-loop integrals \refeq{idiag0subt},
\beqar\label{idiag22sub}
\DDsub{14}(M_{V_1},M_{V_2};r_{ij},r_{kl})
&=&
\DDsub{0}(M_{V_1};r_{ij})\,
\DDsub{0}(M_{V_2};r_{kl})
.
\eeqar

\section{Renormalization}
\label{se:ren}
In this section we discuss the NLL counterterm contributions that
result from the renormalization of the gauge-boson masses,%
\footnote{Note that in \refeq{MASSpertserie1}--\refeq{WFpertserie1}
  we use the expansion parameter $\alphaeps$ defined in
  \refeq{pertserie1}.  }
\beqar\label{MASSpertserie1}
M_{V,0}
&=&
M_V+\sum_{l=1}^\infty
\left(\frac{\alphaeps}{4\pi}\right)^l
\de M^{(l)}_V,
\eeqar
and the electroweak couplings,
\beqar\label{PRpertserie1}
g_{0,i}
&=&
g_i+\sum_{l=1}^\infty
\left(\frac{\alphaeps}{4\pi}\right)^l
\de g^{(l)}_i
,\qquad
e_0
=
e+\sum_{l=1}^\infty
\left(\frac{\alphaeps}{4\pi}\right)^l
\de e^{(l)}
,
\eeqar
as well as from the renormalization constants associated with the 
wave functions of the external fermions $k=1,\dots,n$,
\beqar\label{WFpertserie1}
Z_k
&=&
1+\sum_{l=1}^\infty
\left(\frac{\alphaeps}{4\pi}\right)^l
\de Z^{(l)}_k.
\eeqar
The renormalized one- and two-loop amplitudes are presented in
\refse{se:results}.

\subsection{One-loop contributions}
\label{se:1loopren}
At one loop, the mass counterterms $\de M_V^{(1)}$ can be neglected,
since the gauge-boson mass terms in the Born amplitude give only
contributions of order $M_V^2/Q^2$, which are suppressed in the
high-energy limit, and the same holds for the contributions resulting
from their renormalization.

The electroweak couplings are renormalized in the
$\overline{\mathrm{MS}}$ scheme with an additional subtraction of the
UV singularities as explained in \refse{se:uvsing}.  Assuming that the
renormalization scale%
\footnote{We do not identify the renormalization scale $\muR$ and the
  scale of dimensional regularization $\muD$.}  
$\muR$ is of the order of or larger than $\MW$,
this yields the counterterms
\beqar\label{subcouplingren}
\de g_{i}^{(1)} &\NLLA&
-\frac{g_i}{2}
\frac{1}{\veps}
\betacoeff{i}^{(1)}
\left[\left(\frac{Q^2}{\muR^2}\right)^\veps-1\right]
,\qquad
\de e^{(1)} \NLLA
-\frac{e}{2}
\frac{1}{\veps}
\betacoeff{e}^{(1)}
\left[\left(\frac{Q^2}{\muR^2}\right)^\veps-1\right]
,
\eeqar 
and the one-loop $\be$-function coefficients $\betacoeff{1}^{(1)}$,
$\betacoeff{2}^{(1)}$, and $\betacoeff{e}^{(1)}$ are defined in
\refapp{se:betafunctions}.  The dependence of the counterterms
\refeq{subcouplingren} on the factor $(Q^2/\muR)^\veps$ is due to the
normalization of the expansion parameter $\alphaeps$ in
\refeq{PRpertserie1}.  As explained in \refse{se:uvsing}, the
expressions \refeq{subcouplingren} are obtained by subtracting the UV
poles from the usual $\overline{\mathrm{MS}}$ counterterms.  Since the
same subtraction is performed in the bare loop diagrams, the resulting
renormalized amplitudes correspond to the usual
$\overline{\mathrm{MS}}$ renormalized amplitudes.
The renormalization of the mixing parameters $\cw$ and $\sw$ can be
determined from the renormalization of the coupling constants via
\refeq{neutralhiggs}.  In NLL approximation, this prescription is
equivalent to using the on-shell renormalization condition, \ie
relation \refeq{massrelation},
for the weak mixing angle.

Here and in the following we assume that the Born amplitude
$\mel{0}{}$ is expressed in terms of coupling constants renormalized
at the scale $\muR=Q$.  As a consequence, the contribution of the
counterterms \refeq{subcouplingren} to the one-loop amplitude
$\mel{1}{}$ vanishes.
 
The only one-loop counterterm contribution arises from the on-shell
wave-function renormalization constants $\de Z^{(1)}_{k}$ for the
massless fermionic external legs.  These receive contributions only
from massive weak bosons, whereas the photonic contribution vanishes
owing to a cancellation between UV and mass singularities within
dimensional regularization.  After subtraction of the UV poles we find
\beqar\label{subWFRC}
\de Z^{(1)}_{k}&\NLLA&
-
\frac{1}{\veps}
\left\{
\sum_{V=Z,W^\pm} 
I_k^{\bar{V}}
I_k^{{V}}\,
\left[\left(\frac{Q^2}{M_V^2}\right)^\veps-1\right]
-
I_k^{A}
I_k^{{A}}
\right\}.
\eeqar
Finally, the one-loop counterterm for a process with $n$ external
massless fermions in NLL approximation is obtained as
\beqar\label{WFren}
\nmel{1}{\mathrm{WF}} 
 &=&
\mel{0}{} 
\sum_{k=1}^{n}
\frac{1}{2}
\de Z^{(1)}_{k} .
\eeqar

\subsection{Two-loop contributions}
\label{se:2loopren}

At two loops, the mass renormalization leads to non-suppressed
logarithmic terms only through the insertion of the one-loop
counterterms $\de M_V^{(1)}$ in the one-loop logarithmic corrections.
However, these contributions are of NNLL order and can thus be
neglected in NLL approximation \cite{Pozzorini:2004rm}.  In this
approximation also the purely two-loop counterterms that are
associated with the renormalization of the external-fermion wave
functions and the couplings, \ie $\de Z^{(2)}_k$, $\de g_i^{(2)}$ and
$\de e^{(2)}$, do not contribute.

The only NLL two-loop counterterm contributions are those that result from 
the combination of the one-loop amplitude with the one-loop 
counterterms  $\de Z^{(1)}_k$, $\de g_i^{(1)}$ and $\de e^{(1)}$.
The wave-function counterterms yield
\beqar\label{twoloopWF}
\nmel{2}{\mathrm{WF}}  &=&
\nmel{1}{\rF} \sum_{k=1}^{n}
\frac{1}{2}
\de Z^{(1)}_{k}.
\eeqar
The wave-function renormalization constants $\de Z^{(1)}_{k}$ and the
unrenormalized one-loop amplitude $\nmel{1}{\rF}$ are given in
\refeq{subWFRC} and in \refeq{onelooprepunr}, respectively.  In NLL
approximation the one-loop counterterms $\nmel{1}{\mathrm{WF}}$ do not
contribute, and only the LL part of the one-loop amplitude
$\nmel{1}{\mathrm{F}}$ is relevant for \refeq{twoloopWF}.  This is
easily obtained from \refeq{onelooprepunr} by neglecting the NLL terms
depending on $r_{ij}$ and $\LMZW$, and using global gauge invariance
\refeq{chargeconservation} as
\beqar\label{oneloopLL}
e^2 \nmel{1}{\rF}
&\LLA&
\frac{1}{2}
\mel{0}{}
\sum_{i=1}^{n}
\left\{
\sum_{V=A,Z,W^\pm} 
e^2 I_i^{\bar{V}}I_i^{{V}}
\,
\DDsub{0}(\MW;-Q^2)
+e^2 I_i^{A}I_i^{{A}}\,
\deDDsub{0}(0;-Q^2)
\right\}
\\&\LLA&
\frac{1}{2}
\mel{0}{}
\sum_{i=1}^{n}
\left\{
\left[
 \gb^2 \left(\frac{Y_i}{2}\right)^2+\gw^2 C_{i}
\right]
\,
\DDsub{0}(\MW;-Q^2)
+e^2 Q_i^2\,
\deDDsub{0}(0;-Q^2)
\right\}
.
\label{oneloopLL2}
\eeqar
In the second line we used the identity \refeq{casimir}.
Note that only the LL contributions of $\DDsub{0}$ and $\deDDsub{0}$
contribute in \refeq{oneloopLL} and \refeq{oneloopLL2}.

The remaining NLL two-loop counterterms result from the insertion of
the one-loop coupling-constant counterterms \refeq{subcouplingren} in
the LL one-loop amplitude \refeq{oneloopLL2} and read
\beqar\label{twoloopPR}
e^2 \nmel{2}{\mathrm{PR}}
&\NLLA&
-\frac{1}{2\veps}
\left[\left(\frac{Q^2}{\muR^2}\right)^\veps-1\right]
\mel{0}{}
\sum_{i=1}^{n}
\Biggl\{
\left[
 \gb^2 \betacoeff{1}^{(1)}\left(\frac{Y_i}{2}\right)^2+\gw^2 \betacoeff{2}^{(1)}C_{i}
\right]
\,
\DD{0}(\MW;-Q^2)
\nl&&{}
+e^2 \betacoeff{e}^{(1)}Q_i^2\,
\deDD{0}(0;-Q^2)
\Biggr\}.
\eeqar
The various one-loop $\be$-function coefficients in \refeq{twoloopPR},
$\betacoeff{1}^{(1)}$, $\betacoeff{2}^{(1)}$, and
$\betacoeff{e}^{(1)}$, are defined in \refapp{se:betafunctions}.

\section{Complete one- and two-loop results}
\label{se:results}
In this section we combine the unrenormalized and the counterterm
contributions presented in \refses{se:factcont} and \ref{se:ren} and
provide explicit NLL results for the renormalized one- and two-loop
\mbox{$n$-fermion} amplitudes.  As we have seen in \refses{se:factcont} and \ref{se:ren},
the NLL corrections factorize, \ie they can be expressed through
correction factors that multiply the Born amplitude. Moreover, we find
that the two-loop correction factors can be entirely expressed in
terms of one-loop quantities. In this section
we concentrate on the results. Details on the combination of the
various contributions can be found in \refapps{se:oneloopcomb} and
\ref{se:twoloopcomb}.

\subsection{Renormalized one-loop amplitude}
\label{se:oneloopres}

The renormalized one-loop matrix element for a process with $n$
external massless fermions is given by
\beqar\label{onelooprena}
\nmel{1}{}
&=&
\nmel{1}{\mathrm{F}} 
+
\nmel{1}{\mathrm{WF}} 
,
\eeqar
where 
\beqar\label{oneloopfactorizable}
\nmel{1}{\mathrm{F}} 
=
\frac{1}{2}\sum_{i=1}^{n}\sum_{j=1\atop j\neq i}^{n}
\nmel{1}{ij} 
\eeqar
represents the bare one-loop contribution, which is given by the
factorizable part \refeq{oneloopdiag4}, and $\nmel{1}{\mathrm{WF}}$ is
the wave-function-renormalization counterterm \refeq{WFren}.
Using the explicit results presented in \refse{se:oneloop} we can
write (more details can be found in \refapp{se:oneloopcomb})
\beqar\label{onelooprep}
\nmel{1}{}&\NLLA&
\mel{0}{}
\left[
\FF{1}{\sew}
+\Delta \FF{1}{\elm}
+\Delta \FF{1}{\PZ}
\right].
\eeqar
Here the corrections are split into a symmetric-electroweak (sew) part,
\beqar\label{onelooprepa1}
\FF{1}{\sew}
&=&
-\frac{1}{2}\sum_{i=1}^{n}\sum_{j=1\atop j\neq i}^{n}
\sum_{V=A,Z,W^\pm} 
I_i^{\bar{V}}I_j^{{V}}\,\univfact{\veps}{\MW;-r_{ij}}
,
\eeqar
which is obtained by setting the masses of all gauge bosons,
$A,Z$ and $W^\pm$, equal to $\MW$ in the loop diagrams,
an electromagnetic (em) part
\beqar\label{onelooprepa2}
\Delta \FF{1}{\elm}
&=&
-\frac{1}{2}\sum_{i=1}^{n}\sum_{j=1\atop j\neq i}^{n}
I_i^{A}I_j^{{A}}\,\Delta\univfact{\veps}{0;-r_{ij}}
,
\eeqar
resulting from the mass gap between the W boson and the massless photon,
and an $\MZ$-dependent part
\beqar\label{onelooprepa3}
\Delta \FF{1}{\PZ}
&=&
-\frac{1}{2}\sum_{i=1}^{n}\sum_{j=1\atop j\neq i}^{n}
I_i^{Z}I_j^{{Z}}\,\Delta\univfact{\veps}{\MZ;-r_{ij}}
,
\eeqar
describing the effect that results from the difference between $\MW$
and $\MZ$. For the functions $I$, including contributions
up to the order $\veps^2$, we obtain
\beqar\label{Ifunc}
\univfact{\veps}{\MW;-r_{ij}}&\NLLA&
-L^2
-\frac{2}{3}L^3\Eps{}
-\frac{1}{4}L^4\Eps{2}
+({3}-2l_{ij})
\left(L
+\frac{1}{2}L^2\Eps{}
+\frac{1}{6}L^3\Eps{2}
\right)
+\order(\veps^3)
,\nl
\univfact{\veps}{\MZ;-r_{ij}}&\NLLA&
\univfact{\veps}{\MW;-r_{ij}}+
\LMZW\left(
2L
+2L^2\Eps{}
+L^3\Eps{2}
\right)
+\order(\veps^3)
,\nl
\univfact{\veps}{0;-r_{ij}}&\NLLA&
-2\Epsinv{2}
-(3-2 l_{ij})\Epsinv{1}
,
\eeqar
and the subtracted functions $\Delta I$ are defined as
\beqar\label{subtractionofI}
\Delta\univfact{\veps}{m;-r_{ij}}&=&
\univfact{\veps}{m;-r_{ij}}
-\univfact{\veps}{\MW;-r_{ij}}.
\eeqar
In LL approximation we have
\beq
\univfact{\veps}{m;-r_{ij}}\LLA\DD{0}(m;r_{ij}),
\eeq
such that we can replace $\DD{0}$ and $\deDD{0}$ by $I$ and $\De I$,
respectively, in \refeq{oneloopLL}, \refeq{oneloopLL2}, and
\refeq{twoloopPR}.

\subsection{Renormalized two-loop amplitude}
\label{se:twoloopres}

The renormalized two-loop matrix element reads
\beqar\label{twoloopcomb}
\nmel{2}{}
&=&
\nmel{2}{\mathrm{F}}
+
\nmel{2}{\mathrm{WF}}
+\nmel{2}{\mathrm{PR}}
,
\eeqar
where
\beqar\label{twoloopcomb2}
\nmel{2}{\mathrm{F}}
&=&
\sum_{i=1}^{n}\sum_{j=1\atop j\neq i}^{n}
\left[
\frac{1}{2}
\left(
\nmel{2}{1,ij}
+
\nmel{2}{2,ij}
\right)
+\nmel{2}{3,ij}
+\nmel{2}{4,ij}
+\nmel{2}{5,ij}
\right.\nl&&\left.{}
+\frac{1}{2}\sum_{m=6}^{11}\nmel{2}{m,ij}
+\sum_{k=1\atop k\neq i,j}^{n}
\left(
\nmel{2}{12,ijk}
+\frac{1}{6}
\nmel{2}{13,ijk}
+\frac{1}{8}
\sum_{l=1\atop l\neq i,j,k}^{n}
\nmel{2}{14,ijkl}
\right)\right]
\eeqar
represents the bare two-loop contribution, which is given by the
factorizable terms \refeq{twoloopdiag3}, and $\nmel{2}{\mathrm{WF}}$
and $\nmel{2}{\mathrm{PR}}$ are the wave-function- and
parameter-renormalization counterterms given in \refeq{twoloopWF} and
\refeq{twoloopPR}, respectively.
As shown in
\refapp{se:twoloopcomb}, the two-loop amplitude \refeq{twoloopcomb}
can be expressed in terms of the Born matrix element $\mel{0}{}$ and
the one-loop correction factors
\refeq{onelooprepa1}--\refeq{onelooprepa3} as
\beqar\label{twoloopresult}
\nmel{2}{}&\NLLA&
\mel{0}{}
\Biggl\{
\frac{1}{2}\left[\FF{1}{\sew}\right]^2
+\FF{1}{\sew} \Delta \FF{1}{\elm}
+\frac{1}{2}\left[\Delta \FF{1}{\elm}\right]^2
+\FF{1}{\sew} \Delta \FF{1}{\PZ}
+\Delta \FF{1}{\PZ} \Delta \FF{1}{\elm}
\nl&&{}
+\GG{2}{\sew}
+\Delta \GG{2}{\elm}
\Biggr\}
,
\eeqar
where the additional terms 
\beqar\label{betaterms}
e^2 \GG{2}{\sew}
&=& \frac{1}{2}\sum_{i=1}^{n}\left[
\betacoeff{1}^{(1)} \gb^2 \left(\frac{Y_i}{2}\right)^2
+\betacoeff{2}^{(1)} \gw^2 C_{i}
\right]
J(\veps,\MW,\muR^2)
,\nl
\Delta \GG{2}{\elm}
&=&
\frac{1}{2}\sum_{i=1}^{n}
Q_i^2
\Biggl\{
\betacoeff{e}^{(1)} 
\left[
\Delta J(\veps,0,\muR^2)
-\Delta J(\veps,0,\MW^2)
\right]
+\betacoeff{\QED}^{(1)}
\Delta J(\veps,0,\MW^2)
\Biggr\}
\quad
\eeqar
contain one-loop $\be$-function coefficients, defined in
\refapp{se:betafunctions}, and the combinations
\beqar\label{Jterms}
J(\veps,m,\muR^2)&=&\frac{1}{\veps}\left[
I(2 \veps,m,Q^2)
-\left(\frac{Q^2}{\muR^2}\right)^\veps
I(\veps,m,Q^2)
\right]
,\nl
\Delta J(\veps,m,\muR^2)
&=&
J(\veps,m,\muR^2)-J(\veps,\MW,\muR^2)
\eeqar
of one-loop $I$-functions \refeq{Ifunc} and \refeq{subtractionofI} for
$m=\MW,\MZ,0$.  The relevant $J$-functions read explicitly
\beqar
J(\veps,\MW,\muR^2) &\NLLA& \frac{1}{3}L^3 - \LmuR L^2 +\order(\veps),\nl
\De J(\veps,0,\MW^2) &\NLLA& \frac{3}{2}\Eps{-3} + 2L\Eps{-2} + L^2\Eps{-1} +\order(\veps),\nl
\De J(\veps,0,\muR^2)- \De J(\veps,0,\MW^2) &\NLLA& 
\LmuR\left(-2\Eps{-2}+\Eps{-1}(\LmuR-2L) + \LmuR L
  -\frac{1}{3}\LmuR^2\right) +\order(\veps),
\nln
\eeqar
where 
\beq
\LmuR=\ln\left(\frac{\muR^2}{\MW^2}\right).
\eeq

In order to be able to express \refeq{twoloopresult} in terms of the
one-loop operators \refeq{onelooprepa1}--\refeq{onelooprepa3} it is
crucial that terms up to order $\Eps{2}$ are included in the latter.

The coefficients $\betacoeff{e}^{(1)}$ and
$\betacoeff{\QED}^{(1)}$ describe the running of the electromagnetic coupling 
above and below the electroweak scale, respectively.
The former receives contributions from all charged 
fermions and bosons, whereas the latter 
receives contributions only from light fermions, 
\ie all charged leptons and quarks apart from the top quark.

The couplings that enter the one- and two-loop correction factors%
\footnote{ These are the coupling $\alpha$ in the perturbative
  expansion \refeq{pertserie1a}--\refeq{pertserie1} and the couplings
  $\gb,\gw$ and $e$ that appear in
  \refeq{twoloopresult}--\refeq{Jterms} and enter also
  \refeq{onelooprepa1}--\refeq{onelooprepa3} through the dependence of
  the generators \refeq{genmixing} on the couplings and the mixing
  parameters $\cw$ and $\sw$.  }  are
renormalized at the scale $\muR$.  Instead, as discussed in
\refse{se:1loopren}, the coupling constants in the Born matrix element
$\mel{0}{}$ in \refeq{onelooprep} and \refeq{twoloopresult} are
renormalized at the scale $Q$, \ie
\beq\label{bornrenormalization}
\mel{0}{}\equiv
\mel{0}{}
\Bigg|_{g_i=g_i(Q^2)\atop e=e(Q^2)}
\eeq
with
\beqar
g_{i}^2(Q^2) 
&\NLLA&
g_{i}^2(\muR^2) \left[1
-\frac{\alphaeps}{4\pi}\betacoeff{i}^{(1)}
\frac{1}{\veps}\left[\left(\frac{Q^2}{\muR^2}\right)^\veps-1\right]
\right]
,\nl
e^2(Q^2) 
&\NLLA&
e^2(\muR^2) \left[1
-\frac{\alphaeps}{4\pi}\betacoeff{e}^{(1)}
\frac{1}{\veps}\left[\left(\frac{Q^2}{\muR^2}\right)^\veps-1\right]
\right]
.
\eeqar
Thus, by definition, the Born amplitude $\mel{0}{}$ is independent of
the renormalization scale $\muR$, and the dependence of the one- and two-loop
amplitudes on $\muR$ is described by the terms \refeq{betaterms}.

The contributions \refeq{betaterms} originate from combinations of UV
and mass singularities.  We observe that the term proportional to
$\betacoeff{e}^{(1)}$ vanishes for $\muR=\MW$. Instead, the terms
proportional to $\betacoeff{1}^{(1)}$, $\betacoeff{2}^{(1)}$, and
$\betacoeff{\QED}^{(1)}$ cannot be eliminated through an appropriate
choice of the renormalization scale.  This indicates that such
two-loop terms do not originate exclusively from the running of the
couplings in the one-loop amplitude.

Combining the Born amplitude with the one- and two-loop NLL corrections 
we can write
\beqar\label{factresult1}
\M&\NLLA&\mel{0}{}\, 
F^{\sew}\,
F^{\PZ}\,
F^{\elm},
\eeqar
where we observe a factorization of the symmetric-electroweak
contributions,
\beqar\label{Fsew}
F^{\sew}&\NLLA&
1+
\frac{\alphaeps}{4 \pi}
\FF{1}{\sew}
+\left(\frac{\alphaeps}{4 \pi}\right)^2
\left[
\frac{1}{2}\left(\FF{1}{\sew}\right)^2
+\GG{2}{\sew}
\right]
,
\eeqar
the terms resulting from the difference between $\MW$ and $\MZ$,
\beqar
F^{\PZ}
&\NLLA&
1+
\frac{\alphaeps}{4 \pi}
\Delta \FF{1}{\PZ}
,
\eeqar
and the electromagnetic terms resulting from the mass gap between the
photon and the W~boson,
\beqar\label{eq:Felmnll}
F^{\elm}
&\NLLA&
1+
\frac{\alphaeps}{4 \pi}
\Delta \FF{1}{\elm}
+\left(\frac{\alphaeps}{4 \pi}\right)^2
\left[
\frac{1}{2}\left(
\Delta \FF{1}{\elm}
\right)^2
+\Delta \GG{2}{\elm}
\right]
.
\eeqar
We also observe that the  symmetric-electroweak and electromagnetic terms 
are consistent with the exponentiated expressions
\beqar\label{expresult1}
F^{\sew}&\NLLA&
\exp\left[
\frac{\alphaeps}{4 \pi}
\FF{1}{\sew}
+\left(\frac{\alphaeps}{4 \pi}\right)^2
\GG{2}{\sew}
\right]
,\nl
F^{\elm}&\NLLA&
\exp\left[
\frac{\alphaeps}{4 \pi}
\Delta \FF{1}{\elm}
+\left(\frac{\alphaeps}{4 \pi}\right)^2
\Delta \GG{2}{\elm}
\right].
\eeqar
In particular, these two contributions exponentiate separately.  This
double-exponentia\-ting structure is indicated by the ordering of the
one-loop operators $\FF{1}{\sew}$ and $\Delta \FF{1}{\elm}$ in the
interference term $\FF{1}{\sew} \Delta \FF{1}{\elm}$ in our result
\refeq{twoloopresult}.  It is important to realize that the commutator
of these two operators
yields a non-vanishing NLL two-loop contribution.  This means that the
double-exponentiated structure of the result is not equivalent to a
simple exponentiation, \ie
\beqar\label{expresult1b}
F^{\sew} F^{\elm}\neq
\exp\left[
\frac{\alphaeps}{4 \pi}
\left(\FF{1}{\sew}+\Delta \FF{1}{\elm}\right)
+\left(\frac{\alphaeps}{4 \pi}\right)^2
\left(
\GG{2}{\sew}+\Delta \GG{2}{\elm}
\right)
\right]
.
\eeqar
Instead we observe that in NLL approximation
the commutator of $\Delta \FF{1}{\PZ}$ with all operators
in \refeq{factresult1}--\refeq{eq:Felmnll} vanishes,
and also  $[\Delta \FF{1}{\PZ}]^2$ does not contribute.
Thus, we have 
\beq
F^{\sew}
F^{\PZ}
F^{\elm}
\NLLA
F^{\PZ}
F^{\sew}
F^{\elm}
\NLLA
F^{\sew}
F^{\elm}
F^{\PZ},
\eeq
and, in principle,
$F^{\PZ}$ can be written in exponentiated form,
\beqar
F^{\PZ}&\NLLA&
\exp\left(
\frac{\alphaeps}{4 \pi}
\Delta\FF{1}{\PZ}
\right)
,
\eeqar
or absorbed in one of the other exponentials,
\beqar
F^{\sew}
F^{\PZ}
&\NLLA&
\exp\left[
\frac{\alphaeps}{4 \pi}
\left(
\FF{1}{\sew}
+\Delta\FF{1}{\PZ}
\right)
+\left(\frac{\alphaeps}{4 \pi}\right)^2
\GG{2}{\sew}
\right]
,\nl
F^{\PZ}F^{\elm}&\NLLA&
\exp\left[
\frac{\alphaeps}{4 \pi}
\left(
\Delta \FF{1}{\elm}
+\Delta
\FF{1}{\PZ}
\right)
+\left(\frac{\alphaeps}{4 \pi}\right)^2
\Delta \GG{2}{\elm}
\right].
\eeqar

The one- and two-loop corrections
\refeq{onelooprep}--\refeq{onelooprepa3} and
\refeq{twoloopresult}--\refeq{betaterms} contain various combinations
of matrices $I_i^V$, which are in general non-commuting and
non-diagonal. These matrices have to be applied to the Born amplitude
$\M_{0}$ according to the definition~\refeq{abbcouplings}.  In order
to express the results in a form which is more easily applicable to a
specific process, it is useful to split the integrals
$\univfact{\veps}{M_V;-r_{ij}}$ and $\De\univfact{\veps}{M_V;-r_{ij}}$
in \refeq{onelooprepa1}--\refeq{onelooprepa3} into an
angular-independent part $\univfact{\veps}{M_V;Q^2}$ and
$\De\univfact{\veps}{M_V;Q^2}$, which only involves $\veps$ and $L$,
and an angular-dependent part, which additionally depends on
logarithms of $r_{ij}$.  This permits to eliminate the sum over~$j$
for the angular-independent parts of
\refeq{onelooprepa1}--\refeq{onelooprepa3} using the
charge-conservation relation~\refeq{chargeconservation},
\beqar\label{chargeconversationsum}
\mel{0}{}
\sum_{i=1}^{n}\sum_{j=1\atop j\neq i}^{n}
I_i^{\bar{V}}I_j^{{V}}\,
\univfact{\veps}{M_V;Q^2}
&=&
-\mel{0}{}
\sum_{i=1}^{n}
I_i^{{V}} I_i^{\bar{V}}\,
\univfact{\veps}{M_V;Q^2}.
\eeqar
Moreover, one can easily see that the angular-independent part of
\refeq{onelooprepa1} leads to the Casimir operator \refeq{casimir}.
After these simplifications, all operators that are associated with
the angular-independent parts can be replaced by the corresponding
eigenvalues, and the one- and two-loop results can be written as
\beqar\label{onelooprepb}
\nmel{1}{}&\NLLA&
\mel{0}{}
\left[
\Ff{1}{\sew}
+\Delta \Ff{1}{\elm}
+\Delta \Ff{1}{\PZ}
\right],
\eeqar
and
\beqar\label{twoloopresultb}
\nmel{2}{}&\NLLA&
\mel{0}{}
\Biggl\{
\frac{1}{2}\left[\Ff{1}{\sew}\right]^2
+\Ff{1}{\sew} \Delta \Ff{1}{\elm}
+\frac{1}{2}\left[\Delta \Ff{1}{\elm}\right]^2
+\Ff{1}{\sew} \Delta \Ff{1}{\PZ}
+\Delta \Ff{1}{\PZ} \Delta \Ff{1}{\elm}
\nl&&{}
+\Gg{2}{\sew}
+\Delta \Gg{2}{\elm}
\Biggr\}
,
\eeqar
with
\beqar\label{eigenvalues2}
\Ff{1}{\sew}
&\NLLA& 
-\frac{1}{2} \left(
  L^2
  +\frac{2}{3}L^3\Eps{}
  +\frac{1}{4}L^4\Eps{2}
  -3L
  -\frac{3}{2}L^2\Eps{}
  -\frac{1}{2}L^3\Eps{2}
  \right)
  \sum_{i=1}^n \left[ \frac{\gb^2}{e^2}\left(\frac{y_i}{2}\right)^2 
+
\frac{\gw^2}{e^2}
c_{i}
\right]
\nl&&{}
 +\left( L + \frac{1}{2}L^2\Eps{} + \frac{1}{6}L^3\Eps{2} \right)
 \mathcal{K}^{\mathrm{ad}}_{1}
+ \order(\veps^3)
,\nl
\Delta \Ff{1}{\elm}
&\NLLA&
-\frac{1}{2} \left(
   2\Epsinv{2}
  +3\Epsinv{1}
  -L^2
  -\frac{2}{3}L^3\Eps{}
  -\frac{1}{4}L^4\Eps{2}
  +3L
  +\frac{3}{2}L^2\Eps{}
  +\frac{1}{2}L^3\Eps{2}
  \right)
  \sum_{i=1}^n q_i^2
\nl&&
-\left( 
\Epsinv{1}
+L + \frac{1}{2}L^2\Eps{} + \frac{1}{6}L^3\Eps{2} \right)
  \sum_{i=1}^n \sum_{j=1\atop j\neq i}^n l_{ij} q_i q_j
+ \order(\veps^3)
,\nl
\Delta \Ff{1}{\PZ}
&\NLLA& \left( L + L^2\Eps{} + \frac{1}{2}L^3\Eps{2} \right) \LMZW
  \sum_{i=1}^n \left(\frac{\gw}{e} \cw t^3_i - \frac{\gb}{e} \sw \frac{y_i}{2}\right)^2
+ \order(\veps^3)
,\nl
\Gg{2}{\sew}
&\NLLA& 
\left(
\frac{1}{6}L^3-\frac{1}{2}\LmuR L^2
\right)
\sum_{i=1}^n\left[
 \frac{\gb^2}{e^2}\betacoeff{1}^{(1)}
  \left(\frac{y_i}{2}\right)^2
+ \frac{\gw^2}{e^2}\betacoeff{2}^{(1)}
c_{i}
\right]
+ \order(\veps)
,\nl
\Delta \Gg{2}{\elm}
&\NLLA&
\biggl\{-
\LmuR
\left[
\Epsinv{2}
+\left(L-\frac{1}{2}\LmuR\right)\Epsinv{1}
-\LmuR\left(\frac{1}{2}L-\frac{1}{6}\LmuR\right)
\right]
\betacoeff{e}^{(1)} 
\nl&&{}
+
\left(
\frac{3}{4}\Epsinv{3}
+   L\Epsinv{2}
+ \frac{1}{2}L^2\Epsinv{1}
\right)
\betacoeff{\QED}^{(1)} 
\biggr\} \sum_{i=1}^nq_i^2
+ \order(\veps)
,
\eeqar
where $\LmuR=\ln{(\muR^2/\MW^2)}$, and $c_{i}$, $t^3_{i}$, $y_{i}$,
$q_{i}$, represent the eigenvalues of the operators $C_{i}$,
$T^3_{i}$, $Y_{i}$, and $Q_{i}$, respectively.  The only matrix-valued
expression in \refeq{eigenvalues2} is the angular-dependent part of
the symmetric-electroweak contribution $\Ff{1}{\sew}$,
\beqar\label{angdepmat}
\mathcal{K}^{\mathrm{ad}}_{1}
&=&\sum_{i=1}^n \sum_{j=1\atop j\neq i}^n l_{ij}  \sum_{V=A,Z,W^\pm} I_i^{\bar{V}}I_j^{{V}}
.
\eeqar
The two-loop corrections \refeq{twoloopresultb} involve terms
proportional to $\mathcal{K}^{\mathrm{ad}}_{1}$ and
$[\mathcal{K}^{\mathrm{ad}}_{1}]^2$.  However, the latter
are of NNLL order and thus negligible in NLL approximation.  The
combination of the matrix \refeq{angdepmat} with the Born amplitude,
\beqar\label{angdepmat2}
\mel{0}{}\,
\mathcal{K}^{\mathrm{ad}}_{1}
&=&\sum_{i=1}^n \sum_{j=1\atop j\neq i}^n l_{ij}  \sum_{V=A,Z,W^\pm} 
\mel{0}{\varphi_1\dots\varphi'_i\dots\varphi'_j\dots\varphi_n}
I_{\varphi'_i\varphi_i}^{\bar{V}}I_{\varphi'_j\varphi_j}^{{V}}
\nl
&=&\sum_{i=1}^n \sum_{j=1\atop j\neq i}^n l_{ij}  
\left\{
\mel{0}{\varphi_1\dots\varphi_i\dots\varphi_j\dots\varphi_n}
\left[
\frac{\gb^2}{e^2} \frac{y_i y_j}{4}+
\frac{\gw^2}{e^2} t^3_i t^3_j
\right]
\right.\nl&&\left.{}
+\sum_{V=W^\pm} 
\mel{0}{\varphi_1\dots\varphi'_i\dots\varphi'_j\dots\varphi_n}
I_{\varphi'_i\varphi_i}^{\bar{V}}I_{\varphi'_j\varphi_j}^{{V}}
\right\},
\eeqar
requires the evaluation of matrix elements involving SU(2)-transformed
external fermions $\varphi'_i,\varphi'_j$, \ie isospin partners of the
fermions $\varphi_i,\varphi_j$.

\section{Discussion}
\label{se:disc}

In this section we compare
our results presented in \refse{se:results}
to existing results from the literature and apply them to specific
processes.

\subsection{Extension of previous results}

In \citere{Denner:2003wi} the one- and two-loop LL and
angular-dependent NLL contributions have been calculated for arbitrary
non-mass-suppressed Standard Model processes, using a photon mass and
fermion masses for the regularization of soft and collinear
singularities, respectively.  The purely symmetric-electroweak parts
of our results, made up of $\FF{1}{\sew}$ or $\Ff{1}{\sew}$ and its
square, confirm the corresponding terms involving $\delta_\sew$ in
\citere{Denner:2003wi}, which can be seen e.g. by comparing
\refeq{onelooprepb}, \refeq{twoloopresultb} and \refeq{eigenvalues2}
from our paper with (4.9), (4.10) and (4.24) from
\citere{Denner:2003wi}.  In addition to the existing results we have
added all the remaining (non-angular-dependent) NLL contributions,
including those proportional to $\ln(\MZ^2/\MW^2)$ and the terms
involving $\be$-function coefficients, as well as higher orders in
$\veps$ in the one-loop results. We cannot compare the electromagnetic
parts contained in $\Delta \FF{1}{\elm}$ or $\Delta \Ff{1}{\elm}$ and
$\delta_{\mathrm{sem}}$ because of the different regularization
schemes for the photonic singularities, i.e. for the soft and
collinear divergences resulting from massless photons.

In \citere{Pozzorini:2004rm} the complete one- and two-loop LL and NLL
contributions for the electroweak singlet form factors have been
derived. We can easily reproduce these form factor contributions as a
special case of our results, as for only two external fermions no
angular-dependent logarithms appear and the summation over external
legs, using \refeq{chargeconversationsum}, is trivial.  The functions
$I$, $\Delta I$, $J$, and $\Delta J$ appearing in (4.69), (4.70), and
(4.73) of \citere{Pozzorini:2004rm} correspond to the equally named
functions from this paper for $r_{ij} = -Q^2 = s$.  In the form factor
case, due to the absence of the angular-dependent term
$\mathcal{K}^{\mathrm{ad}}_{1}$ in \refeq{eigenvalues2}, the one-loop
operators $\FF{1}{\sew}$ and $\Delta \FF{1}{\elm}$ commute, so that
the two-loop result can be written as a single exponential and we have
$\nmel{2}{}\NLLA \mel{0}{} \, \{ \frac{1}{2} [\FF{1}{\sew} + \Delta
\FF{1}{\elm} + \Delta \FF{1}{\PZ} ]^2 +\GG{2}{\sew} +\Delta
\GG{2}{\elm} \}$, corresponding to the form
in~\citere{Pozzorini:2004rm}.  Note that we have renormalized all
couplings of the loop corrections at the unique scale
$\mu_{\mathrm{w}} = \mu_e = \muR$, where the last two lines of (4.73)
in \citere{Pozzorini:2004rm} vanish.

\subsection{Comparison to Catani's formula in QCD}
\newcommand{\CatI}{\mbox{\boldmath$I$}}%
\newcommand{\CatT}{\mbox{\boldmath$T$}}%
The structure of our results for electroweak logarithmic corrections
is similar to the singular structure of scattering amplitudes in QCD.
In \citere{Catani:1998bh} the singular part of a QCD one-loop
amplitude, i.e. the pole part with terms $\Epsinv{2}$ and
$\Epsinv{1}$, is obtained from the Born amplitude by applying an
operator $\CatI^{(1)}(\veps,\mu^2;\{p\})$ which, for only fermions and
antifermions as external particles, can be expressed through Eqs.
(12)--(15) of \citere{Catani:1998bh} as
\beqar
\label{eq:Cat1loop}
  \frac{\alpha_{\mathrm{S}}(\muR^2)}{2\pi} \,
  \CatI^{(1)}(\veps,\muR^2;\{p\})
  \NLLA
  \frac{\alphaseps}{4\pi} \left(-\frac{1}{2}\right)
  \sum_i \sum_{j\ne i} \CatT_i\cdot\CatT_j \,
  \univfact{\veps}{0;-r_{ij}}
  ,
\eeqar
where $\univfact{\veps}{0;-r_{ij}}$ is the function defined in
\refeq{Ifunc}, and $\alphaseps$ represents the strong coupling
renormalized at $\muR^2$ and rescaled as in \refeq{pertserie1}.  The
product $\CatT_i\cdot\CatT_j$ of the colour charge operators
corresponds to $\sum_{V=A,Z,W^\pm} I_i^{\bar{V}}I_j^{{V}}$ for
electroweak interactions, so that \refeq{eq:Cat1loop} has exactly the
form of $\FF{1}{\sew}$~\refeq{onelooprepa1} for $\MW=0$, including the
angular-dependent logarithms.  Note that the factor
$(Q^2/\muR^2)^\veps$ from the difference in the definitions of
$\alpha_{\mathrm{S}}$~\cite{Catani:1998bh} and
$\alphaeps$~\refeq{pertserie1},
\beq
\alpha_{\mathrm{S}}(\muR^2)=\alphaseps\left(\frac{Q^2}{\muR^2}\right)^\veps,
\eeq
is multiplied by a factor
$(-\muR^2/r_{ij})^\veps$ from $\CatI^{(1)}(\veps,\muR^2;\{p\})$,
producing the correct angular-dependent logarithms
$l_{ij}=\ln(-r_{ij}/Q^2)$ contained in $\univfact{\veps}{0;-r_{ij}}$.
The $\veps$-dependent pre-factor in the definition of
$\CatI^{(1)}(\veps,\muR^2;\{p\})$, $e^{-\veps\psi(1)}/\Gamma(1-\veps)
= 1 + \order(\veps^2)$, is irrelevant in NLL accuracy, and the same
holds for the corresponding factors in the two-loop result.

In two loops, the operator acting on the Born amplitude is given by
Catani's formula, Eqs. (18)--(21) of \citere{Catani:1998bh},
in NLL accuracy as
\beqar
  \left(\frac{\alpha_{\mathrm{S}}(\muR^2)}{2\pi}\right)^2
  \left\{
    \frac{1}{2} \left[ \CatI^{(1)}(\veps,\muR^2;\{p\}) \right]^2
    + \frac{2\pi\beta_0}{\veps} \left[
      \CatI^{(1)}(2\veps,\muR^2;\{p\})
      - \CatI^{(1)}(\veps,\muR^2;\{p\})
    \right]
  \right\}
  . \quad
\eeqar
The first term in the curly brackets originates from the combination
of $\CatI^{(1)}(\veps,\muR^2;\{p\})$ applied to the one-loop amplitude
and $-\frac{1}{2} [ \CatI^{(1)}(\veps,\muR^2;\{p\}) ]^2$
acting on the Born amplitude in \citere{Catani:1998bh}.
It can be identified with the term $\frac{1}{2}[\FF{1}{\sew}]^2$ from
our two-loop result~\refeq{twoloopresult}.
Using colour conservation, the second term gives
\beqar\label{Catani2loopbeta}
  \lefteqn{\left(\frac{\alpha_{\mathrm{S}}(\muR^2)}{2\pi}\right)^2 \,
  \frac{2\pi\beta_0}{\veps} \left[
    \CatI^{(1)}(2\veps,\muR^2;\{p\})
    - \CatI^{(1)}(\veps,\muR^2;\{p\})
    \right]}\quad
\nl && 
  \NLLA
  \left(\frac{\alphaseps}{4\pi}\right)^2 \,
  4\pi\beta_0 \cdot \frac{1}{2}
  \sum_i \CatT_i^2 \,
  J(\veps,0,\muR^2)
  ,
\eeqar
where $J(\veps,0,\muR^2)$ is the function defined in \refeq{Jterms}.
In the electroweak model, the expression $4\pi\beta_0 \, \CatT_i^2$ in
\refeq{Catani2loopbeta} corresponds
directly to the term
$[\betacoeff{1}^{(1)} g_1^2 (Y_i/2)^2 +
  \betacoeff{2}^{(1)} g_2^2 C_i] / e^2$
in \refeq{betaterms}.
Therefore \refeq{Catani2loopbeta} can be identified with
the contribution of $\GG{2}{\sew}$ for $\MW=0$.

The symmetric-electroweak part of our results can thus be obtained
from Catani's formula by an obvious replacement of gauge-group
quantities together with a simple substitution of massless one-loop
integrals by massive ones.  
The remaining parts of our results, which are due to the differences
between the masses of the photon, $\PZ$~boson, and $\PW$~boson and
which cannot be inferred from Catani's formula, can be expressed as
simple combinations of the same one-loop integrals.

\subsection{Comparison to electroweak resummation results}

In \citeres{Melles:2001gw,Melles:2001ia,Melles:2001mr,Melles:2001dh} a
resummation of electroweak one-loop results has been proposed,
including all LL and NLL corrections apart from the $\ln(\MZ^2/\MW^2)$
terms.

The non-angular-dependent $\order(\veps^0)$-terms of $\Ff{1}{\sew}$
are found in \citere{Melles:2001gw} in the sum $\sum_{k=1}^{n_f}$ of
Eq.~(48) [Eq.~(49) in the hep-ph version] for the contribution of
external fermionic lines above the weak scale.  These expressions are
extended in Eq.~(48) of \citere{Melles:2001mr} in order to include the
$\be$-function terms of $\Gg{2}{\sew}$.  Here care must be taken to
use the \emph{resummed} one-loop running of the couplings, i.e.
$\alpha(s)=\alpha(M^2)/(1+c\,\ln s/M^2)$ with
$c=\alpha(M^2)\beta_0/\pi$, in order to correctly arrive at
\beqar
  \frac{1}{c} \ln\frac{s}{M^2}
    \left(\ln\frac{\alpha(M^2)}{\alpha(s)} - 1\right)
  + \frac{1}{c^2} \ln\frac{\alpha(M^2)}{\alpha(s)}
  = \frac{1}{2} \ln^2\frac{s}{M^2} - \frac{c}{6} \ln^3\frac{s}{M^2}
    + \order(c^2)
  ,
\eeqar
and equivalently for $\alpha\to\alpha'$, $c\to c'$, reproducing the
coefficient of the $\be$-function terms in
$\Gg{2}{\sew}$~\refeq{eigenvalues2} for $\muR=\MW$.

The angular-dependent NLL corrections are treated in
\citere{Melles:2001dh}.  The last line of Eq.~(13) there [Eq.~(11) in
the hep-ph version] matches the $\mathcal{K}^{\mathrm{ad}}_{1}$-term
of \refeq{eigenvalues2} and \refeq{angdepmat} in $\order(\veps^0)$ if
one takes all gauge-boson masses to be $m_{V_a}=\MW$.  The
symmetric-electroweak parts of the one- and two-loop amplitudes in
\citere{Melles:2001dh} are thus in agreement with our results.

To summarize, our explicit one- and two-loop results
\refeq{onelooprepb}, \refeq{twoloopresultb} and \refeq{eigenvalues2}
confirm the symmetric-electroweak parts of the resummed ansatz.
The electromagnetic contributions, i.e. the contributions from below
the weak scale, cannot be compared due to the different regularization
schemes for the photonic singularities.  However, the factorization of
the electromagnetic contributions from the symmetric-electroweak ones
and the fact that the former can be expressed in terms of one-loop QED
corrections are in agreement with the approach proposed in
\citeres{Melles:2001gw,Melles:2001ia,Melles:2001mr,Melles:2001dh}.

\subsection{Four-fermion scattering processes}

We now apply our results to massless four-fermion processes
\beqar\label{4fprocess}
  \varphi_1(p_1) \, \varphi_2(p_2) \to \varphi_3(-p_3) \, \varphi_4(-p_4)
,
\eeqar
where each of the $\varphi_i$ may be a massless fermion, $\varphi_i =
f^{\kappa_i}_{\sigma_i}$, or antifermion, $\varphi_i = \bar
f^{\kappa_i}_{\sigma_i}$, with the notations from
\refse{se:definitions}, provided that the number of fermions and
antifermions in the initial and final state is equal.  The scattering
amplitudes for the processes~\refeq{4fprocess} follow directly from
our results for the generic $n \to 0$ process~\refeq{genproc} by
crossing symmetry. The Mandelstam invariants are given by $s = r_{12}
= r_{34}$, $t = r_{13} = r_{24}$, and $u = r_{14} = r_{23}$ with
$r_{ij} = (p_i+p_j)^2$.

The following discussion focuses on $s$-channel processes of the form
\beqar\label{4fprocess2}
  f^{\kappa}_{\sigma} \, \bar f^{\kappa}_{\rho} \to
  f^{\kappa'}_{\sigma'} \, \bar f^{\kappa'}_{\rho'}
\eeqar
with one fermion and one antifermion in the initial as well as in the
final state, where $f^{\kappa}_{\sigma}$ and $f^{\kappa'}_{\sigma'}$
are neither identical nor isospin partners of each other, and the same
holds for $\bar f^{\kappa}_{\rho}$ and $\bar f^{\kappa'}_{\rho'}$.
Therefore the external fermion lines are always connected between
$f^{\kappa}_{\sigma}$ and $\bar f^{\kappa}_{\rho}$ in the initial
state and between $f^{\kappa'}_{\sigma'}$ and $\bar
f^{\kappa'}_{\rho'}$ in the final state.  The number of independent
chiralities~$\kappa_i$ is thus restricted to two, $\kappa$ and
$\kappa'$, and the particle pairs in the initial and final state must
either be antiparticles of each other or antiparticles of the mutual
isospin partners.  The first case corresponds to neutral-current
four-fermion scattering, which is treated in \refse{se:4fNC}.  The
second case refers to charged-current four-fermion scattering and is
treated in \refse{se:4fCC}.

The scattering amplitudes of all other processes~\refeq{4fprocess} can
be obtained from \refeq{4fprocess2} by crossing symmetry and additive
combinations of amplitudes.  For instance, the full four-quark
amplitude for $\Pu \bar \Pu \to \Pd \bar \Pd$ is given by the sum of the
$s$-channel neutral-current amplitude and the $t$-channel
charged-current amplitude, where the latter follows from the result of
the \mbox{$s$-channel} charged-current amplitude for $\Pu \bar \Pd \to \Pu
\bar \Pd$ by exchanging $p_2$ and $p_3$, and thus $s$ and $t$.  The
neutral- and charged-current results in \refses{se:4fNC} and
\ref{se:4fCC} together are thus sufficient to construct the scattering
amplitudes for all massless four-fermion processes~\refeq{4fprocess}.

\subsubsection{Neutral-current four-fermion scattering}
\label{se:4fNC}

\newcommand{\NC}{\mathrm{NC}}
\newcommand{\NCMz}{\mel{0,\NC}{}}
\newcommand{\NCamp}{\mathcal{A}_\NC}
\newcommand{\NCCz}{C_{0,\NC}}
\newcommand{\NCM}{\mel{\NC}{}}
\newcommand{\NCF}[1]{F_\NC^{#1}}
\newcommand{\NCCsew}{C_{1,\NC}^\sew}
\newcommand{\NCCad}{C_{1,\NC}^\mathrm{ad}}
\newcommand{\NCgsew}{g_{2,\NC}^\sew}
\newcommand{\NCfZ}{\Delta\Ff{1,\NC}{\PZ}}
\newcommand{\NCfem}{\Delta\Ff{1,\NC}{\elm}}
\newcommand{\NCgem}{\Delta\Gg{2,\NC}{\elm}}
\newcommand{\NCMfin}{\mel{\NC}{\mathrm{fin}}}
\newcommand{\NCUQED}{U_\NC^\QED}
\newcommand{\NCfQED}{\Ff{1,\NC}{\QED}}
\newcommand{\NCgQED}{\Gg{2,\NC}{\QED}}
\newcommand{\NCnMfin}[1]{\nmel{#1,\NC}{\mathrm{fin}}}
\newcommand{\Ptop}{\mathrm{top}}

This section deals with the $s$-channel neutral-current four-fermion
processes
\beqar\label{4fNCprocess}
  f^{\kappa}_{\sigma} \, \bar f^{\kappa}_{\sigma} \to
  f^{\kappa'}_{\sigma'} \, \bar f^{\kappa'}_{\sigma'}
\,,
\eeqar
where a fermion--antifermion pair annihilates and produces another
fermion--antifermion pair.  The electromagnetic charge quantum numbers
of the external particles are given by
$q_f = q_{f^{\kappa}_{\sigma}} = -q_{\bar f^{\kappa}_{\sigma}}$
and $q_{f'} = q_{f^{\kappa'}_{\sigma'}} = -q_{\bar f^{\kappa'}_{\sigma'}}$,
the hypercharges by
$y_f = y_{f^{\kappa}_{\sigma}} = -y_{\bar f^{\kappa}_{\sigma}}$
and $y_{f'} = y_{f^{\kappa'}_{\sigma'}} = -y_{\bar f^{\kappa'}_{\sigma'}}$,
the isospin components by
$t^3_f = t^3_{f^{\kappa}_{\sigma}} = -t^3_{\bar f^{\kappa}_{\sigma}}$
and $t^3_{f'} = t^3_{f^{\kappa'}_{\sigma'}} =
  -t^3_{\bar f^{\kappa'}_{\sigma'}}$,
and the isospin by
$t_f = |t^3_f|$, $t_{f'} = |t^3_{f'}|$.
These electroweak quantum numbers depend on the flavour and the
chirality of the fermions.

The Born amplitude reads
\beqar\label{4fBorn}
  \NCMz = \frac{1}{s} \NCamp \, \NCCz
\eeqar
in the high-energy limit, where
\beqar\label{4fspinors}
  \NCamp = \bar v(p_2,\kappa) \gamma^\mu u(p_1,\kappa) \,
           \bar u(-p_3,\kappa') \gamma_\mu v(-p_4,\kappa')
\eeqar
represents the spinor structure of the incoming and outgoing massless
fermions with chiralities $\kappa$ and $\kappa'$, and
\beqar\label{4fresultC0}
  \NCCz = g_1^2(Q^2) \, \frac{y_f y_{f'}}{4} + g_2^2(Q^2) \, t^3_f t^3_{f'}
  \,.
\eeqar
As indicated, the couplings $g_i(Q^2)$ in the Born amplitude are
renormalized at the scale~$Q$, whereas the additional couplings and
mixing angles in the loop corrections below are renormalized at the
scale~$\muR$.

We write the neutral-current amplitude in the form
\beqar
  \NCM \NLLA \NCMz\, \NCF{\sew}\, \NCF{\PZ}\, \NCF{\elm}
\eeqar
according to~\refeq{factresult1}.  Applying the non-diagonal
operator~$F^{\sew}$ \refeq{Fsew} to the Born amplitude, we find
\beqar\label{4fresultsew}
\lefteqn{
  \NCMz \, \NCF{\sew} \NLLA \frac{1}{s} \NCamp \, \Biggl\{
  \NCCz
  + \frac{\alphaeps}{4\pi} \, \Biggl[
    - \left(
            L^2
            +\frac{2}{3}L^3\Eps{}
            +\frac{1}{4}L^4\Eps{2}
            -3L
            -\frac{3}{2}L^2\Eps{}
            -\frac{1}{2}L^3\Eps{2}
      \right)
} \quad
\nl && \qquad\qquad\qquad\qquad\qquad\qquad{} \times
      \NCCz \, \NCCsew
    + \left(
            L + \frac{1}{2}L^2\Eps{} + \frac{1}{6}L^3\Eps{2}
      \right)
      \NCCad
    + \order(\veps^3)
    \Biggr]
\nl && {}
  + \left(\frac{\alphaeps}{4\pi}\right)^2 \Biggl[
    \biggl(\frac{1}{2} L^4
      - 3L^3\biggr) \, \NCCz \left(\NCCsew\right)^2
    - L^3 \, \NCCad \, \NCCsew
    + \NCCz \, \NCgsew
    + \order(\veps)
    \Biggr]
  \Biggr\}
\,,
\nln
\eeqar
where
\beqar\label{4fresultC1sew}
  \NCCsew = \frac{g_1^2}{e^2} \, \frac{y_f^2}{4}
             + \frac{g_2^2}{e^2} \, t_f (t_f + 1)
             + (f \leftrightarrow f')
\eeqar
results from the non-angular-dependent contributions to $\Ff{1}{\sew}$
in \refeq{eigenvalues2}, whereas the application of the
angular-dependent operator $\mathcal{K}^{\mathrm{ad}}_{1}$ on the Born
amplitude yields
\beqar\label{4fresultC1ad}
  \NCCad &=&
  \NCCz \left[
    4\,\ln\left(\frac{u}{t}\right)
      \left( \frac{g_1^2}{e^2} \, \frac{y_f y_{f'}}{4}
        + \frac{g_2^2}{e^2} \, t^3_f t^3_{f'} \right)
    - 2\,\ln\left(\frac{-s}{Q^2}\right) \NCCsew
    \right]
\nl &&{}
  + 2 \, g_2^2(Q^2) \, \frac{g_2^2}{e^2} \left[
      \ln\left(\frac{u}{t}\right) t_f t_{f'}
    - \Biggl( \ln\left(\frac{t}{s}\right) +
        \ln\left(\frac{u}{s}\right) \Biggr) \, t^3_f t^3_{f'}
    \right]
  \,.
\eeqar
The last missing part in \refeq{4fresultsew},
\beqar\label{4fresultg2sew}
  \NCgsew \NLLA \left(\frac{1}{3} L^3 - \LmuR L^2\right)
  \left[
    \frac{g_1^2}{e^2} \, \betacoeff{1}^{(1)} \, \frac{y_f^2}{4}
    + \frac{g_2^2}{e^2} \, \betacoeff{2}^{(1)} \, t_f (t_f + 1)
    + (f \leftrightarrow f')
  \right]
  + \order(\veps)
, \quad
\eeqar
results directly from \refeq{eigenvalues2}, where the values for the
$\be$-function coefficients $\betacoeff{1}^{(1)}$ and
$\betacoeff{2}^{(1)}$ in the electroweak Standard Model are given in
\refeq{betacoeffres}.

The symmetric-electroweak result \refeq{4fresultsew} is
multiplied with the diagonal factors
\beqar\label{4fresultZ}
  \NCF{\PZ} \NLLA 1 + \frac{\alphaeps}{4\pi} \, \NCfZ
\eeqar
and
\beqar\label{4fresultem}
  \NCF{\elm} &\NLLA&
  1 + \frac{\alphaeps}{4\pi} \, \NCfem
  + \left(\frac{\alphaeps}{4\pi}\right)^2 \left[
    \frac{1}{2} \left(\NCfem\right)^2 + \NCgem
    \right]
.
\eeqar
The factors $\NCF{\PZ}$ and $\NCF{\elm}$ follow from
\refeq{eigenvalues2},
\beqar\label{4fresultf1Z}
  \NCfZ &\NLLA&
  2 \left( L + L^2\Eps{} + \frac{1}{2}L^3\Eps{2} \right) \LMZW
  \left[
    \left(\frac{g_2}{e} \cw t^3_f - \frac{g_1}{e} \sw \frac{y_f}{2}
      \right)^2
    + (f \leftrightarrow f')
  \right]
  + \order(\veps^3)
,
\nln
\\
\label{4fresultf1em}
  \NCfem &\NLLA& {}
  - \left(
         2\Epsinv{2}
        +3\Epsinv{1}
        -L^2
        -\frac{2}{3}L^3\Eps{}
        -\frac{1}{4}L^4\Eps{2}
        +3L
        +\frac{3}{2}L^2\Eps{}
        +\frac{1}{2}L^3\Eps{2}
    \right)
    \left( q_f^2 + q_{f'}^2 \right)
\nl && {}
  - \left( 
      \Epsinv{1}
      +L + \frac{1}{2}L^2\Eps{} + \frac{1}{6}L^3\Eps{2}
    \right)
    \left[
      4\,\ln\left(\frac{u}{t}\right) q_f q_{f'}
      - 2\,\ln\left(\frac{-s}{Q^2}\right)
        \left( q_f^2 + q_{f'}^2 \right)
    \right]
\nl && {}
  + \order(\veps^3)
,
\\
\label{4fresultg2em}
  \NCgem &\NLLA&
  \biggl\{
    -\LmuR \left[
      2\Epsinv{2}
      +\left(2L-\LmuR\right)\Epsinv{1}
      -\LmuR\left(L-\frac{1}{3}\LmuR\right)
      \right]
      \betacoeff{e}^{(1)} 
\nl &&\qquad
  + \left(
      \frac{3}{2}\Epsinv{3} + 2L\Epsinv{2} + L^2\Epsinv{1}
    \right)
    \betacoeff{\QED}^{(1)}
    \biggr\}
    \left( q_f^2 + q_{f'}^2 \right)
  + \order(\veps)
,
\eeqar
where the values for the
$\be$-function coefficients $\betacoeff{e}^{(1)}$ and
$\betacoeff{\QED}^{(1)}$ in the electroweak Standard Model are given in
\refeq{betacoeffres} and \refeq{eq:betacoeffQED}.

Our result can be compared to
\citeres{Kuhn:2000nn,Kuhn:2001hz,Jantzen:2005az}, where the
logarithmic two-loop contributions to neutral-current four-fermion
scattering have been determined by resummation techniques.  These
papers use the renormalization scale $\muR=\MW$ for the couplings in
the loop corrections, so that we have to set
$\LmuR=\ln(\muR^2/\MW^2)=0$ in \refeq{4fresultg2sew} and
\refeq{4fresultg2em}.  The large electroweak logarithms in
\citeres{Kuhn:2000nn,Kuhn:2001hz,Jantzen:2005az} are defined with the
choice $Q^2=-s$, such that $\ln(-s/Q^2)=0$ in \refeq{4fresultC1ad} and
\refeq{4fresultf1em}.

In \citeres{Kuhn:2000nn,Kuhn:2001hz,Jantzen:2005az} the photonic
singularities are regularized with a finite photon mass. Therefore we
cannot directly compare their electromagnetic corrections with our
dimensionally regularized result. But
\citeres{Kuhn:2000nn,Kuhn:2001hz,Jantzen:2005az} define a finite scattering
amplitude by factorizing the complete QED corrections,
\beqar\label{4fresultMfin}
  \NCMfin = \frac{\NCM}{\NCUQED}
\,.
\eeqar
The factor $\NCUQED$ represents the full QED corrections from photons
coupling exclusively to fermions, with the photonic singularities
regularized in the same way as in the amplitude~$\NCM$.  The soft and
collinear divergences resulting from massless photons cancel in the
ratio~\refeq{4fresultMfin}, and $\NCMfin$ is independent of the
regularization scheme for the photonic singularities.  Using our
dimensional regularization scheme and $\muR=\MW$, $Q^2=-s$ as above,
$\NCUQED$ is given by
\beqar\label{4fresultUQED}
  \NCUQED \NLLA 1 +
    \frac{\alphaeps}{4\pi} \, \NCfQED
    + \left(\frac{\alphaeps}{4\pi}\right)^2 \left[
      \frac{1}{2} \left(\NCfQED\right)^2 + \NCgQED
      \right]
,
\eeqar
with
\beqar\label{4fresultf1QED}
  \NCfQED &\NLLA&
  - \left(
         2\Epsinv{2}
        +3\Epsinv{1}
    \right)
    \left( q_f^2 + q_{f'}^2 \right)
  - 4 \Epsinv{1}
    \ln\left(\frac{u}{t}\right) q_f q_{f'}
\,,
\\
\label{4fresultg2QED}
  \NCgQED &\NLLA&
  \left[
    \left(
      \frac{3}{2}\Epsinv{3} + 2L\Epsinv{2} + L^2\Epsinv{1}
      + \frac{1}{3}L^3
    \right)
    \betacoeff{\QED}^{(1)}
  + \frac{1}{3} L^3 \, \betacoeff{\Ptop}^{(1)}
  \right]
  \left( q_f^2 + q_{f'}^2 \right)
  + \order(\veps)
. \qquad
\eeqar
In contrast to $\NCF{\elm}$~\refeq{4fresultem}, $\NCUQED$, according
to its definition in \citeres{Kuhn:2000nn,Kuhn:2001hz,Jantzen:2005az},
contains the complete photon contribution without the subtractions at
a photon mass equal to $\MW$. This is why \refeq{4fresultf1QED} and
\refeq{4fresultg2QED} differ from \refeq{4fresultf1em} and
\refeq{4fresultg2em} by finite logarithmic terms.  The expressions
$\NCfQED$ and $\NCgQED$ can be obtained from
$\Delta\FF{1}{\elm}$~\refeq{onelooprepa2} and
$\Delta\GG{2}{\elm}$~\refeq{betaterms} by replacing $\Delta I \to I$
and $\Delta J \to J$ and adding the term proportional to
$\betacoeff{\Ptop}^{(1)}$ in \refeq{4fresultg2QED}, where
$\betacoeff{\Ptop}^{(1)}$ is defined in \refeq{eq:betacoeffQEDtop} and
corresponds to the top-quark contribution to the electromagnetic
$\be$-function.  In our result, this latter term is implicitly
contained in $\NCgsew$~\refeq{4fresultg2sew} and, by construction,
cancels in the subtracted expression $\NCgem$~\refeq{4fresultg2em}.
The factor~$\NCUQED$ in \refeq{4fresultUQED} is valid for $\Mt \sim
\MW$, whereas the corresponding factor in
\citeres{Kuhn:2001hz,Jantzen:2005az} is only valid for a massless top
quark. However, the finite amplitude~$\NCMfin$ is independent of the
top-quark mass in NLL approximation.

The one- and two-loop contributions to
$\NCMfin$~\refeq{4fresultMfin} can be expressed as follows:
\beqar\label{4fresultMfin1}
  \NCnMfin{1} &\NLLA& \frac{1}{s} \NCamp \, \biggl\{
    -\NCCz \left[ \left(L^2 - 3L\right) \NCCsew - \NCfem +
      \NCfQED \right]
\nl && \qquad\qquad{}
    + L \, \NCCad
    + \NCCz \, \NCfZ
    \biggr\}
    + \order(\veps),
\nl
  \NCnMfin{2} &\NLLA& \frac{1}{s} \NCamp \, \Biggl\{
    \frac{1}{2} \NCCz
      \left[ \left(L^2 - 3L\right) \NCCsew - \NCfem +
        \NCfQED \right]^2
\nl && \qquad\qquad{}
    - \left( L \, \NCCad + \NCCz \, \NCfZ \right)
      \left( L^2 \, \NCCsew - \NCfem +
        \NCfQED \right)
\nl && \qquad\qquad{}
    + \NCCz \left( \NCgsew + \NCgem - \NCgQED \right)
    \Biggr\}
    + \order(\veps).
\eeqar
The one-loop result reproduces Eq.~(50) from \citere{Kuhn:2000nn}, and
the two-loop result agrees with Eqs.~(51), (52) of
\citere{Kuhn:2000nn} and Eqs.~(51), (52), (54) of \citere{Kuhn:2001hz}
for $N_g=3$ families of leptons and quarks.  Note that in these papers
$(\alpha/4\pi)(g_2^2/e^2)$ is used as the parameter of the
perturbative expansion.  The corrections proportional to $\NCfZ$,
which account for the mass difference of the heavy electroweak gauge
bosons, $\MZ\ne\MW$, have to be compared to Eq.~(62) of
\citere{Jantzen:2005az}, where the first order of an expansion in the
parameter $\delta_M = \sw^2 = 1-\MW^2/\MZ^2$ is presented.  Using
$\LMZW = \ln(\MZ^2/\MW^2) = \sw^2 + \order(\sw^4)$,
\citere{Jantzen:2005az} gives indeed the first order in~$\sw^2$ of our
result, provided that we set $\sw=0$ and $\cw=1$ in
$\NCfZ$~\refeq{4fresultf1Z}, thus neglecting higher orders in $\sw^2$.

\subsubsection{Charged-current four-fermion scattering}
\label{se:4fCC}

\newcommand{\CC}{\mathrm{CC}}
\newcommand{\CCMz}{\mel{0,\CC}{}}
\newcommand{\CCamp}{\mathcal{A}_\CC}
\newcommand{\CCM}{\mel{\CC}{}}
\newcommand{\CCF}[1]{F_\CC^{#1}}
\newcommand{\CCCsew}{C_{1,\CC}^\sew}
\newcommand{\CCCad}{C_{1,\CC}^\mathrm{ad}}
\newcommand{\CCgsew}{g_{2,\CC}^\sew}
\newcommand{\CCfZ}{\Delta\Ff{1,\CC}{\PZ}}
\newcommand{\CCfem}{\Delta\Ff{1,\CC}{\elm}}
\newcommand{\CCgem}{\Delta\Gg{2,\CC}{\elm}}
\newcommand{\CCCem}{C_{1,\CC}^\elm}
\newcommand{\CCCz}{\frac{g_2^2(Q^2)}{2}}
\newcommand{\CCCadem}{C_{1,\CC}^{\mathrm{ad},\elm}}

In order to complete our predictions for massless four-fermion
scattering, we apply our results to the $s$-channel charged-current
processes
\beqar\label{4fCCprocess}
  f^{\rL}_{\sigma} \, \bar f^{\rL}_{\rho} \to
  f^{\rL}_{\sigma'} \, \bar f^{\rL}_{\rho'}
\,,
\eeqar
where the fermions $f^{\rL}_{\sigma}$ and $f^{\rL}_{\sigma'}$
are the isospin partners of $f^{\rL}_{\rho}$ and $f^{\rL}_{\rho'}$,
respectively. The hypercharge quantum numbers of the external
particles are given by
$y_f = y_{f^{\rL}_{\sigma}} = -y_{\bar f^{\rL}_{\rho}}$
and $y_{f'} = y_{f^{\rL}_{\sigma'}} = -y_{\bar f^{\rL}_{\rho'}}$
and the isospin components by
$t^3 = t^3_{f^{\rL}_{\sigma}} = t^3_{\bar f^{\rL}_{\rho}}
  = \smash{t^3_{f^{\rL}_{\sigma'}} = t^3_{\bar f^{\rL}_{\rho'}}}$.
All external fermions have to be left-handed, so $t = |t^3| = {1}/{2}$.

In the high-energy approximation, the Born amplitude reads
\beqar\label{4fCCBorn}
  \CCMz = \frac{1}{s} \CCamp \, \CCCz
\,
\eeqar
with the spinor structure
\beqar\label{4fCCspinors}
  \CCamp = \bar v(p_2,\rL) \gamma^\mu u(p_1,\rL) \,
           \bar u(-p_3,\rL) \gamma_\mu v(-p_4,\rL)
.
\eeqar
As in the previous section, the amplitude is written in the
form
\beqar
  \CCM \NLLA \CCMz\, \CCF{\sew}\, \CCF{\PZ}\, \CCF{\elm}
\,.
\eeqar
Applying the non-diagonal operator~$F^{\sew}$ \refeq{Fsew} to the
Born amplitude, we find
\beqar\label{4fCCresultsew}
\lefteqn{
  \CCMz \, \CCF{\sew} \NLLA \frac{1}{s} \CCamp \, \Biggl\{
  \CCCz
  + \frac{\alphaeps}{4\pi} \, \Biggl[
    - \left(
            L^2
            +\frac{2}{3}L^3\Eps{}
            +\frac{1}{4}L^4\Eps{2}
            -3L
            -\frac{3}{2}L^2\Eps{}
            -\frac{1}{2}L^3\Eps{2}
      \right)
} \;
\nl && \qquad\qquad\qquad\qquad\qquad\quad\;\;{} \times
      \CCCz \, \CCCsew
    + \left(
            L + \frac{1}{2}L^2\Eps{} + \frac{1}{6}L^3\Eps{2}
      \right)
      \CCCad
    + \order(\veps^3)
    \Biggr]
\nl && {}
  + \left(\frac{\alphaeps}{4\pi}\right)^2 \Biggl[
    \left(\frac{1}{2} L^4 - 3L^3\right) \CCCz \left(\CCCsew\right)^2
    - L^3 \, \CCCad \, \CCCsew
    + \CCCz \, \CCgsew
    + \order(\veps)
    \Biggr]
  \Biggr\}
\,,
\nln
\eeqar
with
\beqar\label{4fCCresultC1sew}
  \CCCsew &=& \frac{g_1^2}{e^2} \, \frac{y_f^2+y_{f'}^2}{4}
             + \frac{3}{2} \, \frac{g_2^2}{e^2}
\,,
\nl
\label{4fCCresultC1ad}
  \CCCad &=&
  \CCCz \left[
    4\,\ln\left(\frac{u}{t}\right)
      \frac{g_1^2}{e^2} \, \frac{y_f y_{f'}}{4}
    - 2 \,
      \Biggl( \ln\left(\frac{t}{s}\right) +
        \ln\left(\frac{u}{s}\right) \Biggr)
      \, \frac{g_2^2}{e^2}
    - 2\,\ln\left(\frac{-s}{Q^2}\right) \CCCsew
    \right]
\nl &&{}
  + 2\,\ln\left(\frac{u}{t}\right)
    g_1^2(Q^2) \, \frac{y_f y_{f'}}{4} \, \frac{g_2^2}{e^2}
\,,
\nl
\label{4fCCresultg2sew}
  \NCgsew &\NLLA& \left(\frac{1}{3} L^3 - \LmuR L^2\right)
  \left(
    \frac{g_1^2}{e^2} \, \betacoeff{1}^{(1)} \, \frac{y_f^2+y_{f'}^2}{4}
    + \frac{3}{2} \, \frac{g_2^2}{e^2} \, \betacoeff{2}^{(1)}
  \right)
.
\eeqar
The diagonal factors
\beqar\label{4fCCresultZ}
  \CCF{\PZ} \NLLA 1 + \frac{\alphaeps}{4\pi} \, \CCfZ
\eeqar
and
\beqar\label{4fCCresultem}
  \CCF{\elm} &\NLLA&
  1 + \frac{\alphaeps}{4\pi} \, \CCfem
  + \left(\frac{\alphaeps}{4\pi}\right)^2 \left[
    \frac{1}{2} \left(\CCfem\right)^2 + \CCgem
    \right]
\eeqar
are expressed through
\beqar\label{4fCCresultf1Z}
  \CCfZ &\NLLA&
  \left( L + L^2\Eps{} + \frac{1}{2}L^3\Eps{2} \right) \LMZW
  \left(
    \frac{g_2^2}{e^2} \cw^2
    + 2 \, \frac{g_1^2}{e^2} \sw^2 \, \frac{y_f^2+y_{f'}^2}{4}
  \right)
  + \order(\veps^3)
,
\nl
\label{4fCCresultf1em}
  \CCfem &\NLLA&
  - \left(
         2\Epsinv{2}
        +3\Epsinv{1}
        -L^2
        -\frac{2}{3}L^3\Eps{}
        -\frac{1}{4}L^4\Eps{2}
        +3L
        +\frac{3}{2}L^2\Eps{}
        +\frac{1}{2}L^3\Eps{2}
    \right)
    \CCCem
\nl &&
  - \left( 
      \Epsinv{1}
      +L + \frac{1}{2}L^2\Eps{} + \frac{1}{6}L^3\Eps{2}
    \right)
    \CCCadem
  + \order(\veps^3)
,
\nl
\label{4fCCresultg2em}
  \CCgem &\NLLA&
  \biggl\{
    -\LmuR \left[
      2\Epsinv{2}
      +\left(2L-\LmuR\right)\Epsinv{1}
      -\LmuR\left(L-\frac{1}{3}\LmuR\right)
      \right]
      \betacoeff{e}^{(1)} 
\nl &&\qquad
  + \left(
      \frac{3}{2}\Epsinv{3} + 2L\Epsinv{2} + L^2\Epsinv{1}
    \right)
    \betacoeff{\QED}^{(1)}
    \biggr\} \,
    \CCCem
  + \order(\veps)
,
\eeqar
with
\beqar
\label{4fCCresultC1em}
  \CCCem &=&
    \frac{1}{2} \left(
      q_{f^\rL_\sigma}^2 + q_{\bar f^\rL_\rho}^2
      + q_{f^\rL_{\sigma'}}^2 + q_{\bar f^\rL_{\rho'}}^2
      \right)
  = \frac{g_1^2}{e^2} \cw^2 \, \frac{y_f^2+y_{f'}^2}{4}
    + \frac{1}{2} \, \frac{g_2^2}{e^2} \sw^2
\,,
\nl
  \CCCadem &=& 2 \, \Biggl[
    \ln\left(\frac{-s}{Q^2}\right) \left(
      q_{f^\rL_\sigma} q_{\bar f^\rL_\rho}
      + q_{f^\rL_{\sigma'}} q_{\bar f^\rL_{\rho'}} \right)
    - \ln\left(\frac{-t}{Q^2}\right) \left(
      q_{f^\rL_\sigma} q_{f^\rL_{\sigma'}}
      + q_{\bar f^\rL_\rho} q_{\bar f^\rL_{\rho'}} \right)
\nl && \qquad{}
    - \ln\left(\frac{-u}{Q^2}\right) \left(
      q_{f^\rL_\sigma} q_{\bar f^\rL_{\rho'}}
      + q_{f^\rL_{\sigma'}} q_{\bar f^\rL_\rho} \right)
    \Biggr]
\nl &=&
      4\,\ln\left(\frac{u}{t}\right)
        \frac{g_1^2}{e^2} \cw^2 \, \frac{y_f y_{f'}}{4}
      - \Biggl( \ln\left(\frac{t}{s}\right)
          + \ln\left(\frac{u}{s}\right) \Biggr) \,
        \frac{g_2^2}{e^2} \sw^2
      - 2\,\ln\left(\frac{-s}{Q^2}\right)
        \CCCem
\,. \qquad
\eeqar

\section{Conclusion}
\label{se:conc}

We have studied the one- and two-loop virtual electroweak corrections
to arbitrary processes with external massless fermions in the Standard
Model. In the high-energy region, where all kinematical invariants are
at an energy scale $Q$ that is large compared to the electroweak
gauge-boson masses, we have calculated mass singularities in
$D=4-2\veps$ dimensions taking into account all leading logarithmic
(LL) and next-to-leading logarithmic (NLL) contributions. This
approximation includes all combinations
$\al^l\veps^{-k}\ln^{j-k}(Q^2/\MW^2)$ of mass-singular logarithms and
$1/\veps$ poles with $j=2l,2l-1$ and $2l-4\le k \le j$. All masses of
the heavy particles have been assumed to be of the same order
$\MW\sim\MZ\sim\MH\sim\Mt$ but not equal, and all light fermions have
been assumed to be massless.

The calculation has been performed in the complete
spontaneously-broken electroweak Standard Model using the
't~Hooft--Feynman gauge.  All contributions have been split into those
that factorize the lowest-order matrix element and non-factorizable
parts. The non-factorizable parts have been shown to vanish in NLL
approximation owing to collinear Ward identities.  All factorizable
contributions have been evaluated using a suitable soft--collinear
approximation and minimal subtraction of the ultraviolet
singularities. Explicit results have been given for all contributing
factorizable Feynman diagrams. The two-loop integrals have been solved
by two independent methods in NLL approximation. One makes use of
sector decomposition to isolate the mass singularities, the other uses
the strategy of regions.  The fermionic wave functions are
renormalized on shell, and coupling-constant renormalization is
performed in the $\overline{\mathrm{MS}}$ scheme, but can be
generalized easily.

In order to isolate the effects resulting from the mass gaps between
the photon, the W~boson, and the Z~boson, all contributions have been
split into parts corresponding to $\MA=\MZ=\MW$ and remaining
subtracted parts associated with the massless photon and the Z~boson.
By combining the results of all diagrams we found that the electroweak
mass singularities assume a form that is analogous to the singular
structure of scattering amplitudes in massless QCD.  The sum of the
two-loop leading and next-to-leading logarithms is composed of terms
that can be written as the second-order terms of exponentials of the
one-loop contribution plus additional NLL contributions that are
proportional to the one-loop $\be$-function coefficients. All terms
can be cast into a product of three exponentials. The first inner
exponential contains the part of the corrections corresponding to
$\MA=\MZ=\MW$, \ie the $\SU(2)\times\U(1)$ symmetric part. The second
exponential contains the part originating from the mass gap between
the Z~boson and the W~boson and contains only terms involving
$\ln(\MZ^2/\MW^2)$. The third outer exponential summarizes the
contributions that originate from the mass gap between the photon and
the W~boson and corresponds to the QED corrections subtracted by the
corresponding corrections with $\MA=\MW$.  While the second
exponential commutes with the other two and does in fact get no
second-order contribution in NLL approximation, the first and the last
exponential do not commute.

If one neglects the NLL contributions proportional to
$\ln(\MZ^2/\MW^2)$, our result confirms the resummation prescriptions
that have been proposed in the literature. These prescriptions are
based on the assumption that, in the high-energy limit, the
electroweak theory can be described by a symmetric, unmixed
$\SU(2)\times\U(1)$ theory, where all electroweak gauge bosons have
mass $\MW$, matched with QED at the electroweak scale.  Indeed, apart
from the terms involving $\ln(\MZ^2/\MW^2)$, in the final result we
observe a cancellation of all effects associated with symmetry
breaking, \ie gauge-boson mixing, the gap between $\MZ$ and $\MW$, and
couplings proportional to the vacuum expectation value.  This simple
behaviour of the two-loop NLL corrections is ensured by subtle
cancellations of mass singularities from different diagrams, where the
details of spontaneous symmetry breaking cannot be neglected but have
to be taken into account properly.

In massless fermionic processes, the symmetry-breaking effects are
restricted to a small subset of diagrams, since the Higgs sector is
coupled to massless fermions only via one-loop insertions in the
gauge-boson self-energies.  Using the techniques developed in this
paper, which are to a large extent process independent, we plan to
extend our study of two-loop NLL mass singularities to processes
involving heavy external particles that are directly coupled to the
Higgs sector.

As an application of our results for general \mbox{$n$-fermion}
processes we have presented explicit expressions for the case of
neutral-current and charged-current 4-fermion reactions. In the former
case we found agreement with existing predictions obtained with the
help of resummations prescriptions.

Our results are also applicable to reactions that involve massless
fermions and (hard) gluons, such as 2-jet production at hadron
colliders, since gluons do not couple to electroweak gauge bosons.

\begin{appendix}

\section{Loop integrals}\label{app:loops}
\newcommand{\lmom}{l}
\newcommand{\slmom}{\lslash}
\newcommand{\mass}{m}
\newcommand{\linea}[1]{k_{#1}}
\newcommand{\lineb}[1]{q_{#1}}
\newcommand{\linec}[1]{r_{#1}}
\newcommand{\slinea}[1]{\ks_{#1}}
\newcommand{\slineb}[1]{\qs_{#1}}
\newcommand{\measure}[1]{\rd \tilde{\lmom}_{#1}}
\newcommand{\propagatorm}[2]{P(#1,#2)}
\newcommand{\propagator}[1]{P(#1)}

In this appendix, we list the explicit expressions for the Feynman
integrals that contribute to the one- and two-loop diagrams discussed
in \refses{se:oneloop} and \ref{se:twoloop}.  In order to keep our
expressions as compact as possible we define the momenta
\beqar\label{momentadef}
\linea{1}&=&p_i+\lmom_1
,\qquad
\linea{2}=p_i+\lmom_2
,\qquad
\linea{3}=p_i+\lmom_1+\lmom_2
,\nl
\lineb{1}&=&p_j-\lmom_1
,\qquad
\lineb{2}=p_j-\lmom_2
,\qquad
\lineb{3}=p_j-\lmom_1-\lmom_2
,\qquad
\lmom_{3}=-\lmom_1-\lmom_2
,\nl
\linec{1}&=&p_k-\lmom_1
,\qquad
\linec{2}=p_k-\lmom_2
,\qquad
\linec{3}=p_k-\lmom_1+\lmom_2
,\qquad
\lmom_{4}=\lmom_1-\lmom_2
.
\eeqar
For massive and massless propagators we use the notation 
\beq\label{propagator}
\propagatorm{q}{m}=q^2-m^2+\ri 0
,\qquad
\propagator{q}=q^2+\ri 0,
\eeq
and for triple gauge-boson couplings we write
\beq\label{YMvrtex}
\Gamma^{\mu_1\mu_2\mu_3}(
\lmom_1,\lmom_2,\lmom_3)
=
g^{\mu_1\mu_2}(\lmom_1-\lmom_2)^{\mu_3}
+g^{\mu_2\mu_3}(\lmom_2-\lmom_3)^{\mu_1}
+g^{\mu_3\mu_1}(\lmom_3-\lmom_1)^{\mu_2}
.
\eeq
The normalization factors occurring in
\refeq{pertserie1} 
are absorbed into the integration measure
\beq\label{measure}
\measure{i}=
{(4\pi)^2}
\left(\frac{4\pi\muD^2}{ \mathrm{e}^{\gamma_{\mathrm{E}}}Q^2}\right)^{D/2-2}
\muD^{4-D}
\frac{\rd^D \lmom_i}{\left(2\pi\right)^D}
=
\frac{1}{\pi^2} \,
\left(
 \mathrm{e}^{\gamma_{\mathrm{E}}}Q^2 \pi\right)^{2-D/2}
\,\rd^D \lmom_i
, 
\eeq
and for the projection introduced in \refeq{projection1} we use the
shorthand
\newcommand{\traceproject}{\Pi} 
\beqar\label{traceproject}
\traceproject_{ij}\left(\Gamma\right)
&=&
\frac{1}{r_{ij}}\Tr\left( \Gamma \om_{\kappa_i} \ps_i \ps_j \right)
=\frac{1}{2 r_{ij}}\Tr\left( \Gamma \ps_i \ps_j \right)
,
\eeqar
where the second equality holds if $\Gamma$ does not involve $\ga_5$
or $\om_{\rR,\rL}$, as it is the case in the following equations.
With this notation we have
\beqar
\label{defint0}
\DD{0}(\mass_1;r_{ij})&=&
\int \measure{1}
\frac{
4 \ri
\linea{1}\lineb{1}
}{
\propagatorm{\lmom_1}{\mass_1}
\propagator{\linea{1}}
\propagator{\lineb{1}}
}
,\nl
\label{defint1}
\DD{1}(\mass_1,\mass_2;r_{ij})&=&
\int \measure{1}  \measure{2}
\frac{
-16
(\linea{1}\lineb{1})
(\linea{3} \lineb{3})
}{
\propagatorm{\lmom_1}{\mass_1}
\propagatorm{\lmom_2}{\mass_2}
\propagator{\linea{1}}
\propagator{\linea{3}}
\propagator{\lineb{1}}
\propagator{\lineb{3}}
}
,\nl
\label{defint2}
\DD{2}(\mass_1,\mass_2;r_{ij})&=&
\int \measure{1}  \measure{2}
\frac{
-16
(\linea{1}\lineb{3})
(\linea{3} \lineb{2})
}{
\propagatorm{\lmom_1}{\mass_1}
\propagatorm{\lmom_2}{\mass_2}
\propagator{\linea{1}}
\propagator{\linea{3}}
\propagator{\lineb{2}}
\propagator{\lineb{3}}
}
,\nl
\label{defint3}
\DD{3}(\mass_1,\mass_2,\mass_3;r_{ij})&=&
\int \measure{1}  \measure{2}
\frac{
-2
\traceproject_{ij}\left(
\slinea{3}
\gamma^{\mu_2}
\slinea{1}
\gamma^{\mu_1}
\right)
\lineb{3}^{\mu_3}
\Gamma_{\mu_1\mu_2\mu_3}(
\lmom_1,\lmom_2,\lmom_3)
}{
\propagatorm{\lmom_1}{\mass_1}
\propagatorm{\lmom_2}{\mass_2}
\propagatorm{\lmom_3}{\mass_3}
\propagator{\linea{1}}
\propagator{\linea{3}}
\propagator{\lineb{3}}
}
,\nl
\label{defint5}
\DD{4}(\mass_1,\mass_2;r_{ij})&=&
\int \measure{1}  \measure{2}
\frac{
2
\traceproject_{ij}\left(
\slinea{1}
\gamma^{\mu_2}
\slinea{3}
\gamma_{\mu_2}
\slinea{1}
\slineb{1}
\right)
}{
\propagatorm{\lmom_1}{\mass_1}
\propagatorm{\lmom_2}{\mass_2}
\left[\propagator{\linea{1}}\right]^2
\propagator{\linea{3}}
\propagator{\lineb{1}}
}
,\nl
\label{defint7}
\DD{5}(\mass_1,\mass_2;r_{ij})
&=&
\int \measure{1}  \measure{2}
\frac{
2
\traceproject_{ij}\left(
\slinea{1}
\gamma^{\mu_2}
\slinea{3}
\slineb{1}
\slinea{2}
\gamma_{\mu_2}
\right)
}{
\propagatorm{\lmom_1}{\mass_1}
\propagatorm{\lmom_2}{\mass_2}
\propagator{\linea{1}}
\propagator{\linea{3}}
\propagator{\linea{2}}
\propagator{\lineb{1}}
}
,\nl
\label{defint10}
\DD{6}(\mass_1,\mass_2,\mass_3,\mass_4;r_{ij})&=&
\int \measure{1}  \measure{2}
\frac{
-4 
\linea{1}^{\mu_1}\lineb{1 \mu_4}
}{
\propagatorm{\lmom_1}{\mass_1}
\propagatorm{\lmom_2}{\mass_2}
\propagatorm{\lmom_3}{\mass_3}
\propagatorm{\lmom_1}{\mass_4}
\propagator{\linea{1}}
\propagator{\lineb{1}}
}
\nl&&{}\times\left[
\Gamma_{\mu_1\mu_2\mu_3}
(\lmom_1,\lmom_2,\lmom_3)
\Gamma^{\mu_4\mu_2\mu_3}
(\lmom_1,\lmom_2,\lmom_3)
+2\lmom_{2\mu_1}\lmom_{3}^{\mu_4}
\right]
,\nl
\label{defint16}
\DD{7}(\mass_1,\mass_2,\mass_3;r_{ij})&=&
\int \measure{1}  \measure{2}
\frac{
-4 
\linea{1}\lineb{1}
}{
\propagatorm{\lmom_1}{\mass_1}
\propagatorm{\lmom_2}{\mass_2}
\propagatorm{\lmom_1}{\mass_3}
\propagator{\linea{1}}
\propagator{\lineb{1}}
}
,\nl
\label{defint15}
\DD{8}(\mass_1,\mass_2,\mass_3,\mass_4;r_{ij})&=&
\int \measure{1}  \measure{2}
\frac{
- 4 
\linea{1}\lineb{1}
}{
\propagatorm{\lmom_1}{\mass_1}
\propagatorm{\lmom_2}{\mass_2}
\propagatorm{\lmom_3}{\mass_3}
\propagatorm{\lmom_1}{\mass_4}
\propagator{\linea{1}}
\propagator{\lineb{1}}
},\nl
\label{defint12}
\DD{9}(\mass_1,\mass_2,\mass_3,\mass_4;r_{ij})&=&
\int \measure{1}  \measure{2}
\frac{
4
\linea{1}^{\mu_1}\lineb{1}^{\mu_4}
(\lmom_{2}-\lmom_{3})_{\mu_1}
(\lmom_{2}-\lmom_{3})_{\mu_4}
}{
\propagatorm{\lmom_1}{\mass_1}
\propagatorm{\lmom_2}{\mass_2}
\propagatorm{\lmom_3}{\mass_3}
\propagatorm{\lmom_1}{\mass_4}
\propagator{\linea{1}}
\propagator{\lineb{1}}
}
,\nl
\label{defint17}
\DD{10}(\mass_1,\mass_2,\mass_3;r_{ij})
&=&
\DD{7}(\mass_1,\mass_2,\mass_3;r_{ij})
,\nl
\label{defint9}
\DD{11,0}(\mass_1,\mass_2,\mass_3,\mass_4;r_{ij})&=&
\int \measure{1}  \measure{2}
\frac{
4
\linea{1}^{\mu_1}\lineb{1}^{\mu_4}
\, \Tr\left(
\gamma_{\mu_1}
\slmom_2
\gamma_{\mu_4}
\slmom_3\right)
}{
\propagatorm{\lmom_1}{\mass_1}
\propagatorm{\lmom_2}{\mass_2}
\propagatorm{\lmom_3}{\mass_3}
\propagatorm{\lmom_1}{\mass_4}
\propagator{\linea{1}}
\propagator{\lineb{1}}
}
,\nl
\label{defint9m}
\DD{11,m}(\mass_1,\mass_2,\mass_3,\mass_4;r_{ij})
&=&
-4 \DD{8}(\mass_1,\mass_2,\mass_3,\mass_4;r_{ij})
,\nl
\label{defint20}
\DD{12}(\mass_1,\mass_2;r_{ik})&=&
\int \measure{1}  \measure{2}
\frac{
-16
(\linea{1}\lineb{1})
(\linea{3} \linec{2})
}{
\propagatorm{\lmom_1}{\mass_1}
\propagatorm{\lmom_2}{\mass_2}
\propagator{\linea{1}}
\propagator{\linea{3}}
\propagator{\lineb{1}}
\propagator{\linec{2}}
}
,\nl
\label{defint21}
\DD{13}(\mass_1,\mass_2,\mass_3;r_{ij},r_{ik},r_{jk})&=&
\int \measure{1}  \measure{2}
\frac{
8
\linea{1}^{\mu_1}
\lineb{2}^{\mu_2}
\linec{3}^{\mu_3}
\Gamma_{\mu_1\mu_2\mu_3}(
-\lmom_1,\lmom_2,\lmom_4)
}{
\propagatorm{\lmom_1}{\mass_1}
\propagatorm{\lmom_2}{\mass_2}
\propagatorm{\lmom_4}{\mass_3}
\propagator{\linea{1}}
\propagator{\lineb{2}}
\propagator{\linec{3}}
}
,\nl
\DD{14}(m_1,m_2;r_{ij},r_{kl})&=&
\DD{0}(m_1;r_{ij})
\DD{0}(m_2;r_{kl})
.
\eeqar

\section{Relations between loop integrals in NLL approximation}
\label{se:looprelations}

In the following we list relations between one- and
two-loop integrals defined in \refapp{app:loops}.  These relations are
valid after subtraction of the UV singularities and in NLL
approximation. They have been obtained from the explicit results
listed in in \refses{se:oneloop} and \ref{se:twoloop} and are employed
in \refapp{se:twoloopcomb} in order to simplify the sum over all NLL
two-loop contributions.

As in \refse{se:twoloop}, also in the following the symbols $m_i$ are
used to denote generic mass parameters, which can assume the values
$m_i=\MW,\MZ,\Mt,\MH$ or $m_i=0$, and the symbols $M_i$ are used to
denote non-zero masses, \ie $M_i=\MW,\MZ,\Mt,\MH$.

Combinations of the 2-leg ladder integrals in \refeq{diagram1} and
\refeq{diagram2} can be expressed as products of one-loop integrals
using
\beqar\label{eq:2legintfac}
\lefteqn{\DDsub{1}(m_1,m_2;r_{ij}) + \DDsub{2}(m_1,m_2;r_{ij})
+\DDsub{1}(m_2,m_1;r_{ij}) + \DDsub{2}(m_2,m_1;r_{ij})}
\quad&&
\phantom{\DDsub{1}(m_1,m_2;r_{ij}) + \DDsub{2}(m_1,m_2;r_{ij})
+\DDsub{1}(m_2,m_1;r_{ij}) \qquad}
\nl
&\NLLA&\DDsub{0}(m_1;r_{ij})\DDsub{0}(m_2;r_{ij})
\nl&\NLLA&
\DDsub{0}(\MW;r_{ij})\DDsub{0}(\MW;r_{ij})
+\deDDsub{0}(m_1;r_{ij})\DDsub{0}(\MW;r_{ij})
\nl&&{}
+\DDsub{0}(\MW;r_{ij})\deDDsub{0}(m_2;r_{ij})
+\deDDsub{0}(m_1;r_{ij})\deDDsub{0}(m_2;r_{ij}),
\eeqar
where the subtracted functions $\deDD{h}{}$ have been defined in
\refeq{eq:subtractedintegral}. For the relations in
\refeq{eq:2legintfac} it is crucial that the results for
$\DDsub{0}(m_k;r_{ij})$ including terms up to order $\Eps{2}$ are
used.  Moreover, $D_1$ and $D_2$ fulfil the relations
\beqar\label{eq:2legintladder}
\DDsub{1}(M_1,m_2;r_{ij}) &\NLLA& \DDsub{1}(M_1,\MW;r_{ij}),\nl
\DDsub{2}(m_1,m_2;r_{ij}) &=& \DDsub{2}(m_2,m_1;r_{ij}).
\eeqar

The loop integrals corresponding to the non-abelian diagrams involving
two external legs \refeq{diagram3} can be replaced as
\beqar\label{eq:2legYM}
\DDsub{3}(M_1,m_2,m_3;r_{ij}) &\NLLA&
\frac{1}{2}\DDsub{2}(m_3,M_1;r_{ij})
-\DDsub{4}(m_3,M_1;r_{ij})
\nl&&{}
-6\DDsub{9}(m_3,M_1,M_1,m_3;r_{ij}),
\nl
\deDDsub{3}(\MW,\MW,m_1;r_{ij}) &\NLLA&
\frac{1}{2}\deDDsub{12}(\MW,m_1;r_{ij})
-\deDDsub{4}(m_1,\MW;r_{ij})
\nl&&{}
-6\deDDsub{9}(m_1,\MW,\MW,m_1;r_{ij}),
\nl
\deDDsub{3}(m_1,\MW,\MW;r_{ij}) &\NLLA& \deDDsub{3}(\MW,m_1,\MW;r_{ij})
+\deDDsub{1}(\MW,m_1;r_{ij})
\nl&&{}
+\deDDsub{2}(\MW,m_1;r_{ij})+\deDDsub{4}(\MW,m_1;r_{ij})
\nl&&{}
-\frac{1}{2}\deDDsub{12}(\MW,m_1;r_{ij}),
\nl
\deDDsub{3}(\MW,m_1,\MW;r_{ij}) &\NLLA&
-\frac{1}{2}\deDDsub{1}(\MW,m_1;r_{ij})
-\deDDsub{4}(\MW,m_1;r_{ij}).
\eeqar
The first of these relations has been verified and is needed only if 
at most one of the masses $m_2$ and $m_3$ is zero.

The loop functions for the diagrams \refeq{diagram5} and
\refeq{diagram7} are equal up to a minus sign
\beqar\label{eq:2legsl}
\DDsub{5}(m_1,m_2;r_{ij})\NLLA
-\DDsub{4}(m_1,m_2;r_{ij}).
\eeqar

Next we give relations for the loop integrals appearing in the
diagrams with self-energy insertions in the gauge-boson line,
\refeq{diagram10}, \refeq{diagram16}, \refeq{diagram15},
\refeq{diagram12}, \refeq{diagram17}, and \refeq{diagram9}.
The loop integrals $D_6$, $D_7$, $D_8$, and $D_{11}$ can be related to 
$D_9$ as
\beqar\label{eq:2legse}
\DD{6}(M_1,M_2,M_3,M_4;r_{ij}) &\NLLA&
10\DD{9}(M_1,M_2,M_3,M_4;r_{ij}),\nl
\DD{11,0}(M_1,M_2,M_3,M_4;r_{ij}) &\NLLA& 
4\DD{9}(M_1,M_2,M_3,M_4;r_{ij}),\nl
\deDD{6}(0,\MW,\MW,0;r_{ij}) &\NLLA&
10\deDD{9}(0,\MW,\MW,0;r_{ij}),\nl
\deDD{11,0}(0,\MW,\MW,0;r_{ij}) &\NLLA& 
4\deDD{9}(0,\MW,\MW,0;r_{ij}),\nl
\deDD{6}(0,M_2,M_3,M_4;r_{ij}) &\NLLA&
\deDD{6}(M_4,M_2,M_3,0;r_{ij})
\nl
&\NLLA&-12\,\frac{M_2^2+M_3^2}{M_4^2}\,\deDD{9}(0,\MW,\MW,0;r_{ij}),\nl
\deDD{7}(0,M_2,M_3;r_{ij}) &\NLLA&
\deDD{7}(M_3,M_2,0;r_{ij})
\nl
&\NLLA&-3\frac{M_2^2}{M_3^2}\deDD{9}(0,\MW,\MW,0;r_{ij}),\nl
\deDD{8}(0,M_2,M_3,M_4;r_{ij}) &\NLLA&
\deDD{8}(M_4,M_2,M_3,0;r_{ij})
\nl
&\NLLA&-\frac{3}{M_4^2}\deDD{9}(0,\MW,\MW,0;r_{ij}),\nl
\deDD{9}(0,M_2,M_3,M_4;r_{ij}) &\NLLA&
\deDD{9}(M_4,M_2,M_3,0;r_{ij})
\nl
&\NLLA&3\,\frac{M_2^2+M_3^2}{M_4^2}\,\deDD{9}(0,\MW,\MW,0;r_{ij}),
\nl
\deDD{11,0}(0,M_2,M_3,M_4;r_{ij}) &\NLLA&
\deDD{11,0}(M_4,M_2,M_3,0;r_{ij})
\nl
&\NLLA&6\,\frac{M_2^2+M_3^2}{M_4^2}\,\deDD{9}(0,\MW,\MW,0;r_{ij}).
\eeqar
The loop integral  $\DDsub{9}$ can be expressed as
\beqar\label{eq:2gse}
3\DDsub{9}(\MW,\MW,\MW,\MW;r_{ij})&\NLLA& -J(\veps,\MW,Q^2),\nl
3\deDDsub{9}(0,\MW,\MW,0;r_{ij})&\NLLA& -\left[
\Delta J(\veps,0,Q^2)-\Delta J(\veps,0,\MW^2)
\right],
\eeqar
and the loop integral  $\DDsub{11}$ as
\beqar\label{eq:2gsef}
3[\deDDsub{11,0}(0,0,0,0;r_{ij})
-\deDDsub{11,0}(0,\MW,\MW,0;r_{ij})] \NLLA -4 \Delta J(\veps,0,\MW^2)
\eeqar
in terms of the functions $J$ and $\De J$ defined in \refeq{Jterms}.

Combinations of 3-leg ladder integrals in \refeq{diagram20} can
be expressed as products of one-loop integrals
\beqar\label{eq:3legintfac}
\lefteqn{\DDsub{12}(m_1,m_2;r_{ij}) + \DDsub{12}(m_2,m_1;r_{ik})
\NLLA\DDsub{0}(m_2;r_{ij})\DDsub{0}(m_1;r_{ik})} \quad
\nl&\NLLA&
\DDsub{0}(\MW;r_{ij})\DDsub{0}(\MW;r_{ik})
+\deDDsub{0}(m_2;r_{ij})\DDsub{0}(\MW;r_{ik})
\nl&&{}
+\DDsub{0}(\MW;r_{ij})\deDDsub{0}(m_1;r_{ik})
+\deDDsub{0}(m_2;r_{ij})\deDDsub{0}(m_1;r_{ik}).
\eeqar
For these relations again the terms of $\order(\Eps{2})$ in 
$\DDsub{0}(m_k;r_{ij})$ are crucial.  Furthermore
\beq\label{eq:antisymmperm}
\sum_{\pi(i,j,k)}\sgn(\pi(i,j,k))\,
\DD{12}(\MW,\MW;r_{ik}) 
\NLLA 0,
\eeq
where the sum runs over all permutations $\pi(i,j,k)$ of $i,j,k$, with
sign $\sgn(\pi(i,j,k))$.
This identity simply follows from the fact that $\DD{12}$ is symmetric
under the interchange of $i$ and $k$ and does not depend on $r_{ij}$
and $r_{jk}$.

The loop functions appearing in the non-abelian diagrams involving
three external legs \refeq{diagram21} can be replaced as
\beqar\label{eq:3legYM}
\DD{13}(M_1,M_2,M_3;r_{ij}) &\NLLA& 0,
\nl
2\deDD{13}(M_1,M_2,m_3;r_{ij},r_{ik},r_{jk}) &\NLLA&
2\deDD{13}(m_1,M_2,M_3;r_{ik},r_{jk},r_{ij}) 
\nl &\NLLA&
2\deDD{13}(M_1,m_2,M_3;r_{jk},r_{ij},r_{ik}) 
\nl&\NLLA&
\deDDsub{12}(\MW,m_3;r_{jk}) -\deDDsub{12}(\MW,m_3;r_{ik}).
\nln
\eeqar

\section[beta-function coefficients]{$\be$-function coefficients}
\label{se:betafunctions}

In this appendix we give relations and explicit expressions for the
one-loop $\be$-function coefficients $\betacoeff{V_1V_2}^{(1)}$,
$\betacoeff{1}^{(1)}$, $\betacoeff{2}^{(1)}$, $\betacoeff{e}^{(1)}$,
$\betacoeff{\QED}^{(1)}$, and  $\betacoeff{\mathrm{top}}^{(1)}$ 
that have been used in the calculation.
For more details we refer to \citeres{Pozzorini:rs,Pozzorini:2004rm}.

The matrix of  $\be$-function coefficients is defined as
\beqar\label{eq:betacoeff}
\betacoeff{V_1V_2}^{(1)} = 
\frac{11}{3} 
\Tr_V(I^{\bar V_1}I^{V_2})
-\frac{1}{6}
\Tr_\Phi(I^{\bar V_1}I^{V_2})  
-\frac{2}{3}\sum_{\Psi}
\sum_{\kappa=\rR,\rL}\Tr_{\Psi^\kappa}(I^{\bar V_1}I^{V_2}),
\eeqar
where $I^{V_i}$ are the generators defined in \refeq{genmixing} and
$\Tr_V$, $\Tr_\Phi$, and $\Tr_\Psi$ denote the traces in the representations
for the gauge bosons, scalars, and fermionic doublets, respectively.
The sum $\sum_{\Psi}$ runs over all doublets of leptons and quarks
including different colours.
The traces read more explicitly%
\footnote{Note that our normalization for the trace of the scalar
  fields differs from the one used in
  \citeres{Pozzorini:rs,Pozzorini:2004rm}.  }
\beqar\label{eq:traceV}
e^2\Tr_V(I^{\bar V_1}I^{V_2}) &=& 
\gw^2\sum_{V_3,V_4=A,Z,W^\pm} \teps^{\bar{V}_1 \bar{V}_3 \bar{V}_4}\teps^{V_{2} {V}_{3}{V}_4},
\\
e^2\Tr_\Phi(I^{\bar V_1}I^{V_2})  &=& 
e^2\sum_{\Phi_{i},\Phi_{j}=H,\chi,\phi^\pm}
I^{\bar{V}_1}_{\Phi_{i}\Phi_{j}}
I^{{V}_2}_{\Phi_{j}\Phi_{i}},
\label{eq:tracePhi}
\\
e^2\Tr_{\Psi^\kappa}(I^{\bar V_1}I^{V_2})&=& 
e^2\sum_{\Psi_{i},\Psi_{j}=u,d}
\,
I^{\bar{V}_1}_{\Psi_{i}^\kappa \Psi_{j}^\kappa}
I^{{V}_2}_{\Psi_{j}^\kappa \Psi_{i}^\kappa}.
\label{eq:tracePsi}
\eeqar
Multiplying $\betacoeff{V_1V_2}^{(1)}$ with generators and summing
over all gauge bosons yields
\beqar\label{eq:betarel}
e^2\sum_{V_1,V_2=A,Z,W^\pm}
\betacoeff{V_1V_2}^{(1)} I_i^{{V}_1} I_i^{\bar{V}_2}
= \gb^2 \betacoeff{1}^{(1)}\left(\frac{Y_i}{2}\right)^2+\gw^2 \betacoeff{2}^{(1)}C_{i},
\eeqar
from which the coefficients corresponding to the weak couplings $\gb$
and $\gw$ can be read off. The coefficient corresponding to the
electric-charge renormalization is given by
\beq\label{eq:betacoeffe}
\betacoeff{e}^{(1)} \equiv \betacoeff{AA}^{(1)} =
\cw^2\betacoeff{1}^{(1)}+ \sw^2\betacoeff{2}^{(1)}.
\eeq

The explicit values in the electroweak Standard Model are ($Y_\Phi=1$)
\beqar\label{betacoeffres}
\betacoeff{1}^{(1)}&=&-\frac{41}{6\cw^2}
,\qquad
\betacoeff{2}^{(1)}=\frac{19}{6\sw^2}
,\qquad
\betacoeff{e}^{(1)}=-\frac{11}{3}
.
\eeqar

The QED $\be$-function coefficient is determined by the light-fermion
contributions only, \ie
\beqar\label{eq:betacoeffQED}
\betacoeff{\QED}^{(1)}&=&-\frac{4}{3}\sum_{f\neq\Pt}N_{\mathrm{c}}^f Q_f^2=-\frac{80}{9},
\eeqar
where $N_{\mathrm{c}}^f$ represents the colour factor, \ie
$N_{\mathrm{c}}^f=1$ for leptons and $N_{\mathrm{c}}^f=3$ for quarks.

The top-quark contribution to the electromagnetic $\be$-function
coefficient reads
\beqar\label{eq:betacoeffQEDtop}
\betacoeff{\mathrm{top}}^{(1)}&=&-\frac{16}{9}.
\eeqar

\section{Summing up the one-loop contributions}
\label{se:oneloopcomb}

In NLL approximation, the contribution of all bare one-loop diagrams
to the matrix element for a process with $n$ external massless
fermions is given by \refeq{oneloopfactorizable}, which results from the
factorizable diagrams \refeq{oneloopdiag4}.  Using the explicit
results presented in \refse{se:oneloop} we can write
\beqar\label{onelooprepunr}
\nmel{1}{\rF}&\NLLA&
\mel{0}{}
\left[
\FF{1}{\rF,\sew}
+\Delta \FF{1}{\rF,\elm}
+\Delta \FF{1}{\rF,\PZ}
\right]
\eeqar
with
\beqar\label{onelooprepu}
\FF{1}{\rF,\sew}
&=&
-\frac{1}{2}\sum_{i=1}^{n}\sum_{j=1\atop j\neq i}^{n}
\sum_{V=A,Z,W^\pm} 
I_i^{\bar{V}}I_j^{{V}}\,\DDsub{0}(\MW;r_{ij})
,\nl
\Delta \FF{1}{\rF,\elm}
&=&
-\frac{1}{2}\sum_{i=1}^{n}\sum_{j=1\atop j\neq i}^{n}
I_i^{A}I_j^{{A}}\,\deDDsub{0}(0;r_{ij})
,\nl
\Delta \FF{1}{\rF,\PZ}
&=&
-\frac{1}{2}\sum_{i=1}^{n}\sum_{j=1\atop j\neq i}^{n}
I_i^{Z}I_j^{{Z}}\,\deDDsub{0}(\MZ;r_{ij})
.
\eeqar
Using the charge-conservation identity \refeq{chargeconservation},
the counterterms \refeq{WFren} can be cast into the form
\beqar\label{oneloopct}
\nmel{1}{\mathrm{WF}} 
 &\NLLA&
\frac{1}{2} 
\mel{0}{} 
\sum_{i=1}^{n}\sum_{j=1\atop j\neq i}^{n}
\Biggl\{
\sum_{V=A,Z,W^\pm} 
I_i^{\bar{V}}
I_j^{{V}}\,
C(\MW;Q^2)
+
I_i^{A}
I_j^{{A}}
\De C(0;Q^2)
\Biggr\}
\eeqar
with
\beqar
C(\MW;Q^2)&\NLLA& L + \frac{1}{2}L^2\Eps{} +\frac{1}{6}L^3\Eps{2} +
\order(\Eps{3}), \nl
\De C(0;Q^2)&\NLLA& -\frac{1}{\veps}
-L - \frac{1}{2}L^2\Eps{} -\frac{1}{6}L^3\Eps{2} + \order(\Eps{3}).
\eeqar

When adding \refeq{onelooprepunr} and \refeq{oneloopct} we find 
\refeq{onelooprep}--\refeq{Ifunc} with
\beqar
\univfact{\veps}{\MW;-r_{ij}} &\NLLA&\DD{0}(\MW;r_{ij})-C(\MW;Q^2),\nl
\De\univfact{\veps}{\MZ;-r_{ij}} &\NLLA&\deDD{0}(\MZ;r_{ij}),\nl
\De\univfact{\veps}{0;-r_{ij}} &\NLLA&\deDD{0}(0;r_{ij})-\De C(0;Q^2).
\eeqar

\section{Summing up the two-loop contributions}
\label{se:twoloopcomb}

In this appendix, the two-loop results listed in \refse{se:twoloop}
are summed and decomposed into reducible contributions, which involve
products of the one-loop integrals $\DD{0}$ \refeq{idiag0subt}, plus
remaining irreducible parts.  To this end, we split the integrals
according to \refeq{eq:subtractedintegral} and use the relations given
in \refapp{se:looprelations} as well as the commutation relations
\refeq{commrela} and, in particular, the fact that $I^A_i$, $I^Z_j$,
and $\sum_{V=A,Z,W^\pm}I_k^{\bar V}I_k^V$ commute with each other.  As
we show, all irreducible contributions cancel apart from those that
can be expressed in terms of the one-loop functions $J$ \refeq{Jterms}
and $\be$-function coefficients.  To start with, we consider four
separate subsets and combine these in a later stage.

\subsection*{Terms related to two external lines not involving
  gauge-boson self-energies}

\newcommand{\nose}{\mbox{\scriptsize no-se}}

We begin by considering the contributions that result from the
diagrams where the soft--collinear gauge bosons couple to two on-shell
external lines and that do not involve self-energy contributions to
the soft--collinear gauge bosons, \ie from the diagrams 1, 2, 3, 4,
and 5 of \refse{se:twoloop},
\beqar\label{Mtwolegsnose}
\nmel{2,\nose}{ij}&=&
\nmel{2}{1,ij}+\nmel{2}{2,ij}
+\left[\nmel{2}{3,ij}+\nmel{2}{4,ij}+\nmel{2}{5,ij}
+ (i\leftrightarrow j)\right]. 
\eeqar
These can be summarized as 
\beqar\label{eq:sigma2nose}
\nmel{2,\nose}{ij}
&=&
\mel{0}{}
\sum_{V_1,V_2=A,Z,W^\pm} 
\Biggl\{ 
I_i^{\bar{V}_2}
I_i^{\bar{V}_1} 
I_j^{{V}_2} 
I_j^{{V}_1}
\DD{1}(M_{V_1},M_{V_2};r_{ij})
\nl&&\qquad{}+
I_i^{\bar{V}_2}
I_i^{\bar{V}_1}
I_j^{{V}_1}
I_j^{{V}_2}
\DD{2}(M_{V_1},M_{V_2};r_{ij})
\nl&&\qquad{}-\biggl[
\ri\frac{\gw}{e}\sum_{V_3=A,Z,W^\pm}
\teps^{V_1 V_2 V_3}
I_i^{\bar{V}_2}
I_i^{\bar{V}_1}
I_j^{\bar{V}_3}
\DD{3}(M_{V_1},M_{V_2},M_{V_3};r_{ij})
\nl&&\qquad{}+
I_i^{{V}_2}
I_i^{\bar{V}_2}
I_i^{{V}_1}
I_j^{\bar{V}_1}
\DD{4}(M_{V_1},M_{V_2};r_{ij})
\nl&&\qquad{}+
I_i^{{V}_2}
I_i^{{V}_1}
I_i^{\bar{V}_2}
I_j^{\bar{V}_1}
\DD{5}(M_{V_1},M_{V_2};r_{ij})
+(i\leftrightarrow j)\biggr]\Biggr\}
\nl
&\NLLA&
\mel{0}{}
\Biggl\{ 
\frac{1}{2} \sum_{V_1,V_2=A,Z,W^\pm} 
I_i^{\bar{V}_2}
I_i^{\bar{V}_1} 
I_j^{{V}_2} 
I_j^{{V}_1}
\DD{0}(\MW;r_{ij})\DD{0}(\MW;r_{ij})
\nl &&{}
+ \sum_{V=A,Z,W^\pm} 
I_i^{\bar{V}}
I_i^{\FA} 
I_j^{{V}} 
I_j^{\FA}
\DD{0}(\MW;r_{ij})\deDD{0}(0;r_{ij})
\nl &&{}
+ \sum_{V=A,Z,W^\pm} 
I_i^{\bar{V}}
I_i^{\FZ} 
I_j^{{V}} 
I_j^{\FZ}
\DD{0}(\MW;r_{ij})\deDD{0}(\MZ;r_{ij})
\nl &&{}
+ \frac{1}{2} 
I_i^{\FA}
I_i^{\FA} 
I_j^{\FA} 
I_j^{\FA}
\deDD{0}(0;r_{ij})\deDD{0}(0;r_{ij})
\nl &&{}
+
I_i^{\FZ}
I_i^{\FA} 
I_j^{\FZ} 
I_j^{\FA}
\deDD{0}(\MZ;r_{ij})\deDD{0}(0;r_{ij})
\nl &&{} +\biggl[
\sum_{V_1,V_2=A,Z,W^\pm} \ri\frac{\gw}{e}\frac{1}{2} 
\teps^{\FA V_1 V_2}
\left(I_i^{\bar{V}_1} I_i^{\FA} I_j^{\bar{V}_2} +
 I_i^{\FA} I_j^{\bar{V}_1} I_j^{\bar{V}_2} \right) 
\deDD{12}(\MW,0;r_{ij})
\nl&&\qquad{}
\nl &&{}\ +
 \sum_{V_1,V_2=A,Z,W^\pm} \ri\frac{\gw}{e}\frac{1}{2}
\teps^{\FZ V_1 V_2}
\left(I_i^{\bar{V}_1} I_i^{\FZ} I_j^{\bar{V}_2} +
 I_i^{\FZ} I_j^{\bar{V}_1} I_j^{\bar{V}_2} \right) 
\deDD{12}(\MW,\MZ;r_{ij})
\nl&&\qquad{}
+ (i\leftrightarrow j)\biggr]
\nl &&{} 
+6 \sum_{V_1,V_2=A,Z,W^\pm} 
I_i^{\bar{V}_1} I_j^{\bar{V}_2} \Tr_V(I^{V_1}I^{V_2}) 
\DD{9}(\MW,\MW,\MW,\MW;r_{ij})
\nl &&{}
+ 3 \sum_{V_1=A,Z,W^\pm} 
\left(  I_i^{\FA} I_j^{\bar{V}_1} + I_i^{\bar{V}_1} I_j^{\FA}\right)
 \Tr_V(I^{\FA}I^{V_1}) 
\deDD{9}(0,\MW,\MW,0;r_{ij})\Biggr\},
\qquad
\eeqar
where we have made use of the identities \refeq{eq:2legintfac},
\refeq{eq:2legintladder}, \refeq{eq:2legYM}, \refeq{eq:2legsl}, and
\refeq{eq:traceV}.

\subsection*{Terms involving gauge-boson self-energy contributions}

\newcommand{\se}{\mbox{\scriptsize se}}

The contributions where a soft--collinear gauge boson connects two
external lines and involves self-energy corrections result from
diagrams 6--11 in \refse{se:twoloop} and read
\beqar\label{Mtwolegsse}
\nmel{2,\se}{ij}&=&
\sum_{m=6}^{11} \nmel{2}{m,ij}.
\eeqar
They can be summarized as
\beqar\label{eq:sigma2se}
\nmel{2,\se}{ij}
&=&
\mel{0}{}
\Biggl\{ 
\frac{1}{2} \frac{\gw^2}{e^2}
\sum_{V_1,V_2,V_3,V_4= A,Z,W^\pm} 
I_i^{\bar{V}_1}
I_j^{\bar{V}_4}\,
\teps^{V_1 \bar{V}_2 \bar{V}_3}
\teps^{V_4 V_2 V_3}\,
\DD{6}(M_{V_1},M_{V_2},M_{V_3},M_{V_4};r_{ij})
\nl&&{}
- \frac{\gw^2}{e^2}
\sum_{V_1,V_2,V_3,V_4= A,Z,W^\pm} 
I_i^{\bar{V}_1}
I_j^{\bar{V}_4}
\teps^{V_1 \bar{V}_2 \bar{V}_3}
\teps^{V_4 V_2 V_3}\,
(D-1) 
\DD{7}(M_{V_1},M_{V_2},M_{V_4};r_{ij})
\nl&&{}
- e^2 \vev^2 
\sum_{V_1,V_3,V_4 = A,Z,W^\pm} 
I_i^{\bar{V}_1}
I_j^{\bar{V}_4}
\sum_{\Phi_{i_2}=
H,\chi,\phi^\pm}
\left\{
I^{{V}_1},
I^{\bar{V}_3}
\right\}_{H\Phi_{i_2}}
\left\{
I^{{V}_3},
I^{{V}_4}
\right\}_{\Phi_{i_2}H}
\nl&&
\quad{}\times\DD{8}(M_{V_1},M_{\Phi_2},M_{V_3},M_{V_4};r_{ij})
\nl&&{}
-\frac{1}{2}
\sum_{V_1,V_4= A,Z,W^\pm} 
I_i^{\bar{V}_1}
I_j^{\bar{V}_4}
\sum_{\Phi_{i_2},\Phi_{i_3}=H,\chi,\phi^\pm}
I^{{V}_1}_{\Phi_{i_3}\Phi_{i_2}}
I^{{V}_4}_{\Phi_{i_2}\Phi_{i_3}}
\nl&&{}
\quad{}\times
\DD{9}(M_{V_1},M_{\Phi_{i_2}},M_{\Phi_{i_3}},M_{V_4};r_{ij})
\nl&&{}
- \frac{1}{2} 
\sum_{V_1,V_4= A,Z,W^\pm} 
I_i^{\bar{V}_1}
I_j^{\bar{V}_4}
\sum_{\Phi_{i_2}
=H,\chi,\phi^\pm
}
\left\{
I^{{V}_1},I^{{V}_4}
\right\}_{\Phi_{i_2}\Phi_{i_2}}
\nl&&{}
\quad{}\times
\DD{10}(M_{V_1},M_{\Phi_{i_2}},M_{V_4};r_{ij})
\nl&&{}
-\frac{1}{2}
\sum_{V_1,V_4= A,Z,W^\pm} 
I_i^{\bar{V}_1}
I_j^{\bar{V}_4}
{}
\nl&&{}
{}\times
\sum_\Psi 
\Biggl\{
\sum_{\Psi_{i_2},\Psi_{i_3}=u,d}
\;
\sum_{\kappa=\rR,\rL}
I^{{V}_1}_{\Psi_{i_3}^\kappa \Psi_{i_2}^\kappa}
I^{{V}_4}_{\Psi_{i_2}^\kappa \Psi_{i_3}^\kappa}
\,
\DD{11,0}(M_{V_1},m_{i_2},m_{i_3},M_{V_4};r_{ij})
\nl&&\qquad{}
{}-
\left(
  I^{{V}_1}_{u^\rR u^\rR} I^{{V}_4}_{u^\rL u^\rL}
+ I^{{V}_1}_{u^\rL u^\rL} I^{{V}_4}_{u^\rR u^\rR}
\right)
m_u^2 \DD{11,m}(M_{V_1},m_u,m_u,M_{V_4};r_{ij})
\Biggr\}\Biggr\}
\nl&\NLLA& -\mel{0}{}\biggl\{
\sum_{V_1,V_2 = A,Z,W^\pm} 
I_i^{{V}_1}
I_j^{\bar{V}_2}
\betacoeff{V_1 V_2}^{(1)} J(\veps,\MW,Q^2)
\nl&&{}+
I_i^{\FA}
I_j^{\FA}
\betacoeff{AA}^{(1)}
\left[
\Delta J(\veps,0,Q^2)
-\Delta J(\veps,0,\MW^2)
\right]
\nl&&{}+
I_i^{\FA}
I_j^{\FA}
\betacoeff{\QED}^{(1)} \De J(\veps,0,\MW^2)
\nl&&{}+
\sum_{V_1,V_2 = A,Z,W^\pm} 
6
I_i^{\bar{V}_1}
I_j^{\bar{V}_2}
\Tr_V(I^{V_1}I^{V_2}) \DD{9}(\MW,\MW,\MW,\MW;r_{ij})
\nl&&{}+
6
I_i^{\FA}
I_j^{\FA}
\Tr_V(I^{\FA}I^{\FA}) \De\DD{9}(0,\MW,\MW,0;r_{ij})
\nl&&{}+
3(I_i^{\FA}I_j^{\FZ}+I_i^{\FZ}I_j^{\FA})
\Tr_V(I^{\FA}I^{\FZ}) \De\DD{9}(0,\MW,\MW,0;r_{ij})
\Biggr\},
\eeqar
where we used the relations \refeq{eq:2legse}, \refeq{eq:2gse},
\refeq{eq:2gsef}, \refeq{eq:betacoeff}, \refeq{eq:traceV},
\refeq{eq:tracePhi}, \refeq{eq:tracePsi}, and \refeq{eq:betacoeffQED}.
For the simplification of the diagram \refeq{diagram15} in addition
\refeq{massmatrix} was employed. We note that in this calculation
terms of the form $\sum_{V_2,V_3= A,Z,W^\pm} I_i^{\FA}I_j^{\FZ}\,
\teps^{\FA \bar{V}_2 \bar{V}_3}\teps^{\FZ V_2 V_3} M_{V_2}^2$ cancel,
where $M_{V_2}^2$ results either from the loop integrals or from
\refeq{massmatrix}.

\subsection*{Terms related to three external lines}
The terms where the soft--collinear gauge bosons couple to three of
the $n$ on-shell external lines result from diagrams 12 and 13 in
\refse{se:twoloop} and can be written as
\beqar\label{diagthreelegs0}
\nmel{2}{ijk}
&=&
\left(\sum_{\pi(i,j,k)}\nmel{2}{12,ijk}\right)
+
\nmel{2}{13,ijk},
\eeqar
where the sum runs over all six permutations $\pi(i,j,k)$ of external lines
$i,j,k$. 
These contributions yield
\beqar\label{eq:sigma3}
\nmel{2}{ijk}
&=&
\mel{0}{}
\Biggl\{ 
\sum_{\pi(i,j,k)}\,
\sum_{V_1,V_2=A,Z,W^\pm} 
I_i^{\bar{V}_2}
I_i^{\bar{V}_1}
I_j^{{V}_1}
I_k^{{V}_2}
\DD{12}(M_{V_1},M_{V_2};r_{ik})
\nl&&{}
-\ri \frac{\gw}{e}
\sum_{V_1,V_2,V_3
=A,Z,W^\pm} 
\teps^{V_1 V_2 V_3}
I_i^{\bar{V}_1}
I_j^{\bar{V}_2}
I_k^{\bar{V}_3}
\DD{13}(M_{V_1},M_{V_2},M_{V_3};r_{ij},r_{ik},r_{jk})
\Biggr\}
\nl&&{}
\nl
&\NLLA&
\mel{0}{}\sum_{\pi(i,j,k)}\,
\Biggl\{ 
\frac{1}{2} \sum_{V_1,V_2=A,Z,W^\pm} 
I_i^{\bar{V}_2}
I_i^{\bar{V}_1} 
I_j^{{V}_2} 
I_k^{{V}_1}
\DD{0}(\MW;r_{ij})\DD{0}(\MW;r_{ik})
\nl &&{}
+ \sum_{V=A,Z,W^\pm} 
I_i^{\bar{V}}
I_i^{\FA} 
I_j^{{V}} 
I_k^{\FA}
\DD{0}(\MW;r_{ij})\deDD{0}(0;r_{ik})
\nl &&{}
+ \sum_{V=A,Z,W^\pm} 
I_i^{\bar{V}}
I_i^{\FZ} 
I_j^{{V}} 
I_k^{\FZ}
\DD{0}(\MW;r_{ij})\deDD{0}(\MZ;r_{ik})
\nl &&{}
+ \frac{1}{2} 
I_i^{\FA}
I_i^{\FA} 
I_j^{\FA} 
I_k^{\FA}
\deDD{0}(0;r_{ij})\deDD{0}(0;r_{ik})
+
I_i^{\FZ}
I_i^{\FA} 
I_j^{\FZ} 
I_k^{\FA}
\deDD{0}(\MZ;r_{ij})\deDD{0}(0;r_{ik})
\nl &&{} +
\sum_{V_1,V_2=A,Z,W^\pm} \ri\frac{\gw}{e}\frac{1}{2} 
\teps^{\FA V_1 V_2}
 I_i^{\FA} I_j^{\bar{V}_2} I_k^{\bar{V}_1}  
\deDD{12}(\MW,0;r_{ij})
\nl &&{} +
\sum_{V_1,V_2=A,Z,W^\pm} \ri\frac{\gw}{e}\frac{1}{2} 
\teps^{\FZ V_1 V_2}
 I_i^{\FZ} I_j^{\bar{V}_2} I_k^{\bar{V}_1}  
\deDD{12}(\MW,\MZ;r_{ij})
\Biggl\},
\end{eqnarray}
where we used 
\refeq{eq:3legintfac}, \refeq{eq:antisymmperm}, and \refeq{eq:3legYM}.

\subsection*{Terms from four external lines}
Finally, we have the contributions where the soft--collinear gauge
bosons couple to four of the $n$ on-shell external lines, \ie diagram
$14$ in \refse{se:twoloop}:
\beqar\label{diagfourlegs}
\nmel{2}{ijkl}{}
&=&
\nmel{2}{14,ijkl}.
\eeqar
These reduce according to \refeq{idiag22sub} directly to products of
one-loop integrals
\beqar\label{eq:sigma4}
\nmel{2}{ijkl}
&=&
\mel{0}{}
\sum_{V_1,V_2
=A,Z,W^\pm} 
I_i^{\bar{V}_1}
I_j^{{V}_1}
I_k^{\bar{V}_2}
I_l^{{V}_2}
\DD{14}(M_{V_1},M_{V_2};r_{ij},r_{kl})
\nl&\NLLA&
\mel{0}{}
\sum_{V_1,V_2=A,Z,W^\pm} 
I_i^{\bar{V}_1}
I_j^{{V}_1}
I_k^{\bar{V}_2}
I_l^{{V}_2}
\DD{0}(M_{V_1};r_{ij})
\DD{0}(M_{V_2};r_{kl}).
\end{eqnarray}

\subsection*{Complete two-loop correction}
\newcommand{\sumtwo}[3]{{\mathop{\smash{\phantom{'}{\sum}'}%
\vphantom{\sum}}^#1_{#2,#3}}}
\newcommand{\sumthree}[4]{{\mathop{\smash{\phantom{'}{\sum}'}%
\vphantom{\sum}}\limits^#1_{#2,#3,#4}}}
\newcommand{\sumfour}[5]{{\mathop{\smash{\phantom{'}{\sum}'}%
      \vphantom{\sum}}\limits^#1_{#2,#3,#4,#5}}} 

The contributions from the above subsets of diagrams can be combined
for an arbitrary process involving $n$ on-shell external massless
fermions according to \refeq{twoloopdiag3} or \refeq{twoloopcomb2} as
\beq\label{twoloopcombunr}
\nmel{2}{\mathrm{F}}
= \sum_{i=1}^{n}\sum_{j=1\atop j\neq i}^{n}
\left[
\frac{1}{2}
\left(
\nmel{2,\nose}{ij} + \nmel{2,\se}{ij} 
\right)
+\sum_{k=1\atop k\neq i,j}^{n}
\left(
\frac{1}{6} \nmel{2}{ijk}
+\sum_{l=1\atop l\neq i,j,k}^{n}
\frac{1}{8}\nmel{2}{ijkl}
\right)\right].
\eeq
As a first step, we show that all terms that cannot be expressed by
the one-loop functions $\DD{0}$ and $J$ cancel.  The terms involving
explicit factors $\Tr_V(I^{V_1}I^{V_2})$ or $\Tr_V(I^{V_1}I^{\FA})$,
\ie the last two lines in \refeq{eq:sigma2nose} and the last three
lines in \refeq{eq:sigma2se} cancel directly if we use the fact that
the latter trace is only non-vanishing for $V_1=\FA,Z$.  The irreducible
terms involving explicit $\teps^{V_1V_2V_3}$ tensors appearing in
\refeq{eq:sigma2nose} and \refeq{eq:sigma3} yield
\beqar\label{eq:s2}
\lefteqn{
\mel{0}{} \sum_{i=1}^{n}\sum_{j=1\atop j\neq i}^{n}
\biggl\{
\sum_{V_1,V_2=A,Z,W^\pm} \ri\frac{\gw}{e}\frac{1}{2} 
\teps^{\FA V_1 V_2}
(I_i^{\bar{V}_1} I_i^{\FA} I_j^{\bar{V}_2} +
 I_i^{\FA} I_j^{\bar{V}_1} I_j^{\bar{V}_2} ) 
\deDD{12}(\MW,0;r_{ij})
}\quad
\nl &&{} +
 \sum_{V_1,V_2=A,Z,W^\pm} \ri\frac{\gw}{e}\frac{1}{2}
\teps^{\FZ V_1 V_2}
(I_i^{\bar{V}_1} I_i^{\FZ} I_j^{\bar{V}_2} +
 I_i^{\FZ} I_j^{\bar{V}_1} I_j^{\bar{V}_2} ) 
\deDD{12}(\MW,\MZ;r_{ij})
\nl&&
+\sum_{k=1\atop k\neq i,j}^{n}\biggl[
\sum_{V_1,V_2=A,Z,W^\pm} \ri\frac{\gw}{e}\frac{1}{2} 
\teps^{\FA V_1 V_2}
 I_i^{\FA} I_j^{\bar{V}_2} I_k^{\bar{V}_1}  
\deDD{12}(\MW,0;r_{ij})
\nl &&{} +
\sum_{V_1,V_2=A,Z,W^\pm} \ri\frac{\gw}{e}\frac{1}{2} 
\teps^{\FZ V_1 V_2}
 I_i^{\FZ} I_j^{\bar{V}_2} I_k^{\bar{V}_1}  
\deDD{12}(\MW,\MZ;r_{ij})
\biggr]\biggr\}.
\eeqar
These terms vanish upon using global gauge invariance
\refeq{chargeconservation}.

The complete two-loop correction is thus given by the contributions to
\refeq{eq:sigma2nose}, \refeq{eq:sigma3}, and \refeq{eq:sigma4}
involving products of $\DD{0}$-functions and the terms involving
$\be$-function coefficients and $J$-functions in \refeq{eq:sigma2se}.
These can be summarized straightforwardly as
\beqar\label{twoloopresultunr0}
\nmel{2}{\mathrm{F}} &\NLLA&
\frac{1}{4}\mel{0}{}
\sum_{i=1}^{n}\sum_{j=1\atop j\neq i}^{n}
\sum_{k=1}^{n}\sum_{l=1\atop l\neq k}^{n}
\Biggl\{ 
\frac{1}{2} \sum_{V_1,V_2=A,Z,W^\pm} 
I_i^{\bar{V}_2}
I_j^{{V}_2} 
I_k^{\bar{V}_1} 
I_l^{{V}_1}
\DD{0}(\MW;r_{ij})\DD{0}(\MW;r_{kl})
\nl &&{}
+ \sum_{V=A,Z,W^\pm} 
I_i^{\bar{V}}
I_j^{{V}} 
I_k^{\FA}
I_l^{\FA} 
\DD{0}(\MW;r_{ij})\deDD{0}(0;r_{kl})
\nl &&{}
+ \sum_{V=A,Z,W^\pm} 
I_i^{\bar{V}}
I_j^{{V}} 
I_k^{\FZ}
I_l^{\FZ} 
\DD{0}(\MW;r_{ij})\deDD{0}(\MZ;r_{kl})
\nl &&{}
+ \frac{1}{2} 
I_i^{\FA}
I_j^{\FA} 
I_k^{\FA} 
I_l^{\FA}
\deDD{0}(0;r_{ij})\deDD{0}(0;r_{kl})
+
I_i^{\FZ}
I_j^{\FZ} 
I_k^{\FA} 
I_l^{\FA}
\deDD{0}(\MZ;r_{ij})\deDD{0}(0;r_{kl})
\Biggr\}\quad
\nl&&{}
-
\frac{1}{2} \mel{0}{}
\sum_{i=1}^{n}\sum_{j=1\atop j\neq i}^{n}
\Biggl\{
\sum_{V_1,V_2 = A,Z,W^\pm} 
I_i^{{V}_1}
I_j^{\bar{V}_2}
\,\betacoeff{V_1 V_2}^{(1)} \,J(\veps,\MW,Q^2)
\nl&&{}
+
I_i^{\FA}
I_j^{\FA}
\,\betacoeff{AA}^{(1)}
\left[
\Delta J(\veps,0,Q^2)
-\Delta J(\veps,0,\MW^2)
\right]
+
I_i^{\FA}
I_j^{\FA}
\,\betacoeff{\QED}^{(1)} \,\De J(\veps,0,\MW^2)
\Biggr\}.
\eeqar

In NLL accuracy, this result can be expressed in terms of the
lowest-order matrix element $\mel{0}{}$ and the one-loop correction
factors \refeq{onelooprepu} as
\beqar\label{twoloopresultunr}
\nmel{2}{\rF}&\NLLA&
\mel{0}{}
\Biggl\{
\frac{1}{2}\left[\FF{1}{\rF,\sew}\right]^2
+\FF{1}{\rF,\sew} \Delta \FF{1}{\rF,\elm}
+\FF{1}{\rF,\sew} \Delta \FF{1}{\rF,\PZ}
\nl&&{}
+\frac{1}{2}\left[\Delta \FF{1}{\rF,\elm}\right]^2
+\Delta \FF{1}{\rF,\PZ} \Delta \FF{1}{\rF,\elm}
+\GG{2}{\rF,\sew}
+\Delta \GG{2}{\rF,\elm}
\Biggr\}
.
\eeqar
Note that in NLL approximation $\FF{1}{\rF,\sew}$ and
$\De\FF{1}{\rF,\elm}$ do not commute, while $\De\FF{1}{\rF,\PZ}$
commutes with the other terms.  Using \refeq{chargeconservation},
\refeq{eq:betarel}, and \refeq{eq:betacoeffe}, the terms resulting
from the last two lines of \refeq{twoloopresultunr0} can be written as
\beqar\label{betatermsunr}
e^2\GG{2}{\rF,\sew}
&=& \frac{1}{2}\sum_{i=1}^{n}
\left[
\betacoeff{1}^{(1)} \gb^2 \left(\frac{Y_i}{2}\right)^2
+\betacoeff{2}^{(1)} \gw^2 C_{i}
\right]
J(\veps,\MW,Q^2)
,\nl
\Delta \GG{2}{\rF,\elm}
&=&
\frac{1}{2}\sum_{i=1}^{n}
Q_i^2
\Biggl\{
\betacoeff{e}^{(1)}
\left[
\Delta J(\veps,0,Q^2)
-\Delta J(\veps,0,\MW^2)
\right]
+\betacoeff{\QED}^{(1)}\,
\Delta J(\veps,0,\MW^2)
\Biggr\}
.\nln
\eeqar

Adding the term \refeq{twoloopPR} resulting from parameter
renormalization with $\DD{0}$ replaced by $I$ and using the definition
of $J$, \refeq{Jterms}, the functions $\GG{2}{\rF,\sew}$ and
$\De\GG{2}{\rF,\elm}$ in \refeq{twoloopresultunr} get replaced by the
functions $\GG{2}{\sew}$ and $\De\GG{2}{\elm}$ as defined in
\refeq{betaterms}.

Finally, when combining the wave-function counterterm
\refeq{twoloopWF} with the parts of \refeq{twoloopresultunr} involving
products of the functions $\FF{1}{\rF}$ these terms can be written in
the form given in \refeq{twoloopresult}. In order to arrive at this
result we write $\nmel{1}{\rF}$ in the form \refeq{oneloopLL} and $\de
Z^{(1)}_{k}$ in the form \refeq{subWFRC}. After arranging the Casimir
operators in an appropriate order, we can use global gauge invariance
\refeq{chargeconservation} to transform the wave-function counterterm
contributions to the form needed.

\end{appendix}

\end{document}